\documentclass[twocolumn,rmp,aps]{revtex4}
\usepackage{graphicx}
\usepackage{dcolumn}
\usepackage{bm}
\usepackage{dcolumn}
    \usepackage{amsmath}
    \usepackage{amssymb}
    \usepackage{fancyhdr}
     \usepackage{ulem}
\usepackage{pb-diagram}
    \usepackage[latin1]{inputenc}
\usepackage{dcolumn}
\newcommand{\mat}[1]{\begin{matrix} #1 \end{matrix}}

\newcommand{\angstrom}{A \kern -.4em\raise 1.4ex \hbox{\tiny °}}
\newcommand{\strikline}[1]{#1 \kern -0.5em\raise 0.5ex \hbox{--}}

\begin{document}                

\title{Exchange bias effect of ferro-/antiferromagnetic heterostructures }

\author{Florin Radu}
\email{florin.radu@rub.de} \affiliation{BESSY GmbH,
Albert-Einstein-Str. 15, 12489 Berlin, Germany}

\author{Hartmut Zabel}
\email{Hartmut.Zabel@rub.de}
 \affiliation{Department of Physics,
Ruhr-University Bochum, D 44780 Bochum, Germany}

\begin{abstract}
The exchange bias (EB) effect was discovered 60 years ago by
Meiklejohn and Bean. Meanwhile the EB effect has become an
integral part of modern magnetism with implications for basic
research and for numerous device applications. The EB effect was
the first of its kind which relates to an interface effect between
two different classes of materials, here between a ferromagnet and
an antiferromagnet. Here we  review fundamental aspects of the
exchange bias effect.
\end{abstract}
\maketitle
\tableofcontents

%
%
%


\section{Introduction}
The exchange bias (EB) effect was discovered 60 years ago by
Meiklejohn and Bean~\cite{bean:1956}. Meanwhile the EB effect has
become an integral part of modern magnetism with implications for
basic research and for numerous device applications. The EB effect
was the first of its kind which relates to an interface effect
between two different classes of materials, here between a
ferromagnet and an antiferromagnet. Later on the interlayer
exchange coupling between ferromagnets interleaved by paramagnet
layers was discovered~\cite{grunberg:1986}, and the proximity
effect between ferromagnetic and superconducting layers was
described~\cite{werthamer:1963,radovic:1991}. Recent reviews on
these topics can be found in Refs:~\cite{reviewIEC,
bergeret:2005,buzdin:2005} and also in this book. The EB effect
manifests itself in a shift of the hysteresis loop to negative or
positive direction with respect to the applied field. Its origin
is related to the magnetic coupling across the common interface
shared by a ferromagnetic (F) and an antiferromagnetic (AF) layer.
Extensive research is being carried out to unveil the details of
this effect, which has resulted in more then 600 publications in
the last five years and since the last comprehensive reviews by
Nogues and Schuller~\cite{nogues:1999}, Kiwi~\cite{kiwi:2001},
Berkowitz~\cite{berkowitz:1999}, and Stamps~\cite{stamps:2000}.

An EB bilayer consists of two key elements, with rather different
magnetic and structural properties: the ferromagnetic layer and
the antiferromagnetic layer. While the ferromagnetic layer can be
studied in detail by using laboratory equipment like SQUID, MOKE
and MFM, this is not the case for the magnetic interface and for
the antiferromagnetic layer. The interface embedded in between the
F and AF layer has low volume, therefore it is difficult to
separate its contribution from the F layer. Still, for the
exchange bias effect the interfacial magnetic properties are
essential for understanding the effect. For this purpose polarized
neutron scattering and soft-xray magnetic scattering techniques
can reveal some of the key magnetic properties of the interface.
The AF layer has in principle  no macroscopic magnetization, even
so the magnetic moment of individual atoms are rather high. The
magnetic properties of the AF materials are traditionally studied
by neutron diffraction. In thin films, due to the reduced AF
volume available for scattering, this method is rather difficult
to apply. Here,  soft-xray magnetic scattering through the linear
dichroism can reveal information about the magnetic properties of
the AF layer, therefore providing useful insights into the origin
of the EB effect.

From the application point of view the situation  appears to be
less complex. The effect is being used in spin valves with one
pinned and one free ferromagnetic layer which are embedded in
devices such as storage media, readout sensors, and magnetic
random access memory(MRAM). Nevertheless, robust, reliable and
easy to control
 functional elements based on the exchange bias phenomenon require more
 understanding of the fundamentals of the effect, which is further
 motivating  research in this field.

Because of recent experimental verifications of the existence of
interfacial layers  by several
groups~\cite{regan:2001,ohldag:2001,radu:2003:1,ohldag:2003,scholl:2004:2,roy:2005,kuch:2006,radu:jmmm:2006,ohldag:2006,hauet:2006,tusche:2006,valev:2006},
earlier models of EB need to be revisited and eventually modified
to take into account the effects, which are introduced by this
layer.

Here we first review some of the basic models for exchange bias.
We focus on numerical calculations and analytical treatment of
those models which are based on the Stoner-Wohlfarth
model~\cite{stoner:1947,stoner:1948}. This has the advantage that
analytical expressions can be derived and a numerical analysis can
be much more efficiently performed as compared to micromagnetic
simulations. It has, however, the disadvantage that only coherent
magnetization reversal processes are described within this
formalism. Nevertheless, for a large fraction of experimental
situations the Stoner-Wolhfarth approach is adequate. The behavior
of the F and AF spins during the magnetization reversal, as well
as the dependence of the critical fields on the parameters of the
F and AF layer are analyzed in detail. The  Meiklejohn and
Bean~\cite{bean:1956,bean:1957,bean:1962} model and the Mauri
models~\cite{mauri:1987:1} are revisited and numerical and
analytical expressions are compared. We continue with describing
the main features of the Random Field model of
Malozemoff~\cite{malozemoff:1987,malozemoff:1988:1,malozemoff:1988:2}
and Domain State(DS)
model~\cite{miltenyi:2000,nowak:2001,nowak:jmmm:2002,nowak:2002,beckmann:2003,misra:2004,scholten:2005,beckmann:2006}.
Then,we review the Kim and Stamp approach~\cite{kim:2005} which
focuses on spring-like behavior of the AF layers and coercivity
enhancement. We continue with one of the most recent models for
exchange bias, the Spin Glass (SG) model~\cite{radu:jpcm:2006}.
Assuming a realistic state of the interface between the F and AF
layers, the SG model describes well most of the important features
of EB heterostructures, including azimuthal dependence of exchange
bias field and coercivity, AF and F thickness dependence, the
inverse linear dependence on the lateral extension, and training
effects. Finally we will discuss recent experiments in the light
of the presented models. However, the main emphasis of this review
is to describe the basic models in a systematic fashion and to
compare them with recent experimental results.

{\section{Stoner-Wohlfarth model}}

The term anisotropy refers to the orientation of the magnetic
moments with respect to given geometrical directions. In bulk
materials the crystal axes are the reference directions, while in
thin films other reference systems become important. In order to
account for the orientation of the magnetic moments in magnetic
materials, the minimum energy state is provided by analysis of the
different contributing terms to the total magnetic energy: Zeeman
term, anisotropy terms, and exchange coupling terms. This
evaluation is performed by minimization of the total magnetic
energy with respect to various parameters.

\begin{figure}[!h]
\begin{center}\includegraphics[clip=true,keepaspectratio=true,width=.5\linewidth]{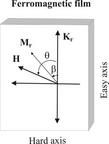}
\end{center}\caption[Definition of angles and vectors used in Stoner-Wohlfarth
model calculations.]{ \label{swfig1} Definition of angles and
vectors used in Stoner-Wohlfarth type model calculations. The
reference direction of the film is along the unidirectional
anisotropy.}
\end{figure}


In the following we use the simplest possible expression for the
total magnetic energy for a ferromagnetic thin film and calculate
the magnetic hysteresis loops.  We assume that all spins  are
confined in the film plane and that the film has a uniaxial
anisotropy. The response of the magnetization to an applied
magnetic field is then uniform. Therefore  the spins will
coherently rotate during the variation of the external field. The
direction of the magnetization can be described by only one
parameter, namely the $(\theta-\beta)$ angle  defining the
direction of the magnetization with respect to the applied field
(see Fig.~\ref{swfig1}). Many complexities of the magnetization
reversal are neglected in this approach. Nevertheless, the
Stoner-Wohlfarth (SW)
model~\cite{stoner:1947,stoner:1948,coehoorn:2001}, named after
the investigators who developed it for treating the magnetization
reversal of a small single domain,  is used with considerable
success for various magnetic thin films and heterostructures.

\begin{figure}[!h]
\begin{center}
\includegraphics[clip=true,keepaspectratio=true,width=1\linewidth]{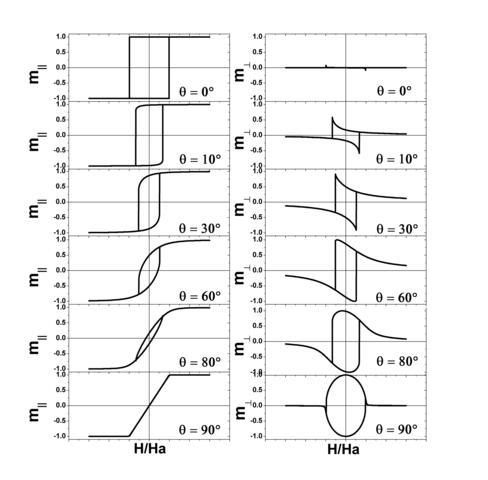}
\end{center}
\caption[The longitudinal and transverse components  of the
magnetization for a film with in-plane
anisotropy.]{\label{figFMHys1}  The longitudinal (left column) and
transverse (right column) components  of the magnetization for a
film with in-plane anisotropy. The curves are generated by
numerical evaluation of Eq.~\ref{sw3}.}

\end{figure}

\begin{figure}[!h]
\begin{center}
\includegraphics[clip=true,keepaspectratio=true,width=1\linewidth]{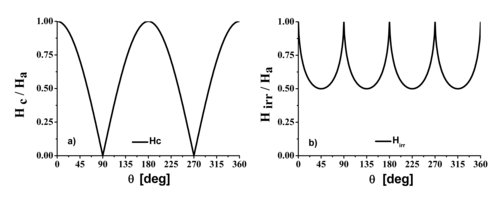}
\end{center}
\caption[The azimuthal dependence of the normalized coercive field
of a ferromagnetic film with uniaxial
anisotropy.]{\label{figFMHys2}  a) The azimuthal dependence of the
normalized coercive field of a ferromagnetic film with uniaxial
anisotropy. The curve is calculated with Eq.~\ref{swhc}. b)  The
azimuthal dependence of the normalized irreversible switching
field  of a ferromagnetic film with uniaxial anisotropy. The curve
is calculated using Eq.~\ref{sw6}.}
\end{figure}

The total magnetic energy per unit volume of a ferromagnetic film
with in-plane uniaxial anisotropy reads:
\begin{equation}
E_V(\beta)=-\mu_0\, H \, M_F \,cos(\theta-\beta)+K_{F} \,
\sin^2(\beta), \label{sw1}
\end{equation}
where the first term is the Zeeman energy contribution and the
second term is the magnetic crystalline anisotropy (MCA), here
assumed to have an uniaxial symmetry. The other parameters are:
$H$ for the applied field, $M_F$ for the saturation magnetization
of the ferromagnet, $K_F$ for the volume anisotropy constant of
the ferromagnet, and $\theta$ for the orientation of the applied
magnetic field with respect to the uniaxial anisotropy direction,
and $\beta$  the orientation of the magnetization vector during
the magnetization reversal.

The minimization of the total magnetic energy with respect to the
angle $\beta$ and the stability equation:
\begin{equation}
\frac{\partial E_V(\beta)}{\partial\beta}=0, \ \ \ \
\frac{\partial^2 E_V(\beta)}{\partial\beta^2} > 0
 \label{sw2}
\end{equation}
leads to the following equations:
\begin{eqnarray}
-\, \mu_0\, H \, M_F \, \sin(\theta-\beta)+K_{F} \,
\sin(2\,\beta)=0
 \label{sw3}\\
\mu_0\, H \, M_F \, \cos(\beta-\theta)+2 K_{F} \, \cos(2\,\beta)\,
> \,0
 \label{sw4}
\end{eqnarray}
By solving Eq.~\ref{sw3} with the condition imposed by the
Eq.~\ref{sw4}  one obtains the $\beta$ angle, which determines the
longitudinal component ($m_{||}=\cos(\beta-\theta)$) and the
transverse component ($m_{\perp}=\sin(\beta-\theta)$) of the
magnetization. Both components are  plotted in
Fig.~\ref{figFMHys1} for different  in-plane orientations
($\theta$). The evolution of the hysteresis loops for different
angles $\theta$ between the applied magnetic field and the
orientation of the uniaxial anisotropy is shown in the left column
of Fig.~\ref{figFMHys1} and reflects the typical behavior of thin
films with in-plane uniaxial anisotropy. Along the easy axis the
hysteresis loop is square shaped and the transverse component is
zero. When the applied field is perpendicular to the anisotropy
axis (hard axis), the hysteresis loop has a linear slope, whereas
the transverse component is ovally shaped.

The expression for the coercive field can easily be inferred from

Eq.~\ref{sw3}:

\begin{equation}H_c\,(\theta) \,=\, \frac{2\, K_F}{ \mu_0 \, M_F}\, |\cos{\theta}|\label{swhc}\end{equation}
For the hysteresis loops  shown in Fig.~\ref{figFMHys1}, the
coercive field   follows in detail the expression above. At the
position of the easy axis ($\theta \,= \,0, \, \pi$) the coercive
field is equal to the anisotropy field $H_a \, =\, 2 \, K_F /
\mu_0 \, M_F$, whereas along the hard axis ($\theta \,= \pm
\pi/2$) the coercive field is zero. Experimentally, this is an
often encountered situation. For instance,  polycrystalline
magnetic films grown on a-plane sapphire substrates show such
uniaxial and growth induced anisotropy due to steps at the
substrate surface.

Aside from the coercive field dependence as a function of the
azimuthal angle $\theta$, another critical field can be recognized
in Fig.~\ref{figFMHys1}. This is the field where the magnetization
changes irreversibly (i.e. where the hysteresis opens). This
irreversible switching field $H_{irr}$ can be extracted by solving
both Eqs.~\ref{sw3} and ~\ref{sw4}. Expressing the applied
magnetic field $\bm{H}$ by its components along the easy and hard
axis directions: $\bm{H}=(\, H_x, \, H_y\,)=(\, H\, \cos{\theta},
\, H\, \sin{\theta} \,)$, the solution of the system of equations
Eq.~\ref{sw3} and \ref{sw4} gives: $H_x=\, -\, H_a \cos^3 \beta$
and $H_y= \, + \, H_a \cos^3 \beta$. Eliminating $\beta$ from the
previous two equations we obtain the asteroid
equation~\cite{slonczewski:1956,thiaville:1998,coehoorn:2001}:
\begin{equation}
{|H_x|}^{2/3}+{|H_y|}^{2/3}=H_a^{2/3}\label{sw5}
\end{equation}
Now, introducing back into the equation above the expression for
the field components we obtain for the irreversible switching
field the following
expression~\cite{slonczewski:1956,thiaville:1998,coehoorn:2001}:
\begin{equation}
{H_{irr}\over H_a}= {1 \over
[(\sin^2{\theta})^{2/3}+(\cos^2{\theta})^{2/3}]^{3/2}}
 \label{sw6}
\end{equation}

\begin{figure}[!ht]
\begin{center}
\includegraphics[clip=true,keepaspectratio=true,width=1\linewidth]{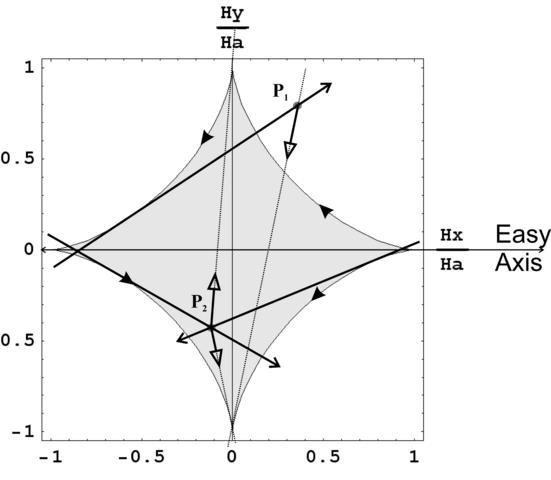}
\end{center}
\caption[The Asteroid]{\label{asteroid}The asteroid curve for a
film with uniaxial anisotropy. Two situations are depicted for
finding a geometrical solution to the Eq~\ref{sw5}: a) The
magnetic field represented as a point P1 lying outside the
asteroid region exhibits one stable solution (solid line with
filled arrow) and one unstable one (solid line with open arrow).
b) A magnetic field P2 within the asteroid curve exhibits four
solutions(see the tangent lines):two of them are stable (solid
line with filled arrow) and the other two  are unstable(solid line
with open arrow).The dotted line show tangents for the unstable
solutions~\cite{slonczewski:1956,thiaville:1998,coehoorn:2001}.}
\end{figure}
This field is plotted in the right panel of Fig.~\ref{figFMHys2}.
At the position of the easy axis ($\theta \,= \,0, \, \pi$) the
irreversible field is equal to the anisotropy field $H_a$, whereas
at $\theta \,=
 \pi/4,\,  3\pi/4$, the irreversible field is equal to half  of
 the anisotropy field ($H_{irr}\, (\pi/4)=H_a/2$). The irreversible switching field
 can be experimentally extracted from the transverse components
 of the magnetization, whereas the coercive field is extracted from
 the longitudinal component of the magnetization(see Fig.~\ref{figFMHys1}).

Fig.~\ref{asteroid} shows the so called asteroid curve
 which defines stability criteria for the
magnetization reversal(Eq.~\ref{sw5}). The asteroid method refers
to an elegant geometrical solution of Eq.~\ref{sw1} introduced by
Slonczewski~\cite{slonczewski:1956}. An extended analysis can be
found in Ref.~\cite{thiaville:1998}. The field measured in units
of H$_a$ appears  as a point  in Fig.~\ref{asteroid}. Given a
field P1 outside  the asteroid curve, two solutions can be found
by drawing tangent lines to the critical curve. Only one is a
stable solution and is given by  the tangent closest to the easy
axis, orienting the magnetization towards the field. For fields
inside the asteroid curve (P2) four tangents leading to four
solutions can drawn. Two solutions are stable and the other two
are unstable. The magnetization is stable oriented along the
corresponding tangent~\cite{slonczewski:1956,thiaville:1998}.

\sectionmark{Discovery of EB effect}
{\section{Discovery of the Exchange Bias effect}}
\sectionmark{Discovery of EB effect}
The exchange bias (EB) effect, also known as unidirectional
anisotropy, was discovered in 1956 by Meiklejohn and
Bean~\cite{bean:1956,bean:1957,bean:1962} when studying Co
particles embedded in  their native antiferromagnetic oxide CoO.
It was concluded from the beginning that the displacement of the
hysteresis loop is brought about by the existence of an oxide
layer surrounding the Co particles. This implies that the magnetic
interaction across their common interface is essential in
establishing the effect. Being recognized as an interfacial
effect, the studies of the EB  effect have been performed mainly
on thin films consisting of a ferromagnetic layer in contact with
an antiferromagnetic one. Recently, however, the lithographically
prepared structures as well as F and AF particles are studied with
renewed vigor.

\begin{figure}[!ht]
\begin{center}
\includegraphics[clip=true,keepaspectratio=true,width=1\linewidth]{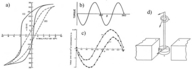}
\caption[Hysteresis loops of Co-CoO particles taken at
77°~K.]{\label{figEBean} a) Hysteresis loops of Co-CoO particles
taken at 77°~K. The dashed line shows the loop after cooling in
zero field. The solid line is the hysteresis loop measured after
cooling the system in a field of 10~kOe. b) Torque curve for Co
particles at 300~K showing uniaxial anisotropy. b) Torque curve of
Co-CoO particles taken at 77°K showing the unusual unidirectioal
anisotropy.
d) The torque magnetometer. The main component is a spring which
measures the torque as a function of the $\theta$ angle on a
sample placed in a magnetic field. \cite{bean:1956,bean:1957}}
\end{center}
\end{figure}

In Fig.~\ref{figEBean}  the original figures from
Ref.~\cite{bean:1956,bean:1957} show the shift of a hysteresis
loop of Co-CoO particles. The system was cooled from room
temperature down to 77~K through the N\'eel temperature of
CoO(T$_N$\, (CoO)\, =\, 291~K). The magnetization curve is shown
in Fig.~\ref{figEBean}a) as a dashed line. It is symmetrically
centered around zero value of the applied field, which is the
general behavior of ferromagnetic materials. When, however, the
sample is cooled in a positive magnetic field, the hysteresis loop
is displaced to negative values (see continuous line of
Fig.~\ref{figEBean}a). Such displacement did not disappear even
when extremely high applied fields of 70~000~Oe were used.

In order to get more insight into this unusual effect, the authors
studied the anisotropy behavior by using a self-made torque
magnetometer schematically shown  in  Fig.~\ref{figEBean}d). It
consists of a spring connected to a sample placed in an external
magnetic field. Generally, torque magnetometry is an accurate
method for measuring the magnetocrystalline anisotropy (MCA) of
single crystal ferromagnets. The torque on a sample is measured as
a function of the angle $\theta$ between certain crystallographic
directions and the applied magnetic field. In strong external
fields, when the magnetization of the sample is almost parallel to
the applied field (saturation), the torque is equal to:
$$T=-\frac{\partial E(\theta)}{\partial \theta},$$
where $E(\theta)$ is the MCA energy. In the case of Co, which has
a hexagonal structure, the torque about an axis perpendicular to
the c-axis follows a $\sin(2\theta)$ function as seen in
Fig.~\ref{figEBean}b). The torque and the energy density can then
be written as:
$$T=-K_1 \sin(2\theta)$$
$$E_V=\int K_1 \sin(2\theta)\,d\theta=K_1 \sin^2(\theta)+K_0,
$$
where $K_1$ is the MCA anisotropy and $K_0$ is an integration
constant. It is clearly seen from the energy expression that along
the c-axis, at $\theta=0$ and $\theta=180°$, the particles are in
a stable equilibrium. This typical case of a uniaxial anisotropy
is seen for the Co particles at room temperature, where the CoO is
in a paramagnetic state. At 77°K, after field cooling, the CoO is
in an antiferromagnetic state. Here, the torque curve of the
Co-CoO system looks completely different as seen in
Fig.~\ref{figEBean}c). The torque curve is a function of
$\sin(\theta)$:
 \begin{equation}
T=-K_u \sin(\theta),\label{torque1}\end{equation}
 hence,
\begin{equation}
E_V=\int K_u \sin(\theta)\,d\theta=-K_u \cos(\theta)+K_0,
\label{torque2}
\end{equation}
The energy function shows that the particles are in equilibrium
for one position only, namely $\theta=0$. Rotating the sample to
any angle, it tries to return to the original position. This
direction is parallel to the field cooling direction and such
anisotropy was named {\textit{unidirectional anisotropy}}.

Now, one can analyze whether the same unidirectional anisotropy
observed by torque magnetometry is also responsible for the loop
shift. In Fig.~\ref{swfig1} are shown schematically the vectors
involved in writing the energy per unit volume for a ferromagnetic
layer with uniaxial anisotropy having the magnetization oriented
opposite to the field. It reads:
\begin{equation}
E_V=-\mu_0\, H \, M_F \,  \cos(-\beta)+K_{F} \, \sin^2(\beta)
\label{eqebiferro}
\end{equation}
where $H$ is the external field, $M_F$ is the  saturation
magnetization of the ferromagnet per unit volume, and $K_F$ is the
MCA of the F layer. The two terms entering in the formula above
are the Zeeman interaction energy of the external field with the
magnetization of the F layer and the MCA energy of the F layer,
respectively. Now, writing the stability conditions and assuming
that the field is parallel to the easy axis, we find that the
coercive field is:
\begin{equation}
H_c= 2 \, K_F/\mu_0 \, M_F
\end{equation}
Next step is to cool the system down in an external magnetic field
and to introduce in Eq.~\ref{eqebiferro} the unidirectional
anisotropy term. The expression for the energy density then
becomes:
\begin{equation}
E_V=-\mu_0\, H \, M_F \, \cos(-\beta)+K_{F} \, \sin^2(\beta)-K_u\,
\cos(\beta)
\label{eqebideal}
\end{equation}
 We notice that the solution is identical to the previous case
 (Eq.~\ref{eqebiferro}) with the substitution of an effective field: $H'=H+K_u/M_F$.
 This causes the hysteresis loop to be shifted by $-K_u/ \mu_0 \, M_F$. Thus,
 Meiklejohn and Bean concluded that the loop displacement is equivalent to the
 explanation for the unidirectional anisotropy.

Besides the shift of the magnetization curve and the
unidirectional anisotropy, Meiklejohn and Bean have observed
another effect when measuring the torque curves. Their experiments
revealed an appreciable hysteresis of the torque (see Fig.~9 and
Fig.~10 of Ref~\cite{bean:1957} and Fig.~2 of Ref~\cite{bean:1962}
), indicating that irreversible changes of the magnetic state of
the sample take place when rotating the sample in an external
magnetic field. As the system did not display any rotational
anisotropy when the AF was in the paramagnetic state, this
provided evidence for the coupling between the AF CoO shell and
the F Co core. Such irreversible changes were suggested to take
place in the AF layer.

\sectionmark{Phenomenology}
 {\section{Ideal model of the exchange bias: Phenomenology}}
\sectionmark{Phenomenology}

\begin{figure}[!ht]
\begin{center}
\includegraphics[clip=true,keepaspectratio=true,width=1\linewidth]{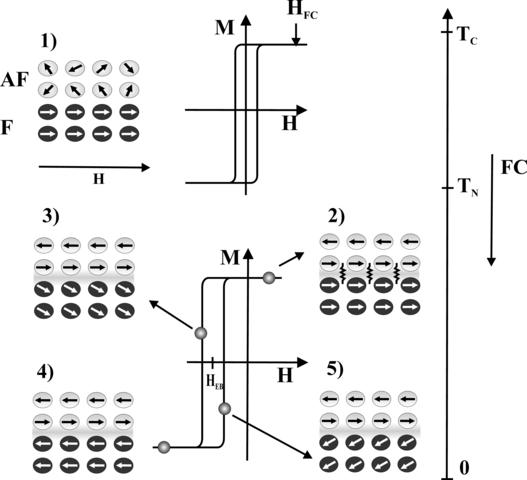}
\caption[Phenomenological model of exchange bias for an AF-F
bilayer.]{\label{figPheno1} Phenomenological model of exchange
bias for an AF-F bilayer. 1) The spin configuration at a
temperature which is higher than $T_N$ and smaller than $T_C$. The
AF layer is in a paramagnetic state while the F layer is ordered.
Its magnetization curve (top-right) is centered around zero value
of the applied field. Pannel 2): the spin configuration of the AF
and F layer after field cooling the system through $T_N$ of the AF
layer in a positive applied magnetic field ($H_{FC}$). Due to
uncompensated spins at the AF interface, the F layer is coupled to
the AF layer. Panel 4): the saturated state at negative fields.
Panel 3) and 5) show the configuration of the spins during the
remagnetization, assuming that this takes place through in-plane
rotation of the F spins. The center of  magnetization curve is
displaced at negative values of the applied field by $H_{eb}$.
(The description is in accordance with
Ref.~\cite{bean:1956,bean:1957,nogues:1999})}
\end{center}
\end{figure}

The macroscopic observation  of the magnetization curve shift due
to unidirectional anisotropy of a F/AF bilayer can qualitatively
be understood by analyzing the microscopic magnetic state of their
common interface. Phenomenologically, the onset of exchange bias
is depicted in Fig.~\ref{figPheno1}. A ferromagnetic layer is in
close contact to an antiferromagnetic one. Their critical
temperatures should satisfy the condition: $T_C>T_N$, where $T_C$
is the Curie temperature of the ferromagnetic layer and $T_N$ is
the N\'eel temperature of the antiferromagnetic layer. At a
temperature which is higher than the N\'eel temperature of the AF
layer and lower than the Curie temperature of the ferromagnet
($T_N<T<T_C$), the F spins align along the direction of the
applied field, whereas the AF spins remain  randomly oriented in a
paramagnetic state (see Fig~\ref{figPheno1}~1)). The hysteresis
curve of the ferromagnet is centered around zero, not being
affected by the proximity of the AF layer. Next, we saturate the
ferromagnet by applying a high enough external field $H_{FC}$ and
then, without changing the magnitude or direction of the applied
field, the temperature is decreased to a finite value lower then
$T_N$ ({\textit{field cooling procedure}}).

After field cooling the system, due to exchange interaction at the
interface the first monolayer of the AF layer will  align parallel
(or antiparallel) to the F spins. The next monolayer of the
antiferromagnet will align antiparallel to the previous layer as
to complete AF order, and so on (see Fig~\ref{figPheno1}-2). Note
that the spins at the AF interface are uncompensated, leading to a
finite net magnetization of this monolayer. It is assumed that
both the ferromagnet and the antiferromagnet are in a single
domain state and that they will remain in this single domain state
during the magnetization reversal process. When reversing the
field, the F spins will try to rotate in-plane to the opposite
direction. Being coupled to the AF spins,  it takes a bigger force
and therefore a stronger external field to overcome this coupling
and to rotate the ferromagnetic spins. As a result, the first
coercive field is higher than it used to be at $T>T_N$, where the
F/AF interaction is not yet active. On the way back from negative
saturation to positive field values (Fig~\ref{figPheno1}~4)), the
F spins require a smaller external force in order to rotate back
(Fig~\ref{figPheno1}~5)) to the original direction. A torque is
acting on the F spins for  all other angles, except the stable
direction which is along the field cooling direction
(unidirectional anisotropy). As a result, the magnetization curve
is shifted to negative values of the applied field. This
displacement of the center of the hysteresis loop is called
{\textit{exchange bias field}}, and is negative in relation to the
orientation of the F spins after field cooling ({\textit{negative
exchange bias}}). It should be noted that in this simple
description the AF spins are considered to be rigid and fixed to
the field cooling direction during the entire reversal process.

\sectionmark{The ideal M\&B model}
{\section{The ideal Meiklejohn-Bean model: Quantitative
Analysis\label{MBmodel1}}}
\sectionmark{The ideal M\&B model}

Based on their observation about the rotational anisotropy,
Meiklejohn and Bean proposed a model to account for the magnitude
of the hysteresis shift. The assumptions made are the
following~\cite{bean:1956,nogues:1999,coehoorn:2001}:

\begin{itemize}
\item{The F  layer rotates rigidly, as a whole} \item{Both the F
and AF are in a single domain state} \item{The AF/F interface is
atomically smooth.} \item{The AF layer is magnetically rigid,
meaning that the AF spins remain unchanged
 during the rotation of the F spins }
\item{The spins of the AF interface are fully uncompensated: the
interface layer has a net magnetic moment } \item{The F and the AF
layers are coupled by an exchange interaction across the F/AF
interface. The parameter assigned to this interaction is the
interfacial exchange coupling energy per unit area $J_{eb}$}
\item{The AF layer has an in-plane uniaxial anisotropy}

\end{itemize}

\begin{figure}[!ht]
\begin{center}
\includegraphics[clip=true,keepaspectratio=true,width=.5\linewidth]{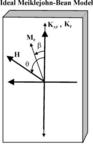}
\caption[Schematic view of the angles and vectors used in the
ideal Meiklejohn and Bean model.]{ Schematic view of the angles
and vectors used in the ideal Meiklejohn and Bean model. The AF
layer is assumed to be rigid and no deviation from its initially
set orientation is allowed. $\bm{K}_{AF}$ and $\bm{K}_{F}$ are the
anisotropy of the AF layer and F layer, respectively, which are
assumed to be parallel oriented to the field cooling direction.
$\beta$ is the angle between magnetization vector $\bm{M}_F$ of
the F layer and the anisotropy direction of the F layer. This
angle is variable during the magnetization reversal. $\bm{H}$ is
the external magnetic field which can be applied at any direction
$\theta$ with respect to the field cooling direction at $\theta=0$
(see
Ref.~\cite{bean:1956,bean:1957,nogues:1999}).}\label{figMBrot1}
\end{center}
\end{figure}

In general, for describing the coherent rotation of the
magnetization vector the
Stoner-Wohlfarth~\cite{stoner:1947,stoner:1948} model is used.
Different energy terms can be added as needed and to best account
for the quantitative and qualitative behavior of the macroscopic
magnetization reversal. In Fig.~\ref{figMBrot1} is shown
schematically the geometry of the vectors involved in the ideal
Meiklejohn and Bean model. ${H}$ is the applied magnetic field,
which makes an angle $\theta$ with respect to the field cooling
direction denoted by  $\theta=0$, ${K}_{F}$ and ${K}_{AF}$ are the
uniaxial anisotropy directions of the F and the AF layer,
respectively. They are assumed to be oriented parallel to the
field cooling direction. ${M}_F$ is the magnetization orientation
of the F spins during the magnetization reversal.
It is assumed that the AF spins are fixed to their orientation
defined during the field cooling procedure (rigid AF). In the
analysis below the angle ($\theta\, =\, 0$) for the applied field
is assumed to be parallel to the field cooling direction. This
condition refers to the direction along which the hysteresis loops
is measured , whereas $\theta\ne0$ is used for torque measurements
or for measuring the azimuthal dependence of the exchange bias
field.

Within this model the energy per unit area assuming coherent
rotation of the magnetization, can be written
as~\cite{bean:1956,nogues:1999,stamps:2000}:
\begin{equation}
E_A=-\mu_0\, H \, M_F \, t_F \, \cos(-\beta) + K_{F} \, t_{F}\,
\sin^2(\beta) -J_{eb} \, \cos(\beta)
\label{eqebi3}
\end{equation}
where $J_{eb}\, [J/m^2]$ is the interfacial exchange energy per
unit area, and $M_F$ is the saturation magnetization of the
ferromagnetic layer. The interfacial exchange energy  can be
further expressed in terms of pair exchange interactions:
$E_{int}=\sum_{ij} J_{i\, j} \bm{S}_i^{AF} \bm{S}_j^{F} $, where
the summation includes all interactions within the range of the
exchange
coupling~\cite{malozemoff:1987,malozemoff:1988:2,gokemeijer:1997,takano:1997}.

 The stability condition
$\partial E_A\, / \, \partial \theta=0$ has two types of
solutions: one is $\beta=\cos^{-1}[(J_{eb}-\mu_0 \, H \, M_F \,
t_F)/(2 \, K_F)]$ for $\mu_0 \, H \, M_F \, t_F-J_{eb} \le 2 \,
K_F$,
; the other one is $\beta=0,\pi$ for $\mu_0 \, H \, M_F \,
t_F-J_{eb} \ge 2 \, K_F$, corresponding to positive and negative
saturation, respectively. The coercive fields $H_{c1}$ and
$H_{c2}$ are extracted from the stability equation above for
$\beta=0, \pi$:
\begin{equation}
H_{c1} =- \frac{2 \, K_F t_F+J_{eb} }{\mu_0 M_F \, t_F}
\label{eqebi4}
\end{equation}
\begin{equation}
H_{c2} = \frac{2 \, K_F t_F-J_{eb} }{ \mu_0 M_F \, t_F}
\label{eqebi5}
\end{equation}

Using the expressions above, the coercive field $H_{c}$ of the
loop and the displacement $H_{eb}$
 can be calculated according to:
\begin{equation}
H_{c}=\frac{-H_{c1}+H_{c2}}{2}\, ~\textrm{and}\,\,
H_{eb}=\frac{H_{c1}+H_{c2}}{2} \label{EXPRHcEB}
\end{equation}
which further gives:
\begin{equation}
H_{c}=\frac{2 \, K_F}{\mu_0 M_F}
\label{eqebi6}
\end{equation}
and
\begin{equation}
H_{eb}=\, - \, \frac{J_{eb}}{\mu_0 M_F \, t_F}
\label{eqebi7}
\end{equation}

The equation Eq.~\ref{eqebi7} is the master formula of the EB
effect. It gives the expected characteristics of  the hysteresis
loop for an ideal case, in particular the linear dependence on the
interfacial energy $J_{eb}$ and the inverse dependence on the
ferromagnetic layer thickness. Therefore this equation serves as a
guideline to which experimental values are compared. In the next
section we will discuss some predictions of the model above.

\subsection{The sign of the exchange bias}

Eq.~\ref{eqebi7} predicts that the sign of the exchange bias is
negative. Almost all hysteresis loops shown in the literature are
shifted oppositely to the field cooling direction. The positive or
negative exchange coupling across the interface produces the same
sign of the exchange bias field. There are, however, exceptions.
Positive exchange bias was observed  for CoO/Co,
Fe$_x$Zn$_{1-x}$F$_2$/Co and Cu$_{1-x}$Mn$_{x}$/Co bilayers when
the measuring temperature was close to the blocking
temperature~\cite{radu:2003:1,gredig:2002,prados:2002,shi:2005,ali:2007}.
At low temperatures positive exchange bias was observed in
Fe/FeF$_2$~\cite{nogues:1996} and Fe/MnF$_2$~\cite{leighton:1999}
bilayers. Specific of the last two systems is the low anisotropy
of the antiferromagnet and the antiferromagnetic type of coupling
between the F and AF layers. It was proposed that, at high cooling
fields, the interface layer of the antiferromagnet aligns
ferromagnetically with the external applied field and therefore
ferromagnetically with the F itself. As  the preferred orientation
between the interface spins of the F layer and AF layer is the
antiparallel one (AF coupling), the EB becomes positive. Further
theoretical and experimental details of the positive exchange bias
mechanism are presented in
Ref.~\cite{hong:1998,kagerer:2000,kirk:2002}. In the original
Meiklejohn and Bean model the interaction of the cooling field
with the AF spins is not taken into account. However this
interaction can be easily introduced in their model. The positive
exchange bias could also be accounted for  in the M\&B model by
simply changing the sign of $J_{eb}$ in Eq.~\ref{eqebi3} from
negative to positive.

 \subsection{The magnitude of the EB}\label{magnitudeEB}

Often the exchange coupling parameter $J_{eb}$ is identified with
the exchange constant of the AF layer ($J_{AF}$). For various
calculations a value ranging from $J_{AF}$ to J$_{F}$ was assumed.
For CoO,  $J_{AF}=21.6K=1.86$~meV~\cite{rechtin:1972}. Using this
value, the expected exchange coupling constant $J_{eb}$ of a
CoO(111)/F layer can be estimated
as~\cite{malozemoff:1987,miltenyiphd:2000}:
\begin{equation}
J_{eb}=N J_{AF}/A= 4 \, mJ/m^2, \label{jebest}
\end{equation}
where $N\,=\,4$ is the number of  Co$^{2+}$ ions at the
uncompensated CoO interface per unit area A=$\sqrt(3)\, a^2$, and
$a=4.27~\angstrom$ is the CoO lattice parameter. With this number
we would expect for a $100$~\AA \, thick Co layer, which shares an
interface with a CoO AF layer, an exchange bias of:
\begin{equation}
H_{eb}\, [Oe]=\frac{J_{eb}\, [J/m^2]}{M_F \, [kA/m] \, t_F\,
[\angstrom]}\, 10^{11}
\end{equation}
$$H_{eb} =\frac{0.004} {1460 \times 100} \ 10^{11}= 2740\ Oe $$

This exchange bias field is by far bigger than experimentally
observed. So far an ideal magnitude of the EB field as predicted
by the Eq.~\ref{jebest} has not yet been observed, even so for
some bilayers high EB fields were measured (see
Table~\ref{ebitable1}). We encounter here two problems: first, we
do not know how to evaluate the real coupling constant $J_{eb}$ at
the interfaces with variable degrees of complexity, and the
second, in reality interfaces are never atomically smooth. The
unknown interface was nicely labelled by Kiwi~\cite{kiwi:2001} as
"a hard nut to crack". Indeed, the features of the interfaces may
be complex regarding the structure, the roughness, the magnetic
properties, and domain state of the AF and F layers.

\onecolumngrid
\begin{table}[!ht]
 \caption[Experimental values related to Co/CoO exchange bias systems.]{\label{ebitable1}
 Experimental values related to Co/CoO exchange bias systems. 
 The symbols used in the table  are:
 {\textit{ebe}}-electron-beam evaporation,
 {\textit{rsp}}-reactive sputtering,
 {\textit{msp}}-magnetron sputtering,
 {\textit{mbe}}-molecular beam epitaxy,
F-ferromagnet, AF antiferromagnet, $t_{AF}$-the thickness of the
AF, $t_{F}$-the thickness of the F, $H_{eb}$-measured exchange
bias field, $H_{c}$-measured coercive field,
 $T_{mes}$-measured blocking temperature,
 $T_{mes}$-the measuring temperature, $J_{eb}$-the coupling energy
 extracted from the experimental value of exchange bias field ($J_{eb}\,=\, H_{eb}\, (\mu_0 M_F t_F)$). }
\begin{center}
 \begin{tabular}{|l|l|l|l|l|l|l|l|l|l|l|} 
\hline
AF& F&$t_{AF}$ & $t_{F}$ & $H_{eb}$& $H_{c}$& $T_B$ & $T_{mes}$ & $J_{eb}$ & Ref\\
& & [\AA]&[\AA] &[Oe] &[Oe]&[K]&[K]&[mJ/m$^2$]& \\
\hline
CoO (air)&Co(rsp)&20&40&-3000&NA&-&4.2&1.75& \cite{miller:1996}\\
CoO (air)&Co(rsp)&25&27&-2321&3683&180&10&0.91& \cite{radu:2003:1}\\
CoO (air)&Co(rsp)&25&56&-1073&1751&180&10&0.88& \cite{radu:2003:1}\\
CoO (air)&Co(rsp)&25&87&-675&1315&180&10&0.86& \cite{radu:2003:1}\\
CoO (air)&Co(rsp)&25&119&-557&901&180&10&0.97& \cite{radu:2003:1}\\
CoO (air)&Co(rsp)&25&153&-443&789&180&10&0.99& \cite{radu:2003:1}\\
CoO (air)&Co(rsp)&25&260&-251&427&180&10&0.95& \cite{radu:2003:1}\\
CoO (air)&Co(rsp)&25&320&-202&346&180&10&0.94& \cite{radu:2003:1}\\
CoO (air)&Co(rsp)&25&398&-174&290&180&10&1.00& \cite{radu:2003:1}\\
CoO (air)&Co(msp)&33&139&-145&325&-&5&0.29& \cite{velthuis:2000}\\
CoO (air)&Co(msp)&33&139&-50&NA&-&30&0.1& \cite{welp:2003}\\
CoO (rsp)&Co(rsp)&20&150&-25&295&-&20&0.055& \cite{gredig:2002}\\
$[$CoO (rsp)&Co(rsp)$]_{x 25}$&70&37&-2500&5000&-&5&13.5!& \cite{radu:2002:2}\\
CoO (in-situ)&Co(ebe)&20&160&-220&330&180&10&0.51& \cite{gruyters:2000}\\
CoO(111)(mbe)&Fe(110)(mbe)&200&150&-150&520&291&10&0.4& sec.~\cite{gnowak:2007,RaduDiss}\\
\hline
 \end{tabular}
 \end{center}
 \end{table}
\twocolumngrid

In Table~\ref{ebitable1} are listed some EB data of systems with
CoO as the AF layer. We focus on experimentally determined
interfacial exchange coupling constants using $J_{eb}= H_{eb}\,
\mu_0\, M_F \, t_F$. The observed exchange coupling constant is
usually smaller then the expected value of $4~mJ/m^2$ for CoO/Co
bilayers by a factor ranging from 3 to several orders of
magnitude. One anomaly is seen for the multilayer system Co/CoO
which is actually 3 times higher then the expected value of $4 \,
mJ/m^2$, and to our knowledge is the highest value observed
experimentally~\cite{radu:2002:2}. Such a variation of the
experimental values for the interfacial exchange coupling constant
is motivating further considerations of the mechanisms controlling
the EB effect.

\subsection{The 1/t$_F$ dependence of the EB field}

Eq.~\ref{eqebi7}  predicts that the variation of the EB field is
proportional to the inverse thickness of the ferromagnet:
\begin{equation}
  H_{eb}\, \approx\,\frac{1}{t_F}
\label{tlaw}
\end{equation}
This dependence was subject of a large number of experimental
investigations \cite{nogues:1999}, because it is associated with
the interfacial nature of the exchange bias effect. For the CoO/Co
bilayers   no deviation was observed \cite{gruyters:2000}, even
for very low thicknesses (2~nm) of the Co layer\cite{radu:2003:1}.
For other systems with thin F layers of the order of several
nanometers it was observed that the $1/t_F$ law is not closely
obeyed~\cite{nogues:1999}. It was suggested that the F layer is no
longer laterally continuous \cite{nogues:1999}. Deviations from
$1/t_F$ dependence for the other extreme when the F layer is very
thick were observed as well~\cite{nogues:1999}. For this regime it
is assumed that for F layers thicker than the domain wall
thickness (500nm for permalloy), the F spins may vary appreciably
across the film upon the magnetization reversal~\cite{tsang:1981}.

\subsection{Coercivity and exchange bias}

According to Eq.~\ref{eqebi6} the coercivity of the magnetic layer
is the same with and without exchange bias effect. This
contradicts experimental observations. Usually an increase of the
coercive field is observed.

\section{Realistic Meiklejohn and Bean model}\label{MBmodel2}

\begin{figure}[!ht]
\begin{center}
\includegraphics[clip=true,keepaspectratio=true,width=.5\linewidth]{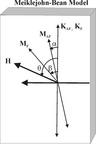}
\caption[Schematic view of the angles and vectors used for the
 Meiklejohn and Bean model.]{\label{figMBrot2} Schematic view of the angles and
vectors used for the
 Meiklejohn and Bean model, allowing a rotation $\alpha$ of the AF layer
 as a whole with respect to the initially set orientation.
 $\bm{M}_{AF}$ is the sublattice magnetization of the AF
layer. $\bm{K}_{AF}$ and $\bm{K}_{F}$ are the anisotropy of the AF
layer and F layer, respectively, which are assumed to be parallel
oriented to the field cooling direction. $\beta$ is the angle
between F magnetization vector $\bm{M}_F$ and the anisotropy
direction of the F layer. This angle is variable during the
magnetization reversal. $\bm{H}$ is the external magnetic field
which can be applied at any direction $\theta$ with respect to the
field cooling direction at $\theta=0$
(Ref.~\cite{bean:1956,bean:1957, bean:1962,nogues:1999}).}
\end{center}
\end{figure}

In Ref.~\cite{bean:1957} a new degree of freedom for the AF spins
was introduced: the AF is still rigid, but it can slightly rotate
during the magnetization reversal as a whole as indicated in
Fig.~\ref{figMBrot2}. This parameter was introduced in order to
account for the rotational hysteresis observed during the torque
measurements. Allowing the AF layer to rotate is not in
contradiction to the rigid state of the AF layer, because it is
allowed only to rotate as a whole. Therefore, the fourth
assumption of the ideal M\&B model in section~\ref{MBmodel1} is
removed. The  new condition for the AF spins is: $\alpha \ne0$.
With this new assumption, the equation Eq.~\ref{eqebi3}
reads~{\cite{bean:1962,nogues:1999}}:

\begin{eqnarray}
E_A&=&-\mu_0\, H \, M_F \, t_F \, \cos(\theta-\beta)\nonumber\\
& &+ K_{F} \, t_{F}\, \sin^2(\beta) + K_{AF} \, t_{AF}\, \sin^2(\alpha)\nonumber\\
& &-J_{eb} \, \cos(\beta-\alpha),\label{GMBeq2}
\end{eqnarray}
where $t_{AF}$ is the thickness of the antiferromagnet, and
$K_{AF}$ is the MCA of the AF layer per unit area. The new energy
term in the equation above as compared to Eq.~\ref{eqebi3} is the
anisotropy energy of the AF layer.

The Eq.~\ref{GMBeq2}  above can be analyzed numerically by
minimization of the energy in respect to the $\alpha$ and the
$\beta$ angles. Below we will perform a numerical analysis of
Eq.~\ref{GMBeq4} and highlight a few of the conclusions discussed
in Ref.~\cite{bean:1957,bean:1962,coehoorn:2001}. The minimization
with respect to $\alpha$ and $\beta$ leads to a system of two
equations:

\begin{equation}\label{GMBeq4}
  \lbrace
 \begin{matrix}
 \frac{H}{H_{eb}^{\infty}} \, \sin(\theta-\beta) \,
 + \, \sin(\beta-\alpha)=0 \\  R \, \sin(2\, \alpha)-\sin(\beta-\alpha) = 0
 \end{matrix}
\end{equation}
where
\begin{equation}
H_{eb}^{\infty}\, \equiv \, -\frac{J_{eb}}{\mu_0\, M_F \, t_F \,}
\label{H_infi}
\end{equation}
is the value of the exchange bias field when the anisotropy of the
AF is infinitely large, and
\begin{equation}
R\, \equiv \, \frac{K_{AF} \, t_{AF}}{J_{eb}},
\label{Rvalue}
\end{equation}
is the parameter defining the ratio between the AF anisotropy
energy and the interfacial exchange energy $J_{eb}$. As we will
see further below, exchange bias is only observed, if the AF
anisotropy energy is bigger than the exchange energy. The unknown
variables $\alpha$ and $\beta$ are numerically extracted as a
function of the applied field $H$. Note that for clarity reasons
the anisotropy of the ferromagnet was neglected ($K_F=0$) in the
system of equation above. As a result the coercivity, which will
be discussed further below, is not related to the F layer anymore,
but to the AF layer alone. Also, in order to simplify the
discussion we consider first the case $\theta=0$, which
corresponds to measuring a hysteresis loop parallel to the field
cooling direction.

\begin{figure}[!ht]
\begin{center}
\includegraphics[clip=true,keepaspectratio=true,width=.48\linewidth]{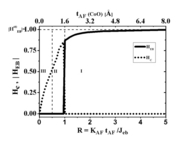}
\includegraphics[clip=true,keepaspectratio=true,width=.48\linewidth]{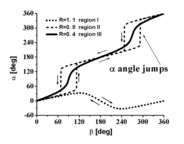}
\caption[The phase diagram of the exchange bias field and the
coercive fields given by the
 Meiklejohn-Bean formalism.]{\label{figMBPhaseDiagram1} Left: The phase diagram of
the exchange bias field and the coercive fields as a function of
the Meiklejohn-Bean parameter $R$. Right: Typical behavior of the
antiferromagnetic  angle $\alpha$ for the three different regions
of the phase diagram. Only region I can lead to a shift of the
hysteresis loop. In the other two regions a coercivity is observed
but no exchange bias field. }
\end{center}
\end{figure}

\begin{figure}[!ht]
\begin{center}
\includegraphics[clip=true,keepaspectratio=true,width=1\linewidth]{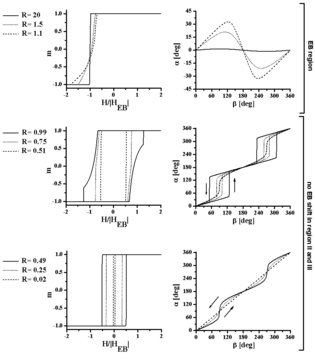}
\caption[Several hysteresis loops and antiferromagnetic spin
orientation  during the magnetization
reversal.]{\label{figMBalphaPhaseDiagram1} Simulation of several
hysteresis loops and antiferromagnetic spin orientations  during
the magnetization reversal. For the simulation we used the
Meiklejohn-Bean formalism. Top row shows three hysteresis loops
calculated for different R ratios within region I. The graph in
the last column to the right shows the $\alpha$ angle of the AF
layer for the three R values. The middle raw shows corresponding
hysteresis loops and $\alpha$ angles for R values in region II.
The bottom row shows simulations for region III. Note that the
scales for the $\alpha$ angle in the top panel is enlarged
compared to those in the lower two panels.}
\end{center}
\end{figure}

Numerical evaluation of the Eqs.~\ref{GMBeq4} yields the angles:
\begin{itemize}
\item{$\alpha$ of the AF spins as a function of the applied field
during the hysteresis measurement} \item{$\beta$ of the F spins
which rotate coherently during their reversal}
\end{itemize}
The $\beta$ angle defines completely the hysteresis loop and at
the same time the coercive fields $H_{c1}$ and $H_{c2}$. These
fields, in turn, define the coercive field $H_c$ and the exchange
bias field $H_{eb}$ (see Eq.~\ref{EXPRHcEB}). The $\alpha$ angle
influences the shape of the hysteresis loops when the $R$-ratio
has low values, as we will see below. For high R values the
rotation angle of the antiferromagnet is close to zero, giving a
maximum exchange bias field equal to $H_{eb}^{\infty}$.

The properties of the EB system originate from  the properties of
the AF layer, which are accounted for by  one parameter, the
$R$-ratio. We will consider the effect of the $R$-ratio on the
angels  $\beta$ and $\alpha$ which, as stated above, define the
macroscopic behavior and the critical fields of the EB systems.

Numerical simulations of Eq.~\ref{GMBeq4} as a function of
$R$-ratio are shown in Fig.~\ref{figMBPhaseDiagram1} and
Fig.\ref{figMBalphaPhaseDiagram1}. We distinguish three physically
distinct
regions~\cite{bean:1957,bean:1962,fujiwara:1999,coehoorn:2001}:


\begin{itemize}
\item{I. $R\ge1$\\ In this region the coercive field is zero and
the exchange bias field is finite, decreasing from the asymptotic
value $H_{eb}^\infty$ to the lowest finite value at $R=1$. The AF
spins rotate reversibly during the complete reversal of the F
spins. The $\alpha$ angle has a maximum value as a function of the
$R$-ratio, ranging from approximatively zero for $R=\infty$ to
$\alpha=45°$ at $R=1$.  Notice that as the maximum angle of the AF
spins increases, a slight decrease of the exchange bias field is
observed. When the $R$-ratio approaches the critical value of
unity, the exchange bias has a minimum. In the phase diagram, only
range I can cause a shift of the magnetization curve. Simulated
hysteresis loops for three different values of the $R$-ratio are
shown in Fig.~\ref{figMBalphaPhaseDiagram1}. One notices that not
only the size of the exchange bias field decreases when the
$R$-ratio approaches unity, but also the shape of the hysteresis
curve is changing. At high  $R$-ratios  the reversal is rather
sharp, whereas for $R$-ratios close to unity it become more
extended, almost resembling a spring-like behavior.}
\item{II. $0.5\le R<1$\\
Characteristic for this region is that the AF spins are no more
reversible. They follow the F spins and they change direction
irreversibly, causing a coercive field at the expense of the
exchange bias field, which becomes zero. Furthermore, depending on
the field sweeping direction, there is a hysteresis-like behavior
of the AF spin rotation. At a critical angle $\beta$ of the F spin
rotation, the AF spins cannot withstand the torque by the coupling
to the F spins and they jump in a discontinuous fashion to another
angle ({\textit{jump angle}}). The hysteresis loops corresponding
to this region (see Fig.~\ref{figMBalphaPhaseDiagram1}) are
drastically different from the previous case. The coercivity shows
strong dependence as function of the $R$-ratio and they are not
shifted at all. Moreover, the AF {\textit{jump angles}} are
clearly visible as kinks in the hysteresis shape during the
reversal.}

\item{III. $R<0.5$\\
This region preserves the features of the previous one with one
exception, namely that the AF spins follow reversibly the F spins,
without any jumps. Therefore, no hysteresis-like behavior of the
$\alpha$ angle is seen. The exchange bias field is zero and the
coercive field is finite, depending on the $R$-ratio. The
hysteresis loops shown in Fig.~\ref{figMBalphaPhaseDiagram1} are
quite similar to a ferromagnet with uniaxial anisotropy.  Within
the Stoner-Wolhfarth model the resultant coercive field can be
roughly approximated  as~\cite{coehoorn:2001}: $H_c \approx 2\,
K_{AF}\, t_{AF}/(\mu_0 \, M_F \, t_F)$. }
\end{itemize}


It is easy to recognize that allowing the AF to rotate as a whole
leads to an impressively rich phase diagram of the EB systems as a
function of the parameters of the AF layer (and F layer). The
$R$-ratio can be varied across the whole range from zero to
infinity by changing the thickness of the AF
layer~\cite{nogues:1999}, by varying its anisotropy (dilution of
the AF layers with non-magnetic
impurities~\cite{miltenyi:2000,keller:2002,hong:2006}) , or by
varying the interfacial exchange energy $J_{eb}$ (low dose ion
bombardment~\cite{chappert:1998,mougin:2001,hoink:2005}).
Recently~\cite{urazhdin:2005}, an almost ideal M\&B behavior has
been observed in Ni$_{80}$Fe$_{20}$/Fe$_{50}$Mn$_{50}$ bilayers.
At high thicknesses of the AF layer the hysteresis loop is shifted
to negative values and the coercivity is almost zero, whereas for
reduced AF thicknesses a strong increase of the coercive field is
observed together with a drastic decrease of exchange bias.

{\subsection{Analytical expression of the exchange bias field }

First we calculate analytically the expression of the exchange
bias field for $\theta=0$ and $K_F=0$. The exact analytical
solution is obtained by solving the system of  Eqs.~\ref{GMBeq4}
for $\beta=0$
 , which leads to:
\begin{equation}
 H_{eb}=\lbrace  \mat{H_{eb}^{\infty}
 \sqrt{1-\frac{1}{4 R^2}
}\, \qquad R \ge1 \\\\0 \, \qquad
 R <1}
\label{GMBebAExpr}
\end{equation}
This equation retains the $1/t_F$ dependence of the exchange bias
field, but at the same time provides new features. The most
important one is an additional term, which effectively lowers the
exchange bias field when the R-ratio approaches the critical value
of one.  The R-ratio has three terms. One of them includes the
thickness of the AF layer. The analytical expression for the
exchange bias field (Eq.~\ref{GMBebAExpr}) predicts that there is
a critical AF thickness $t_{AF}^{cr}$ below which the exchange
bias cannot exist. This is~\cite{mauri:1987:2}:
\begin{equation}
t_{AF}^{cr}=\frac{J_{eb}}{K_{AF}}
\end{equation}
Below this critical thickness the interfacial energy is
transformed into coercivity. Above the critical thickness the
exchange bias increases as a function of the AF layer thickness,
reaching the asymptotic (ideal) value $H_{eb}^\infty$ when
$t_{AF}$ is infinite. Most recent observation of  an AF critical
thickness can be found in Ref.~\cite{kohlhepp:2006}.

A similar expression as Eq.~\ref{GMBebAExpr} was derived by Binek
et al.~\cite{binek:2001} using a series expansion of
Eq.~\ref{GMBeq2} with respect to $\alpha$=0 : $H_{eb}\, \approx \,
H_{eb}^{\infty} (1-\frac{1}{8 R^2})$ for 1/R$\ge$0. Note that
close to the critical value of $R$=1 the $\alpha$ angle could
reach high values up to 45°. Therefore, the series expansion with
respect to $\alpha$=0 is a good approximation for R$>$5.

{\subsection{Azimuthal dependence of the exchange bias field}}

In the following we consider the exchange bias field in region I,
where it acquires non vanishing values. 
%
The coercive fields and the exchange bias field are extracted from
the condition $\beta=\theta+\pi/2$ for both $H_{c1}$ and $H_{c2}$.
This gives $H_{c1}=H_{c2}=(-J_{eb}/\mu_0 M_F t_F) \cos(\alpha(R,\,
\theta+\pi/2)-\theta)$, where $\alpha(R, \, \theta+\pi/2)$ is the
value of the rotation angle of the AF spins at the coercive field.
With the notation: $\alpha_0\equiv\alpha(R, \, \theta+\pi/2)$, and
using the expression~\ref{EXPRHcEB}, the angular dependence of the
exchange bias field becomes:
\begin{equation}
 H_{eb}(\theta)=\frac{-J_{eb}}{\mu_0 \, M_F \,
 t_F}\cos(\, \alpha_0-\theta)
\label{GMBebhcrot1}
\end{equation}
The equation above can be also written as:
\begin{equation}
 H_{eb}(\theta)=-\frac{K_{AF} \, t_{AF}}{\mu_0 \, M_F \,
 t_F}\ \sin(2\, \alpha_0)
\label{GMBebhcrot3}
\end{equation}
Interestingly, the exchange coupling parameter $J_{eb}$ in
Eq.~\ref{GMBebhcrot1} is missing in Eq.~\ref{GMBebhcrot3}, leaving
instead an explicit dependence of the exchange bias field on the
parameters of the antiferromagnet and the ferromagnet. The
exchange coupling constant and the $\theta$ angle are  accounted
for by the AF angle $\alpha_0$.

The Eq.~\ref{GMBebhcrot1} and  Eq.~\ref{GMBebhcrot3} are the most
general expressions for an exchange bias field. They include both,
the influence of the rotation of the AF layer and the influence of
the azimuthal orientation of the applied field. Moreover, the
anisotropy and the thickness of the AF layer are explicitly shown
in Eq.~\ref{GMBebhcrot3}. To illustrate their generality we
consider below two special cases for the Eq.~\ref{GMBebhcrot1}:
\begin{itemize}
\item{$\theta=0$}\\
In this case the hysteresis loop is measured along the field
cooling direction ($\theta=0$) and Eq.~\ref{GMBebhcrot1} becomes
equivalent to the Eq.~\ref{GMBebAExpr}.

\item{$R\rightarrow \infty$}\\
When $R$ is very large ($R\gg1$), $\alpha$ approximates zero, i.e.
the rotation of the AF - layer becomes negligible. This is
actually the original assumption of the Meiklejohn and Bean model.
Such a condition ($R\rightarrow \infty$) is approximately
satisfied for large thicknesses of  the AF layer. Then the
exchange bias field as function of $\theta$ can be written
as~\cite{kim:2000,binek:2001,binekbook}:
\begin{equation}
 H_{eb}^{\alpha=0}(\theta)=\frac{-J_{eb}}{\mu_0 \, M_F \,
 t_F}\cos(\theta)
\label{GMBebhcrot2}
\end{equation}
\end{itemize}



In order to get more insight into the azimuthal dependence of the
exchange bias field, we show in Fig.~\ref{figMBAngDep}  the
normalized exchange bias field $H_{eb}(\theta)\ /
|H_{eb}^{\infty}(0)|$ as a function of the $\theta$ angle,
according to  Eq.~\ref{GMBebhcrot1} and Eq.~\ref{GMBebhcrot3}, and
for three different values of the $R$ ratio ($R=1.1, \ R=1.5, \
R=20$). The $\alpha_0$ angle (see
Fig.~\ref{figMBalphaPhaseDiagram1}) was obtained by numerically
solving the system of Eqs.~\ref{GMBeq4}.  For large values of $R$,
the azimuthal dependence of the exchange bias field follows
closely a $\cos(\theta)$ unidirectional dependence. When, however,
the R-ratio takes small values  but larger then unity, the
azimuthal behavior of the EB field deviates from the ideal
unidirectional characteristic. There are two distinctive features:
one is that at $\theta=0$ the exchange bias field is reduced, and
the other one is that the maximum of the exchange bias field is
shifted from zero towards negative azimuthal angle values.
According to Eq.~\ref{GMBebhcrot1} this shift angle is equal to
$\alpha_0$.  In other words, the exchange bias field is not
maximum along the field cooling direction. Another striking
feature is that the shifted maximum of the exchange bias field
with respect to the azimuthal angle $\theta$ does not depend on
thickness and anisotropy of the AF layer:
\begin{equation}
H_{eb}^{MAX}=-\frac{J_{eb}}{\mu_0 \, M_F \, t_F}. \label{GMBemax}
\end{equation}

Summarizing we may state that, within the  Meiklejohn and Bean
model, a reduced exchange bias field is observed along the field
cooling direction depending on the parameters of the AF layer
($K_{AF}$ and $t_{AF}$). However, for $R\ge1$ the maximum value
for the exchange bias field which is reached at $\theta\ne0$ does
not depend on the anisotropy constant ($K_{AF}$) and thickness
($t_{AF}$) of the AF layer. The azimuthal characteristic of the
exchange bias allows to extract all three essential parameters
defining the exchange bias field: $J_{eb}, K_{AF}$ and, $t_{AF}$.

The condition for extracting the $H_{c1}$ and $H_{c2}$ from the
same $\beta$ angle hides an important property of the
magnetization reversal  of the ferromagnetic layer. This will be
 described next.

\begin{figure}[!ht]
\begin{center}
\includegraphics[clip=true,keepaspectratio=true,width=1\linewidth]{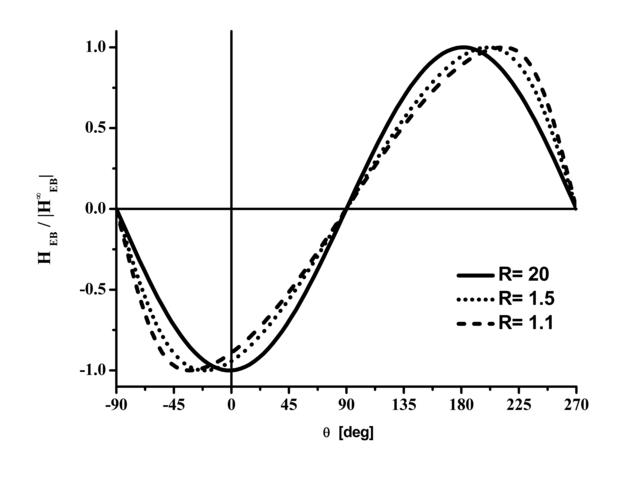}
\caption[Azimuthal dependence of exchange
bias.]{\label{figMBAngDep} Azimuthal dependence of exchange bias
as a function of the $\theta$ angle. The curves are calculated by
the Eq.\ref{GMBebhcrot1} and Eq.\ref{GMBebhcrot3}.}
\end{center}
\end{figure}

\subsection{Magnetization reversal}

A distinct feature of exchange bias phenomena is the magnetization
reversal mechanism. In Fig.~\ref{figMBreversal} is shown the
parallel component of the magnetization $m_{||}=\cos(\beta)$
versus the perpendicular component $m_{\perp}=\sin(\beta)$ for
several R-ratios and for $\theta=30°$. The geometrical conventions
are the ones shown in Fig.~\ref{figMBrot2}. We see that for $R<1$
the reversal of the F spins is symmetric, similar to typical
ferromagnets with uniaxial anisotropy. Although  the regions
$0.5\le R<1$ and $R<0.5$ exhibit different reversal modes of the
AF spins (see Fig.\ref{figMBalphaPhaseDiagram1}), there is little
difference with respect to the F spin rotation. For both regions
$0.5\le R<1$ and $R<0.5$ the F spins do make a full rotation.
Similar to the uniaxial ferromagnets. At the steep reversal
branches one would expect magnetic domain formation.
\begin{figure}[!ht]
\begin{center}
\includegraphics[clip=true,keepaspectratio=true,width=1\linewidth]{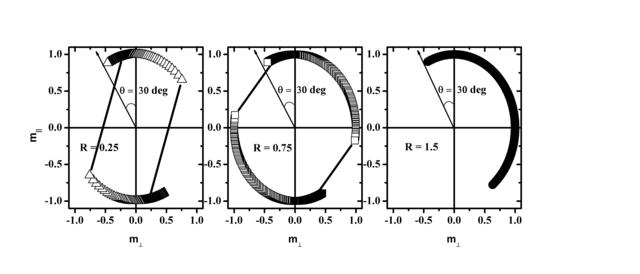}
\caption[Magnetization reversal or several values of the $R$
ratio.]{\label{figMBreversal} Magnetization reversal for several
values of the $R$ ratio. The parallel component of the
magnetization vector $m_{||}=\cos(\beta)$ is plotted as a function
of the perpendicular component of the magnetization
$m_{\perp}=\sin(\beta)$. The reversal for $R<1$ resembles the
typical reversal of ferromagnets with uniaxial anisotropy. For
$R\ge1$ the reversal proceeds along the same path for the
increasing and decreasing branch of the hysteresis loop. The angle
$\theta=30$° is chosen arbitrarily.}
\end{center}
\end{figure}

When $R\ge1$ another reversal mechanism is observed. The
ferromagnetic spins  first rotate towards the unidirectional axis
as lowering the field from positive to negative values, and then
the rotation proceeds continuously until the negative saturation
is reached. On the return path, when the field is swept from
negative to positive values, the ferromagnetic spins follow the
same path towards the positive saturation. The rotation is
continuous without any additional steps or jumps. A similar
behavior was observed theoretically within the domain state
model~\cite{beckmann:2003,beckmann:2006}. The magnetization
reversal modes can be accessed experimentally by using the
Vector-MOKE
technique~\cite{daboo:1995,SchmitteDiss,camarero:2005,radu:jpcm:2006}
which allows to follow both the magnetization vector and its angle
during the reversal process.

\subsection{Rotational hysteresis}
We briefly discuss again the rotational hysteresis deduced from
torque measurements~\cite{bean:1956, bean:1957, bean:1962}, now in
the light of the analysis provided above. The torque measurements
were carried out in a strong applied magnetic field H. Therefore
the applied field H and the magnetization $M_F$ can be assumed to
be parallel ($\beta=\theta$). The torque is given by:
$$
T=-\frac{\partial E(\theta)}{\partial \theta}= J_{eb}
\sin(\theta-\alpha(\theta))
$$
This expression differs from Eq.~\ref{torque1} for the ideal model
by the rotation of the AF spins through the $\alpha$ angle.
However, this does not explain the energy loss during the torque
measurements, as observed in the experiment (\ref{figEBean}(b)).
The torque curve would only be a bit distorted but completely
reversible. The integration of the energy curve predicts a
rotational hysteresis $W_{rot}=0$. In order to account for a
finite rotational hysteresis, one can assume that a fraction $p$
of particles at the F-AF interface are uniaxially coupled behaving
as in region II, whereas the remaining fraction $(1-p)$ of the
F-AF particles are coupled unidirectionally, having the ideal
behavior as described in regime I. As seen in
Fig.~\ref{figMBPhaseDiagram1}(left), when the R-ratio of the
uniaxial particle is in the range $0.5 \, \le \, R \, < \,1$, the
AF spins will rotate irreversibly, showing hysteresis-like
behavior due to $\alpha$ jumps indicated in the
Fig.~\ref{figMBPhaseDiagram1} (right). A rotational hysteresis is
not expected for unidirectional particles with $R\ge 1$ because
the AF structure changes reversibly with $\theta$. With this
assumption the uniaxial particles will contribute to the energy
loss during the torque measurements, while the unidirectional
particles are responsible for the unidirectional feature of the
torque curve. This argument was used by Meiklejohn and
Bean~\cite{bean:1962} when studying the exchange bias in
core-Co/shell-CoO. A fraction $p=0.5$ was inferred from the torque
curves shown in Fig.~\ref{figEBean}.





\sectionmark{N\'eel's AF domain wall}
{\section{N\'eel's AF domain wall - Weak coupling}}
\sectionmark{N\'eel's AF domain wall}

The rigid AF spin state and rigidly rotating AF spins concepts
impose a restriction on the behavior of the antiferromagnetic
spins, namely that the AF order is preserved during the
magnetization reversal. Such restriction implies that the
interfacial exchange coupling is found almost entirely  in the
hysteresis loop either as a loop shift or as coercivity.
Experimentally, however, the size of the exchange bias does not
agree with the expected value, being several orders of magnitude
lower then predicted. In order to cope with such loss of coupling
energy, one can assume that a partial domain wall develops in the
AF layer during the magnetization reversal. This concept was
introduced by N\'eel~\cite{neel:1967,kiwi:2001} when considering
 the coupling between a ferromagnet and a low anisotropy antiferromagnet.
The AF partial domain wall will store an important fraction of the
exchange coupling energy, lowering the shift of the hysteresis
loop.

N\'eel has calculated the magnetization orientation of each layer
through a differential equation. The weak coupling is consistent
with a partial AF domain wall which is parallel to the interface
(N\'eel domain wall). His model predicts that a minimum AF
thickness is required to produce hysteresis shift. More
importantly the partial domain wall concept forms the basis of
further models which incorporate either N\'eel wall or Bloch wall
formation as a way to reduce the observed magnitude of exchange
bias.

\sectionmark{Malozemoff model}
{\section{Malozemoff Random Field Model }}
\sectionmark{Malozemoff model}

Malozemoff (1987) proposed a novel mechanism for exchange
anisotropy postulating a random nature of exchange interactions at
the F-AF
interface~\cite{malozemoff:1987,malozemoff:1988:1,malozemoff:1988:2}.
He assumed that the chemical roughness or alloying at the
interface, which is present for any realistic bilayer system,
causes lateral variations of the exchange field acting on the F
and AF layers. The resultant random field causes the AF to break
up into magnetic domains due to the energy minimization. By
contrast with other theories, where the unidirectional anisotropy
is treated either
microscopically~\cite{koon:1997,schulthess:1998,schulthess:1999}
or macroscopically~\cite{bean:1956,bean:1957,mauri:1987:1}, the
Malozemoff approach  belongs to models on the mesoscopic scale for
surface magnetism.

\begin{figure}[!ht]
\begin{center}
\includegraphics[clip=true,keepaspectratio=true,width=.5\linewidth]{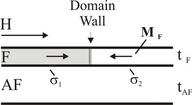}
\caption[Schematic side view of a F/AF bilayer with a
ferromagnetic wall driven by an applied field
$\textbf{H}$.]{\label{figmalozemoff1} Schematic side view of a
F/AF bilayer with a ferromagnetic wall driven by an applied field
$\textbf{H}$~\cite{malozemoff:1987}.}
\end{center}
\end{figure}

The general idea for estimating the exchange anisotropy is
depicted in Fig.~\ref{figmalozemoff1}, where a domain wall in an
uniaxial ferromagnet is driven by an applied in-plane magnetic
field {{H}}~\cite{malozemoff:1987}. Assuming that the interfacial
energy in one domain ($\sigma_1$) is different from the energy in
the neighboring domain $\sigma_2$, then the exchange field can be
estimated by the equilibrium condition between the applied field
pressure $2\, H \, M_F \,t_F$ and the effective pressure from the
interfacial energy $\Delta  \sigma$:
\begin{equation}
H_{eb}=\frac{\Delta \sigma}{2\, M_F\, t_F}, \label{maloeq1}
\end{equation}
where $M_F$ and $t_F$ are the magnetization and thickness of the
ferromagnet. When the interface is treated as ideally
"compensated", then the exchange bias field is zero. On the other
hand, if the AF/F  interface is ideally uncompensated there is an
interfacial energy difference $\Delta \sigma=2 J_i/a^2$, where
$J_i$ is the  exchange coupling constant across the interface, and
a is the lattice parameter of a simple cubic structure assigned to
the AF layer. The EB field is $H_{eb}=J_i/a^2 \, M_F \, t_F$ (see
Fig~\ref{figmalozemoff2}) \footnote{For this example the energy is
calculated as $E_{kl}=-J_i \bm{S}_k\bm{S}_l$ per pair of nearest
neighbor spins $kl$ at the interface }.

Estimating numerically the size of the EB field using the equation
above for an ideally uncompensated interface, results in a
discrepancy of several orders of magnitude with respect to the
experimental observation . Therefore, a novel mechanism based on
random fields at the interface acting on the AF layer is proposed
as to drastically reduce the resulting exchange bias field.

\begin{figure}[!ht]
\begin{center}
\includegraphics[clip=true,keepaspectratio=true,width=1\linewidth]{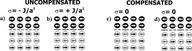}
\caption[Schematic view of possible atomic configurations in a
F-AF bilayer with ideal interfaces.]{\label{figmalozemoff2}
Schematic view of possible atomic configurations in a F-AF bilayer
with ideal interfaces. Frustrated bonds are indicated by crosses.
Compensated configuration a) will result in configuration b) by
reversing the ferromagnetic spins through domain wall movement. It
gives an exchange bias field of $H_{eb}= J_i/ a^2 M_F t_F$. The
compensated configuration c) will result in the compensated
configuration d). The exchange bias field for this case is zero
($H_{eb}=0$). ~\cite{malozemoff:1987}. }
\end{center}
\end{figure}

By simple and schematic arguments  Malozemoff describes how
roughness on the atomic scale of a "compensated" AF interface
layer can lead to  uncompensated spins required for the loop to
shift. An atomic rough interface depicted in
Fig~\ref{figmalozemoff3}a) containing a single mono-atomic bump in
a  cubic interface gives rise to six net antiferromagnetic
deviations from a perfect compensation.  A bump shifted by one
lattice spacing as shown in Fig.~\ref{figmalozemoff3}b), which is
equivalent to reversing the F spins, provides six net
ferromagnetic deviations from perfect compensation. Thus a net
energy difference of $z_i J_i$ with $z_i=12$ acts at the interface
favoring one domain orientation over the other. Note that for an
ideally uncompensated interface the energy difference is only $8
J_i$ when reversing the F spins. This implies that an atomic step
roughness at a compensated interface  leads to a higher exchange
bias field as compared to the ideally compensated interface.

The estimates of this local field can be further refined assuming
a more detailed model. For example, by inverting the spin in the
bump shown in Fig.~\ref{figmalozemoff3}c, the interfacial energy
difference is reduced by $5\times 2 J_i$ at the cost of generating
one frustrated pair in the AF layer just under the bump. This
frustrated pair increases the energy difference by $2 J_A$, where
$J_A$ is the AF exchange constant. Thus the energy difference
between the two domains becomes $2J_i+2 J_A$ or roughly $4 J$ if
$J_i \approx J_A \approx J$. If one allows localized canting of
the spins, one expects the energy difference to be reduced
somewhat further.

\begin{figure}[!ht]
\begin{center}
\includegraphics[clip=true,keepaspectratio=true,width=1\linewidth]{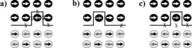}
\caption[Schematic side of possible atomic moment configurations
for non-planar interface.]{\label{figmalozemoff3} Schematic side
view of possible atomic moment configurations for a non-planar
interface. The bump should be visualized on a two-dimensional
interface. Configuration c) represents the lower energy state of
a). The configuration b) is energetically equivalent to flipping
the ferromagnetic spins of a). The x signs represent frustrated
bonds. ~\cite{malozemoff:1987} }
\end{center}
\end{figure}

Each interface irregularity will give a local energy difference
between domains whose sign depends on the particular location of
the irregularity and whose magnitude is on the average $2 z J$,
where $z$ is a number of order unity. Furthermore, for an
interface which is random on the atomic scale, the local
unidirectional interface energy $\sigma_l=\pm zJ/a^2$ will also be
random and its average $\sigma$ in a region $L^2$ will go down
statistically as $\sigma\approx\sigma_l/\sqrt{N}$, where
$N=L^2/a^2$ is the number of sites projected onto the interface
plane. Therefore the effective AF-F exchange energy per unit area
is given by:
$$J_{eb}\approx \frac{1}{\sqrt{N}} J_{i}\approx \frac{1}{L} J_{i},$$
where $J_{i}$ is the exchange energy of a fully uncompensated
AF-surface.

Given a random field provided by the interface roughness and
assuming a region with a single domain of the ferromagnet, it is
energetically favorable for the AF to break up into magnetic
domains, as shown schematically  in Fig.~\ref{figmalozemoff4}. A
perpendicular domain wall is the most preferable situation. This
perpendicular domain wall is permanently present in the AF layer.
It should be distinguished from a domain wall parallel to AF/F
interface, which according to the Mauri model~\cite{mauri:1987:1}
develops temporarily during the rotation of the F layer.
\begin{figure}[!ht]
\begin{center}
\includegraphics[clip=true,keepaspectratio=true,width=1\linewidth]{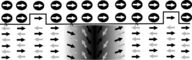}
\caption[Schematic view of a vertical domain wall in the AF
layer.]{\label{figmalozemoff4} Schematic view of a vertical domain
wall in the AF layer. It appears as an energetically favorable
state of F/AF systems with rough interfaces
~\cite{malozemoff:1987,coehoorn:2002}. }
\end{center}
\end{figure}

By further analyzing the stability of the magnetic domains in the
presence of random fields, a characteristic length $L$ of the
frozen-in AF domains and their characteristic height are obtained:
$L\approx \pi \sqrt{A_{AF}/K_{AF}}$ and $h=L/2$, where $A_{AF}$ is
the exchange stiffness and $h$ is the characteristic height of the
AF domains. Once these domains are fixed, flipping the
ferromagnetic orientation causes an energy change per unit area of
$\Delta \sigma=4 z J/\pi a L$, which further leads to the
following expression for the EB field:
\begin{equation}
H_{eb}=\frac{2\, z\, \sqrt{A_{AF} K_{AF}}}{\pi^2\, M_F
t_F}\label{maloEB}
\end{equation}
Assuming a CoO/Co$(100 \, \angstrom)$ film, the calculated
exchange bias using the Eq.~\ref{maloEB} is:
\begin{eqnarray}
H_{eb}&=& \frac{2 \times 1 \sqrt{\frac{0.0186\times 1.6\times
10^{-19}
 [J]}{4.27\times10^{-10}[m]} 2.5\times10^7 \, [J/m^3]}}{\pi^2\times 1460 \,
 [kA/m]\, 100\times10^{-10}}\times10\nonumber\\
 &=& 580\,Oe
\label{maloEBest}
 \end{eqnarray}
 For the estimations above we used for  the exchange
stiffness the following value: $A_{AF}=J_{AF}/a$, where $a$ is the
lattice parameter of CoO ($a=4.27 \angstrom$) and
$J_{AF}=1.86$~meV is the exchange constant for CoO
~\cite{rechtin:1972}.

The characteristic length of  the AF domains is for CoO:
\begin{eqnarray}
L&=&\pi \sqrt{A_{AF}/K_{AF}}\nonumber\\
&=&\pi\times\sqrt{\frac{0.0186\times 1.6\times
10^{-19}[J]}{4.27\times10^{-10}[m]\times 2.5\times10^7 \, [J/m^3]}
 }\nonumber\\
 &=&\, 16.6 \, \angstrom
 \end{eqnarray}
 The height of the AF domains is $h=L/2=8.3\,\angstrom$. Comparing
this value to the experimental data on CoO$(25\angstrom)$/Co
studied in ~\cite{radu:2003:2}, we notice that the calculated EB
 field agrees well with the value observed experimentally. For
example, the exchange bias field for CoO$(25\, \angstrom)$/Co$(119
\,\angstrom)$ is 557~Oe and the theoretical value calculated with
Eq.~\ref{maloEB} is 487~Oe. Also, the length and the height of the
AF domains have enough space to develop. The difference between
theory and experiment is, however, that experimentally AF domains
occur after the very first magnetization reversal, whereas within
the Malozemoff model the AF domains are assumed to develop during
the field cooling procedure. Nevertheless, the agreement appears
to be excellent.


\sectionmark{DS model}
\section{Domain State Model}
\sectionmark{DS model}
The Domain State model (DS) introduced by Usadel and
coworkers~\cite{miltenyi:2000,nowak:2001,nowak:2002,beckmann:2003,misra:2003,misra:2004,scholten:2005}
is a microscopic model in which disorder is introduced  via
magnetic dilution not only at the interface but also in the bulk
of the AF layer as in Fig.~\ref{figNowak2}. The key element in the
model is that the AF layer is a diluted Ising  antiferromagnet in
an external magnetic field (DAFF) which exhibits a phase diagram
like the one shown in Fig.~\ref{figNowak1}~\cite{nowak:2002}. In
zero field the system undergoes a phase transition from a
disordered, paramagnetic state to a long-range-ordered
antiferromagnetic phase at the dilution dependent N\'eel
temperature. In the low temperature region, for small fields, the
long-range interaction phase is stable in three dimensions. When
the field is increased at low temperature the diluted
antiferromagnet develops a domain state phase with a
spin-glass-like behavior. The formation of the AF domains in the
DS phase  originates from the statistical imbalance of the number
of impurities of the two AF sublattices within any finite region
of the DAFF. This imbalance leads to a net magnetization which
couples to the external field. A spin reversal of  the region, i.
e., the creation of a domain, can lower the energy of the system.
The formation of a domain wall can be minimized if the domain wall
passes through nonmagnetic defects at a minimum cost of exchange
energy.
\begin{figure}[!h]
\begin{center}
\includegraphics[clip=true,keepaspectratio=true,width=1\linewidth]{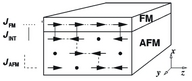}
\caption[Sketch of the DS model with one ferromagnetic layer
and...]{Sketch of the domain state model with one ferromagnetic
layer and three diluted antiferromagnetic layers. The dots mark
defects~\cite{nowak:2002}. }\label{figNowak2}
\end{center}
\end{figure}

\begin{figure}[!h]
\begin{center}
\includegraphics[clip=true,keepaspectratio=true,width=1\linewidth]{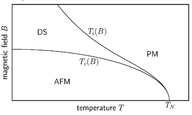}
\caption[Schematic phase diagram of a three-dimentional diluted
antiferromagnet.]{Schematic phase diagram of a three-dimentional
diluted antiferromagnet. AF is the antiferromagnetic phase, DS the
domain state phase, and PM the paramagnetic
phase~\cite{nowak:2002}.}\label{figNowak1}
\end{center}
\end{figure}

Nowak et al.~\cite{nowak:2002} further argue that during the field
cooling below the irreversibility line $T_i(B)$, in an external
field and in the presence of the interfacial exchange field of the
ferromagnet, the AF develops a frozen domain state with an
irreversible surplus of magnetization. This irreversible surplus
magnetization controls then the exchange bias.

The F layer is described by a classical Heisenberg model with the
nearest-neighbor exchange constant $J_{F}$. The AF is modelled as
a magnetically diluted Ising system with an easy axis parallel to
that of the F. The Hamiltonian of the system is given
by~\cite{nowak:2002}:

\begin{eqnarray}
H&=&-J_{F}\sum_{<i, j> \epsilon F}{\bm{S}_i . \bm{S}_j}-\sum_{i \epsilon F}{(d_z S^2_{iz} +d_x S^2_{ix}+\mu \bm{B} \bm{S}_i)}\nonumber\\
& &-J_{AF}\sum_{<i, j> \epsilon AF}{\epsilon_i \epsilon_j \bm{\sigma}_i  \bm{\sigma}_j}-\sum_{i \epsilon AF}{\mu B_z \epsilon_i \sigma_i}\nonumber\\
& &-J_{INT}\sum_{<i \epsilon AF, j \epsilon F>}{\epsilon_i\sigma_i
S_{jz}} ,\label{DSMeq}
 \end{eqnarray}
where the $\bm{S_i}$ and $\bm{\sigma_i}$ are the classical spin
vectors at the {\textit i}th site of the F and AF, respectively.
The first line contains the energy contribution of the F, the
second line describes the diluted AF layer, and the third line
includes the exchange coupling across the interface between F and
DAFF, where it is assumed that the Ising spins in the topmost
layer of the DAFF interact with the $z$ component of the
Heisenberg spins of the F layer.

In order to obtain the hysteresis loop of the system, the
Hamiltonian in Eq.~\ref{DSMeq} is treated by Monte Carlo
simulations. Typical hysteresis loops are shown in
Fig.~\ref{figNowak3}~\cite{misra:2004}, where both the
magnetization curve of the F layer and of the interface monolayer
of the DAFF are shown. The coercive field extracted from the
hysteresis curve depends on the anisotropy of the F layer, but it
is also influenced by the DAFF. It, actually, depends on the
thickness and anisotropy of the DAFF layer. The coercive field
decreases with the increasing thickness of the DAFF layer
\cite{misra:2004}, which can be understood as follows: the
interface magnetization tries to orient the F layer along its
direction. The coercive field has to overcome this barrier, and
the higher the interface magnetization of the DAFF, the stronger
is the field required to reverse the F layer. The interface
magnetization decreases with increasing DAFF thickness due to a
coarsening of the AF domains accompanied by smoother domain walls.

\begin{figure}[!ht]
\begin{center}
\includegraphics[clip=true,keepaspectratio=true,width=1\linewidth]{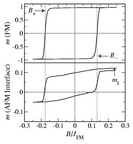}
\caption[Simulated hysteresis loops within the DS
model.]{Simulated hysteresis loops within the domain state model.
The top hysteresis belongs to the ferromagnetic layer and the
bottom hysteresis belongs to an  AF interface
monolayer~\cite{misra:2004}. }\label{figNowak3}
\end{center}
\end{figure}

The strength of the exchange bias field can be estimated from the
Eq.~\ref{DSMeq} using simple ground state arguments. Assuming that
all spins in the F remain parallel during the field reversal and
some net magnetization of the interface layer of the DAFF remains
constant during the reversal of the F, a simple calculation gives
the usual estimate for the bias field~\cite{nowak:2002}:
\begin{equation}
l \mu B_{eb}=J_{INT} m_{INT},
\end{equation}
where $l$ is the number of the F layers and $m_{INT}$ is the
interface magnetization of the AF per spin. $B_{eb}$ is the
notation  for the exchange bias field in Ref.~\cite{nowak:2002}
and is equivalent to $H_{eb}$ in this chapter. For an ideal
uncompensated interface ($m_{INT}=1$) the exchange bias is too
high, whereas for an ideally compensated interface the exchange
bias is zero. Within the DS model the interface magnetization
$m_{INT}<1$ is neither a constant nor is it a simple
quantity~\cite{nowak:2002}. Therefore, it is replaced by
$m_{IDS}$, which is a measure of the irreversible domain state
magnetization of the DAFF interface layer and is responsible for
the EB field. With this, an estimate of exchange bias field for
$l=9$, $J_{INT}=-3.2\times 10^{-22} J$, and $\mu=1.7 \, \mu_B$
gives a value of about 300~Oe.

The exchange bias field depends also on the bulk properties of the
DAFF layer as shown  by Milt\'enyi et al.~\cite{miltenyi:2000}.
There the AF layer was diluted by substituting non-magnetic Mg in
the bulk part and away from the interface. The representative
results are shown in Fig.~\ref{figmilt1}. It was shown
experimentally that the EB field depends strongly on the dilution
of the AF layer.  As a function of concentration of the
non-magnetic Mg impurities, the EB evolves as following: at zero
dilution the exchange bias has finite values, whereas by
increasing the Mg concentration, the EB field increases first,
showing a broad peak-like behavior, and then, when the dilution is
further increased the EB field decreases again. Simulations within
the DS model showed an overall good qualitative agreement. The
peak-like behavior of the EB as a function of the dilution is
clearly seen in the simulations (see Fig.~\ref{figmilt1}).
However, it appears that at zero dilution, the DS gives vanishing
exchange bias whereas experimentally finite values are observed.
The exchange bias is missing at low dilutions because the domains
in the AF cannot be formed as they would cost too much energy to
break the AF bonds. This
discrepancy~\cite{nowak:2002,papusoi:2006} is thought to be
explained by other imperfections, such as grain boundaries in the
AF layer which is similar to dilution and which can also reduce
the domain-wall energy, thus leading to domain formation and EB
even without dilution of the AF bulk.

An important property of the kinetics of the DAFF is the slow
relaxation of the remanent magnetization, i.e., the magnetization
obtained after switching off the cooling field~\cite{nowak:2002}.
It is known that the remanent magnetization of the DS relaxes
nonexponentially on extremely long-time scales after the field is
switched off or even within the applied field. In the DS model the
EB is related to this remanent magnetization. This implies a
decrease of EB due to slow relaxation of the AF domain state. The
reason for the training effect can be understood within DS model
from Fig.~\ref{figNowak3} bottom panel, where it is shown that the
hysteresis loop of the AF interface layer is not closed on the
right hand side. This implies that the DS magnetization is lost
partly during the hysteresis loop due to a rearrangement of the AF
domain structure. This loss of magnetization clearly leads to a
reduction of the EB.

\begin{figure}[!ht]
\begin{center}
\includegraphics[clip=true,keepaspectratio=true,width=1\linewidth]{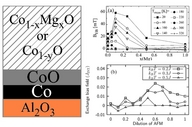}
\caption[Simulated hysteresis loops within the DS model.]{In the
right side of the figure is shown the film structure used to study
the dilution influence on the exchange bias field.  a) EB field as
function of the Mg concentration x in the Co$_{1-x}$Mg$_x$O layer
for several temperatures. b) EB field as a function of different
dilutions of the AF volume~\cite{miltenyi:2000}. }\label{figmilt1}
\end{center}
\end{figure}

The blocking temperature~\footnote{The blocking temperature of an
exchange bias system is the temperature where  the hysteresis loop
acquires a negative or positive shift with respect to the field
axis. It is always lower then the N\'eel temperature of the AF
layer.} within the DS model can be understood by considering the
phase diagram of the DAFF shown in Fig~\ref{figNowak2}}. The
frozen DS of the AF layer occurs after field cooling the system
below the irreversibility temperature $T_i(b)$. Within this
interpretation, the blocking temperature corresponds to $T_i(b)$.
 Since $T_i(b) < T_N$, the blocking temperature
 should be always  below the Ne\'el
temperature and should be dependent on the strength of the
interface exchange field. The simulations within the DS model
shows that  EB depends linearly on the temperature, as observed
experimentally in some Co/CoO systems, but no reason is given for
this behavior~\cite{nowak:2002}. In Ref.~\cite{chen:2001} the
blocking temperature of a DAFF system
(Fe$_{1-x}$Zn$_x$F$_2$(110)/Fe/Ag with x=0.4)  exhibits a
significant enhancement with respect to the global ordering
temperature T$_N$=46.9~K, of the bulk antiferromagnet
Fe$_0.6$Zn$_0.4$F$_2$.

Overall, it is believed that strong support for the DS model is
given by experimental observations where  nonmagnetic impurities
are added to the AF layer in a systematic and controlled
fashion~\cite{miltenyi:2000,mewes:2000, mougin:2001,
shi:2002,keller:2002,papusoi:2006,hong:2006}. Also,  good
agreement has been observed  in Ref.~\cite{ali:2003}, where the
dependence of the EB as a function of AF thickness and temperature
for IrMn/Co was analyzed. The asymmetry of the magnetization
reversal mechanisms~\cite{beckmann:2003,beckmann:2006} is shown to
 be dependent on the angle between the easy axis of the F and DAFF
layers. It was found that either identical or different F reversal
mechanisms (domain wall movement or coherent rotation) can occur
as the relative orientation between the anisotropy axis of the F
and AF is varied. This is discussed in more detail in
sections~\ref{depexbi}~and~\ref{expassym}.

\sectionmark{Mauri model}
{\section{Mauri model}}
\sectionmark{Mauri model}

The model of Mauri~{\textit et al.}~\cite{mauri:1987:1} renounces
the assumption of a rigid AF layer and proposes that the AF spins
develop a domain wall parallel to the interface. The motivation to
introduce such an hypothesis was to explore a possible reduction
of the exchange bias field resulting from the Meiklejohn and Bean
model.

The assumptions of the Mauri model are:

\begin{itemize}
\item{both the F and AF are in a single domain state;} \item{the F
layer rotates rigidly, as a whole;} \item{the AF layer develops a
domain wall parallel to the interface;} \item{the AF interface
layer is uncompensated (or fully compensated);} \item{the AF layer
has a uniaxial anisotropy;} \item{the cooling field is oriented
parallel to the uniaxial anisotropy of the AF layer;} \item{the AF
and F spins rotate coherently, therefore the Stoner-Wohlfarth
model is used to describe the system.}
\end{itemize}

\begin{figure}[!ht]
\begin{center}
\includegraphics[clip=true,keepaspectratio=true,width=1\linewidth]{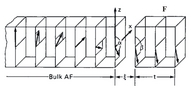}
\caption[The Mauri model]{ Mauri model for the interface of a thin
ferromagnetic film on a antiferromagnetic
substrate~\cite{mauri:1987:1}.}\label{figMauri}
\end{center}
\end{figure}

Schematically the spin configuration within the Mauri model is
shown in Fig.~\ref{figMauri}. The F spins rotate coherently, when
the applied magnetic field is swept as to measure the hysteresis
loop. The first interfacial AF monolayer is oriented away from the
F spins making an angle $\alpha$ with the field cooling direction
and with the anisotropy axis of the AF layer. The next AF
monolayers are oriented away from the interfacial AF spins as to
form a domain wall parallel to the interface. The spins of only
one AF sublattice are depicted, the spins of the other sublattice
being oppositely oriented for completing the AF order. At a
distance $\xi$ at the interface, a ferromagnetic layer of
thickness $t_{F}$ follows. Using the Stoner-Wohlfarth model, the
total magnetic energy can be written as~\cite{mauri:1987:1}:
\begin{eqnarray}
E&=&-\mu_0\, H \, M_F \, t_F \, \cos(\theta-\beta)\nonumber\\
& &+ K_{F} \, t_{F}\, \sin^2(\beta)\nonumber\\
& &-J_{eb} \,\cos(\beta-\alpha)\nonumber\\
& &-2 \sqrt{A_{AF} K_{AF}} (1-\cos(\alpha)), \label{Maurieq1}
\end{eqnarray}
where the first term is the Zeeman energy of the ferromagnet in an
applied magnetic field, the second term is the anisotropy term of
the F layer, the third term is the interfacial exchange energy
and, the forth term is the energy of the partial domain wall. The
new parameter in the equation above is the exchange stiffness
$A_{AF}$. As in the case of the Meiklejohn and Bean model, the
interfacial exchange coupling parameter $J_{eb}$~$[J/m^2]$ is
again undefined, assuming that it ranges between the exchange
constant of the F layer to the exchange constant of the AF layer
divided by an effective area (see section~\ref{magnitudeEB}).

The total magnetic  energy can be written in units of $2
\sqrt{A_{AF} K_{AF}}$, which is the energy per unit surface of a
90° domain wall in the AF layer:

\begin{eqnarray}
e&=& k \, (1-\cos(\beta)) + \mu \, \cos(\beta)^2 \nonumber\\
& &+\lambda \,[1- \cos(\alpha-\beta)]+(1-\cos(\alpha)),
\label{Maurieq2}
\end{eqnarray}
where $\lambda=J_{eb}\, /  (\, 2 \,\sqrt{A_{AF} K_{AF}})$, is the
interface exchange, with $J_{eb}$ being redefined as $J_{eb}\equiv
A_{12}\, / \xi$, where  $A_{12}$ is the interfacial exchange
stiffness and $\xi$ is the thickness of the interface (see
Fig.~\ref{figMauri}), $\mu=K_F \, t_F \, / \,2 \sqrt{A_{AF}
K_{AF}}$ is the reduced ferromagnet anisotropy, and $k=\mu_0\, H
\, M_F \, t_F \, / \, 2 \sqrt{A_{AF} K_{AF}}$ is the reduced
external magnetic field.

Mauri~{\textit{et~al.}}~\cite{mauri:1987:1} have calculated
 the magnetization curves by numerical minimization of
 the reduced total magnetic  energy Eq.~\ref{Maurieq2}.
 Several values of the
$\lambda$ and $\mu$ parameters were considered providing quite
realistic hysteresis loops. Their analysis highlights two limiting
cases with the following expressions for the exchange bias field:
\begin{equation}
H_{eb}=
\lbrace{\mat{- \, (A_{12}\, / \,\xi) / \, \mu_0 \, M_F \, t_F
\qquad for \ \lambda \ll 1 \\ \\ - \, 2 \sqrt{A_{AF} K_{AF}} / \,
\mu_0 \, M_F \, t_F \qquad for \  \lambda \gg 1}}
\label{Maureeq3}
\end{equation}
In the strong coupling limit $\lambda \ll 1$, the expression for
the  exchange bias field is similar to the value given by the
Meiklejohn and Bean model. For this situation, practically no
important differences between the predictions of the two models
exist. When the coupling is weak ($\lambda \gg 1$), the Mauri
model delivers a reduced exchange bias field which is practically
independent  of the interfacial exchange energy. It depends on the
domain wall energy and the parameters of the F layer. In either
case the "$1/t_F$" dependence is preserved by the Mauri-model.

{\subsection{Analytical expression of exchange bias field }

In order to compare the predictions of the Mauri model and the
Meiklejohn and Bean approach, we reconsider the analysis of the
free energy. Starting from the expression of the free energy
Eq.~\ref{Maurieq1}, the minimization with respect to the $\alpha$
and $\beta$ angle leads to the following system of equations:

\begin{equation}
\lbrace{\mat{\frac{H}{-\frac{J_{eb}}{\mu_0 \, M_F \, t_F }} \,
\sin(\theta-\beta)\, +\, \sin(\beta-\alpha)=0\\\\
\frac{2\,\sqrt{A_{AF} \,K_{AF}}}{J_{eb}}\,\sin(\alpha)-\sin(\beta-\alpha)=0}}
\label{Maurieq3}
\end{equation}
Similar to the Meiklejohn and Bean model we define the parameters
$P \equiv \frac{2 \sqrt{A_{AF} \,K_{AF}}}{J_{eb}}$ and
$H_{eb}^\infty \equiv -\frac{J_{eb}}{\mu_0 \, M_F \,t_F}$. Also we
set $\theta=0$, meaning that the applied field is swept along the
easy axis of the AF layer. Also, we  do not take into account the
anisotropy of the ferromagnet ($K_F=0$), for two reasons. For one
the  coercive fields of the exchange bias systems are usually much
higher then the coercive field of the isolated F layer, and
secondly it is easier to compare the results of the Mauri model
and the M\&B model when the anisotropy of the F layer is
disregarded. From inspection of the equations Eq.~\ref{Maurieq3}
and Eq.~\ref{GMBeq4} one can clearly see that the first equations
of the two systems are identical, while the second ones are
different in two respects. The first difference is related to the
term $P$, which includes the domain wall energy instead of the AF
anisotropy term in the $R$ ratio. The second difference is that
instead of a $\sin(2\, \alpha)$ term in the second equation of
Eq.~\ref{GMBeq4}, the Mauri model has a $\sin( \alpha)$ term,
which influences strongly the phase diagram shown in
Fig.~\ref{figMauryPhaseDiagram1}.

\begin{figure}[!h]
\begin{center}
\includegraphics[clip=true,keepaspectratio=true,width=1\linewidth]{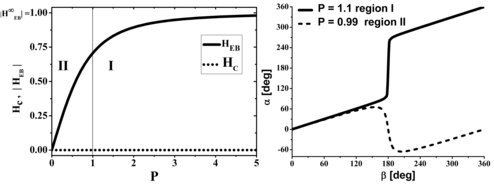}
\caption[The phase diagram of the exchange bias field and the
coercive fields given by the Mauri
formalism.]{\label{figMauryPhaseDiagram1}Left: The phase diagram
of the exchange bias field and the coercive fields given by the
Mauri formalism. Right: Typical behavior of the antiferromagnetic
angle $\alpha$ for the two different regions of the phase diagram.
In both regions I and II  a shift of the hysteresis loop can
exist. The coercive field is zero in both regimes.}
\end{center}
\end{figure}

\begin{figure}[!h]
\begin{center}
\includegraphics[clip=true,keepaspectratio=true,width=1.0\linewidth]{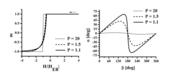}
\includegraphics[clip=true,keepaspectratio=true,width=1.0\linewidth]{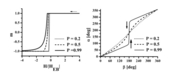}
\caption[Several hysteresis loops and antiferromagnetic spin
orientation during the magnetization reversal within the Mauri
model. ]{\label{figMauryhys1} Several hysteresis loops and
antiferromagnetic spin orientations as plotted during the
magnetization reversal. For the simulation we used the Mauri
formalism. Top row shows three hysteresis loops calculated for
different $P$ ratios of the region I shown in
Fig.~\ref{figMauryPhaseDiagram1}. The right hand panel in the top
row shows the $\alpha$ angle of the antiferromagnetic layer for
the three $P$ parameters of the hysteresis loop. The bottom raw
are the corresponding hysteresis loops and $\alpha$ angles for the
$P$ values within region II. }
\end{center}
\end{figure}

The analytical expression for the exchange bias are obtained by
solving the system of Eqs.~\ref{Maurieq3}. First step in solving
the Eq.~\ref{Maurieq3} is to extract the $\alpha$ angle
($\alpha=\pm\, \arccos (\pm \, \frac{P -\cos (\beta)}{\sqrt{1 +
P^2 - 2\, P \,\cos (\beta)}})$) from the second equation and to
introduce it in the first equation. Next we use the condition that
at the coercive field $\beta=-\pi/2$ to obtain both coercive
fields $H_{c1}=H_{c2}$. Then, inserting them into the general
expressions of Eq.~\ref{EXPRHcEB}, the coercive field $H_c$ is
zero and the exchange bias field becomes~\cite{stamps:2000}:
\begin{eqnarray}
 H_{eb}&=&-\frac{J_{eb}}{\mu_0 \, M_F \, t_F} \frac{2 \sqrt{A_{AF} \, K_{AF}}}{\sqrt{J_{eb}^2+4\,A_{AF} \,
K_{AF}}}\nonumber\\
&=&-\frac{J_{eb}}{\mu_0 \, M_F \, t_F} \frac{P}{\sqrt{1+P^2}}
\label{Maurieq4}
\end{eqnarray}
This equation is plotted as a function of $H/|H_{EB}|$ and for
different  $P$ values ranging from $P=0$ to $P=5$. ( compare
Fig.~\ref{figMauryPhaseDiagram1}~left panel). The behavior of the
EB field according to the Eq.~\ref{Maurieq4} is monotonic with
respect to the stiffness and anisotropy of the AF spins. At
$P\gg1$ the exchange bias is equal to $H^{P\rightarrow
\infty}_{eb}=-\frac{J_{eb}}{\mu_0 \, M_F \, t_F}$, which is the
well known expression given by M\&B model. When, however, the
$P$-ratio approaches low values, the exchange bias decreases,
vanishing at $P=0$, provided that the thickness of the AF layer is
sufficiently thick to allow a 180° wall. With some analytical
analysis of the Eq.~\ref{Maurieq4} one can easily reach the
limiting cases of weak coupling ($P\ll1$) and strong coupling
($P\gg1$) discussed by Mauri~{\textit{et al.}}~\cite{mauri:1987:2}
(see Eq.~\ref{Maureeq3}). In Fig.~\ref{figMauryPhaseDiagram1}
right column is shown the representative behavior of the $\alpha$
angle
 of the first interfacial AF monolayer as
a function of the $\beta$ orientation of the F spins during the
magnetization reversal and for two representative $P$ values (see
the  discussion below).


In Fig.~\ref{figMauryhys1}   the  hysteresis loops
($m_{||}=\cos(\beta)$) and the corresponding AF angle rotation
during the magnetization reversal are plotted for several
$P$-ratios. They were obtained by solving numerically the system
of Eqs.~\ref{Maurieq3}. For all the values of the $P$-ratio the
magnetization curves are shifted to  negative values of the
applied magnetic field. We distinguish two different regions with
respect to the behavior of the $\alpha$ angle of the first AF
monolayer. In the first region , for $P\ge1$ (region I), the AF
monolayer in the proximity of the F layer behaves similar to the
Meiklejohn and Bean, namely the $\alpha$ angle deviates reversibly
from the anisotropy direction as function of $\beta$. The maximum
value of the $\alpha$ angle acquired during the rotation of the F
layer is two times higher for the Mauri model as compared to the
M\&B model, reaching a maximum value of 90° at $P=1$. The coercive
field in this region is zero. The angle $\alpha$ in the region II
where $P<1$ has a completely different behavior. It rotates with
the ferromagnet  following the general behavior depicted in
Fig.~\ref{figMauryhys1}. Notice that $\alpha$ follows
monotonically the rotation of the F spins, with  no jumps or
hysteresis-like behavior in contrast to the M\&B model(see
Fig.~\ref{figMBPhaseDiagram1}).  Very importantly, the exchange
bias field does not vanish in this region and therefore no
additional coercive field related to the AF  is observed, provided
that the AF layer is sufficiently thick to allow for a domain wall
as shown in Fig.~\ref{figMauri}. In this region (II) the EB field
is smaller as compared to the M\&B model. This reduction is more
clearly seen further below, when analyzing the azimuthal
dependence of the EB field within the Mauri model.

Comparing the phase diagram of the Mauri model
Fig.~\ref{figMauryPhaseDiagram1}~left to the corresponding one
given by the M\&B model Fig.~\ref{figMBPhaseDiagram1}~left one can
clearly see that region I of both models is very similar with
respect to the  qualitative behavior of the exchange bias field as
a function of the $R$-ratio and, respectively, $P$-ratio. However,
we can compare those curves only when accounting for the variation
of the EB field as a function of the anisotropy of the AF layer.
Both models predict that the EB field depends on the anisotropy of
the AF layer in a similar qualitative manner. Additionally, within
the M\&B model the EB field includes also the dependence on the
thickness of the AF layer, which is not visible in the Mauri
model. The other regions of both phase diagrams are completely
different. Within the Mauri model, the exchange bias does not
vanish at $P<1$, but it continuously decreases, whereas the M\&B
model predicts that the exchange bias field vanishes for $R<1$
leading to enhanced coercivity. Also note that for the weak
coupling region (II) of the Mauri model, the exchange bias would
strongly  depend on the temperature through the anisotropy
constant of the AF layer~\cite{fitzsimmons:2001}.

 {\subsection{Azimuthal dependence of the exchange
bias field}}

Next we analyze the azimuthal dependence of the EB field by
deriving an analytical expression of the EB field as a function of
the rotation angle $\theta$. By solving the second equation of the
system of equation Eqs.~\ref{Maurieq3} with respect to $\beta$,
one finds the angle $\alpha$ as function of $\beta$. Using the
condition for the coercive field as $\beta=\theta-\pi/2$, and
introducing it in the first  line of Eq.~\ref{Maurieq3}, one
obtains the coercive fields $H_{c1}=H_{c2}$. It follows that the
coercive field $H_c (\theta)=0$ and the exchange bias field as
function of the azimuthal angle is~\cite{stamps:2000}:

\begin{equation}
H_{eb}(\theta)=-\frac{J_{eb}}{\mu_0 \, M_F \, t_F} \frac{2
\sqrt{A_{AF} K_{AF}} \cos(\theta)}{\sqrt{J_{eb}^2+4 A_{AF} K_{AF}-
4 J_{eb}\sqrt{A_{AF} K_{AF}}\sin(\theta)}} \label{MauriAzimuth}
\end{equation}

In Fig.~\ref{figMauriAngDep} is plotted the EB bias field
calculated by the expression above for different values of the
$P$-ratio, which was also confirmed numerically. In region I the
EB field is maximum parallel to the field cooling directions
($\theta=0$) only for very large $P$-ratios. When $P$ approaches
the unity, the maximum of the EB field is shifted away from
$\theta=0$, to higher azimuthal angles and has the value:
\begin{equation}
H^{MAX, P\ge1}_{eb}=-\frac{J_{eb}}{\mu_0 \, M_F \,t_F}
\end{equation}
This expression is identical to Eq.~\ref{GMBemax} of  the M\&B
model and shown in Fig.~\ref{figMBAngDep}. The shape of the curves
evolves from an ideal unidirectional shape at $P \rightarrow
\infty$ to a skewed shape at $P\ge1$.

\begin{figure}[!h]
\begin{center}
\includegraphics[clip=true,keepaspectratio=true,width=1\linewidth]{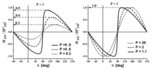}
\caption[Azimuthal dependence of exchange bias within the Mauri
model.]{\label{figMauriAngDep} Azimuthal dependence of exchange
bias as a function of the $\theta$ angle. The dotted line  for
$P$~=~20 can be considered an "ideal" case. The curves are plotted
according to Eq.~\ref{MauriAzimuth}. }
\end{center}
\end{figure}

In region II (Fig.~\ref{figMauriAngDep} left) a drastic change as
compared to region I is seen  for the maximum of the exchange bias
field as a function of azimuthal angle. Its value decreases
monotonically towards zero according to the following expression:
\begin{equation}
H^{MAX, P<1}_{eb}=\, -P \, =\, -\frac{2 \sqrt{A_{AF} \,
K_{AF}}}{J_{eb}}.
\end{equation}
In this region the shape of the curves is also skewed for
$P$-ratios close to unity, but as $P$ decreases towards zero, the
curves acquire a more ideal unidirectional behavior. Similar to
region I, the maximum EB value is shifted away from the field
cooling direction to higher azimuthal angles, whereas in M\&B it
is shifted to lower azimuthal angles.

In the limit of strong $R$ and $P$-ratios ($R,P \gg 1$) the Mauri
and M\&B models give similar results. The differences appear for
the $R$ and $P$-ratios which are close to but higher than one.
This region ($0<R, P \le 5$) can be experimentally explored in
order to decide in favor of one or the other model. In order to
distinguish between the Mauri and the M\&B model, the azimuthal
dependence of the exchange bias offers an excellent tool because
it is visibly different for the two models (shift to higher or
lower azimuthal angles).

The presence of the planar domain wall as described by Mauri
\textit{et al.} appears to have been confirmed experimentally for
first time in Ref.~\cite{scholl:2004:2}. Although the Co/NiO
system studied by Scholl \textit{et al.} did not exhibit a
hysteresis loop shift, the rotation angle $\alpha$ of the AF
planar wall was deduced as function of field. Also, in a recent
publication  by Gornakov \textit{et al.}~\cite{gornakov:2006} show
experimental results that are  similar to the characteristic
curves for the $\alpha$ angle shown in
Fig.~\ref{figMauryPhaseDiagram1}. An
 AF critical thickness  which is often observed experimentally do
 not appear explicitly in the Mauri model. To account for the AF
 thickness dependence, Xi and White~\cite{xi:2000} proposed a
 model which assumes a helical structure for the AF spins during
 the magnetization reversal. The temperature dependence of EB is
 accounted for by the Mauri model  through the anisotropy of the
 AF (H$_{EB}\ \approx \ \sqrt{K_{AF}}$). Since
 the anisotropy constant is rather difficult to  measure for
 thin films, the evidence for the predicted temperature dependence
 remains elusive~\cite{fitzsimmons:2001}. One insufficiency of the
 Mauri model is the inability to predict any changes in
 coercivity. The  domain wall produces only shifted reversible
 magnetization curves~\cite{kim:2005}. Therefore, further refinements of the model
 were introduced which is described in the next section.

\sectionmark{Kim-Stamps approach}
\section{Kim-Stamps Approach - Partial domain wall}
\sectionmark{Kim-Stamps approach}

The approach of Kim and
Stamps~\cite{stamps:2000,stamps:2000prb,kim:2000,kim:2001:1,kim:2001:2,kim:2005}
follows from the work of N\'eel and Mauri~{\textit{et al.}}
extending the model of an extended planar domain wall to the
concept of a partial domain wall in the AF layer.  Biquadratic
(spin-flop) and bilinear coupling energies are used to describe
exchange biased system, where the bias is created by the formation
of partial walls in the AF layer. The model applies to
compensated, partially compensated, and uncompensated interfaces.

\begin{figure}[!h]
\begin{center}
\includegraphics[clip=true,keepaspectratio=true,width=1\linewidth]{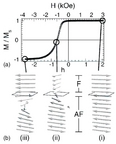}
\caption[Magnetization curve for the ferromagnet/antiferromagnet
system (Kim and Stamps).]{\label{figKimStamps1} (a) Magnetization
curve for the ferromagnet/antiferromagnet system. (b) Calculated
spin structure at three different points of the magnetization
curve. The creation of a partial antiferromagnet domain wall can
be seen in (iii). Only the spins close to the F/AF interface are
shown. From Ref.~\cite{kim:2005}. }
\end{center}
\end{figure}

A typical hysteresis loop indicative for a  partial domain wall in
the AF layer is shown in Fig.~\ref{figKimStamps1}~\cite{kim:2005}.
In saturation at $h > 0$, the F spins are aligned with the
external field while the AF spins are in a perfect N\'eel state,
collinear with the easy axis. The interface spins are antiparallel
to the F layer due to a presumably antiparallel coupling. As the
field is reduced and reversed, the AF pins the F layer by
interfacial exchange coupling until the critical value of the
reversal field is reached at h$_{c}$, where the magnetization
begins to rotate. When it is energetically more favorable  to
deform the AF, rather than breaking the interfacial coupling, a
partial wall twists up as the F rotates. The winding and unwinding
of the partial domain wall in the AF  is reversible, therefore the
magnetization is reversible (no coercivity). This mechanism is
only possible if the AF is thick enough to support a partial wall.
The magnitude of the exchange bias is similar to the one  given by
the Mauri model.
 Neiter the partial-wall theory nor the Mauri model
account, however, for the coercivity enhancement that accompanies
the hysteresis loop shift in single domain materials, which is
usually observed in experiments.

The enhanced coercivity  observed experimentally, is proposed to
be related to the domain wall pinning at magnetic defects. The
presence of an attractive domain-wall potential in the AF layer,
arising from magnetic impurities can provide an energy barrier for
domain-wall processes that controls coercivity. Following the
treatment of pinning in magnetic materials by Braun et.
al.~\cite{braun:1997}, Kim and Stamps examined the influence of a
pointlike impurity at an arbitrary position in the AF layer. As a
result,  the AF energy acquires, besides the domain wall energy,
another term which depends on the concentration of the magnetic
defects. These defects decrease the anisotropy locally and lead to
an overall reduction of  the AF energy. This reduction of the AF
energy gives rise to a local energy minimum for certain defect
positions relative to the interface. The domain walls can be
pinned at such positions and contribute to the coercivity.

\begin{figure}[!h]
\begin{center}
\includegraphics[clip=true,keepaspectratio=true,width=1\linewidth]{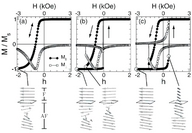}
\caption[Defect-induced asymmetry in hysteresis loops (Kim and
Stamps).]{\label{figKimStamps2} Defect-induced asymmetry in
hysteresis loops. The hysteresis loops are shown for a reduced
exchange defect at $x_L=5$ for three concentrations: (a)
$\rho_J=0.15$,. (b) $\rho_J=0.45$, and (c) $\rho_J=0.75$. The
components of magnetization parallel ($M_{||}$) (dots) and
perpendicular ($M_{\perp}$) (open circles) to the field direction
are shown. The arrows indicate the directions for reversal and
remagnetization. The spin configuration near the interface is
shown for selected field values below the hysteresis
curves~\cite{kim:2005}. }
\end{center}
\end{figure}

Kim and Stamps argue that irreversible rotation of the ferromagnet
due to a combination of wall pinning an depinning transitions,
give rise to asymmetric hysteresis loops. Some examples are given
in Fig.~\ref{figKimStamps2}~\cite{kim:2005}. The loops are
calculated with an exchange defect at $x_L=5$, for three different
values of defect concentration $\rho_J$, where $x_L$ denotes the
defect positions  in the antiferromagnet, with $x_L=0$
corresponding to the interface layer and $x_L=t_{AF}-1$ being the
free surface. At low defect concentrations, the pinning potential
is insufficient to modify partial-wall formation. The resulting
magnetization curve, as shown in Fig.~\ref{figKimStamps2}(a), is
reversible and resembles the curve obtained with the absence of
impurities. Pinning of the partial wall occurs during reversal for
moderate concentrations, which appears as a sharp rotation of the
magnetization at negative fields, as shown in
Fig.~\ref{figKimStamps2}(b). During remagnetization the wall is
released from the pinning center at a different field, thus
resulting in an asymmetry in the hysteresis loop. The release of
the wall is indicated by a sharp transition in M. The energy
barrier between wall pinning and release increases with defect
concentration, resulting in a larger coercivity and reduced bias
as in Fig.~\ref{figKimStamps2}(c).

Within this model, the asymmetry of the hysteresis loops is
interpreted in terms of domain-wall pinning processes in the
antiferromagnet. This explanation appears to be consistent with
some recent work by Nikitenko~\textit{et al.} and
Gornakov~\textit{et al.} on a NiFe/FeMn
system~\cite{nikitenko:2000,gornakov:2006} who concluded that the
presence of an antiferromagnetic wall at the interface is
necessary to explain their hysteresis measurements.
\newpage

\sectionmark{Spin-Glass model}
\section{The Spin Glass model of exchange bias\label{sgmodel}}
\sectionmark{Spin-Glass model}

To overcame the theoretical difficulties in explaining
interconnection between the exchange bias and coercivity, in
Ref.~\cite{radu:jpcm:2006,RaduDiss} is considered a magnetic state
of the interface between F and AF layer which is magnetically
disordered behaving similar to a spin glass system. The
assumptions of the spin glass(SG) model are:

\begin{itemize}
\item{the  F/AF interface is a frustrated spin system (spin-glass
like) } \item{frozen-in uncompensated  AF spins are responsible
for the EB shift  } \item{low anisotropy interfacial AF spins
contribute to the coercivity }
\end{itemize}

Within this model, the AF layer is assumed to contain, in a first
approximation, two types of AF statesv(see Fig.~\ref{figModel}).
One part has a large anisotropy with the orientation ruled by  the
AF spins and another part with a weaker anisotropy which allows
some spins to rotate together with the F spins. This interfacial
part of the AF is a frustrated region (spin-glass-like) and gives
rise to an increased coercivity. The presence of a low anisotropy
AF region can be rationalized as follows: the interface between
the F and AF layer is never perfect, therefore one may assume
chemical intermixing, deviations from stoichiometry, structural
inhomogeneities, low coordination, etc, at the interface to take
place. This leads to the formation of a transition region from the
pure AF state to a pure F state. On average, the anisotropy of
such an interfacial region is reduced. In addition, structural and
magnetic roughness can provide a weak AF interface region.
Therefore, we assume that a  fraction of the frustrated
interfacial spins do rotate almost in phase with the F spins and
that they mediate enhanced coercivity. We describe them by an
effective uniaxial anisotropy $K^{eff}_{SG}$, because they are
coupled to the presumably uniaxial AF layer.

\begin{figure}[!ht]
\begin{center}
\includegraphics[clip=true,keepaspectratio=true,width=1\linewidth]{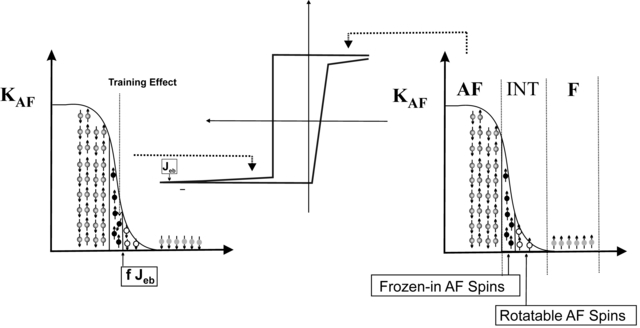}
\caption{ Schematic view of the SG Model. At the interface between
the AF and the F layer the AF anisotropy is assumed to be reduced
leading to two types of AF states after field cooling the system:
frozen-in AF spins and rotatable AF spins. After reversing the
magnetic field, the rotatable spins follows the F layer rotation
mediating coercivity. The frozen-in spins remain largely unchanged
in moderate fields. But some of them will also deviate from the
original cooling state. This could lead to training effects and
also to an open loop in the right side of the hysteresis loop. At
larger applied fields in  the negative direction, the frustrated
frozen-in spins can further reverse  leading to a slowly
decreasing slope of the hysteresis loop. A more complex
antiferromagnetic state consisting of frozen magnetic domains
or/and AF grains can be also reduced to the basic concepts
depicted here. \label{figModel} }
\end{center}
\end{figure}

Generally, one can visualize a spin glass
system~\cite{binder:1986} as a collection of spins which remains
in a frozen disordered state even at low-temperatures. In order to
achieve such a state, two ingredients are necessary: a) there must
be a competition among the different interactions between the
moments, in the sense that no single configuration of the spins is
uniquely favored by all interactions (this is commonly called
`frustration'); b) these interactions must be at least partially
random. This partial random state will be introduced in the M\&B
model as an effective uniaxial anisotropy.

Adding this effective anisotropy to the M\&B model, the free
energy reads~\cite{radu:jpcm:2006}:
\begin{eqnarray}
E&=&-\mu_0\, H \, M_F \, t_F \, \cos(\theta-\beta)\nonumber\\
&+&K_F \,t_F \, \sin^2(\beta)\nonumber\\
&+&K_{AF}(t_{AF}) \, t_{AF} \, \sin^2(\alpha)\nonumber\\
&+& K^{eff}_{SG} \, \sin^2(\beta-\gamma) -J_{eb}^{eff}
\,\cos(\beta-\alpha)
\label{SG1}
\end{eqnarray}
 where,  $K^{eff}_{SG}$ is an effective uniaxial SG
anisotropy related to the frustrated AF spins with reduced
anisotropy at the interface, $J_{eb}^{eff}$ is the reduced
interfacial exchange energy, and $\gamma$  is the average angle of
the effective SG anisotropy. $K_{AF}(t_{AF})$ is the anisotropy
constant of AF layer. To avoid further complications for the
numerical simulations, we neglect  the thickness dependence of the
$K_{AF}$ anisotropy ($K_{AF}(t_{AF})\equiv K_{AF}$). We mention
though, that this dependence could become important for low AF
thicknesses due to finite size effects and due to structurally
non-ideal very thin layers. From now on, the MCA anisotropy of the
ferromagnetic layer ($K_F=0$) will also be  neglected as to
highlight more clearly the influences of AF layer and the SG
interface onto the general properties of the EB systems. Note that
the Zeeman energies of the ferromagnetic-like AF interfacial spins
are neglected in the model since they are usually much smaller as
compared to Zeeman energy of the F layer. Nevertheless, they can
be seen as a  vertical shift of the hysteresis loop (frozen-in AF
spins in Fig.~\ref{figModel}) and as an additional contribution to
the total magnetization (rotatable AF spins in
Fig.~\ref{figModel}).

The model is depicted schematically in Fig.~\ref{figModel}. At the
interface two rather distinct AF phases are assumed to occur in an
EB system: the rotatable AF spins, depicted as open circles   and
frozen-in AF spins shown as filled circles. After field cooling, a
presumably collinear arrangement is depicted in the right hand
panel. After reversing the magnetic field, the rotatable AF spins
follow the F layer rotation mediating coercivity. The frozen-in
spins remain largely unchanged in moderate fields. But some of
them could also deviate from the original cooling state.
Irreversible changes of the frustrated AF spins  lead to training
effects and also to an open loop in the right side of the
hysteresis loop. At larger negative applied fields, the frustrated
frozen-in spins could further reverse leading to a slowly
degreasing slope of the hysteresis loop.A more complex
antiferromagnetic state consisting of frozen magnetic domain state
or/and AF grains can be also reduced to the basic two spin
components depicted in Fig.~\ref{figModel}. Basically, additional
frozen-in spins can occur in the AF layer extending to the
interface and, therefore, leading to an even more interfacial
disorder.

Next, we evaluate numerically the resulting hysteresis loops and
azimuthal dependence of the exchange bias within the SG model.
When the $K^{eff}_{SG}$ parameter is zero, the system behaves
ideally as described by the M\&B model discussed in
~Sec.\ref{MBmodel2}: the coercive field is zero and the exchange
bias is finite. In the other case, when the interface is
disordered we relate the SG effective anisotropy to the available
interfacial coupling energy as follows:
\begin{eqnarray}
& &K_{eff}= (1-f)\, J_{eb}\nonumber\\
& &J_{eb}^{eff}=f \, J_{eb}
\end{eqnarray}
where $J_{eb}$ is the total available exchange energy and $f$ is a
conversion factor describing the fractional order at the
interface, with $f\,=\,1$ for a perfect interface and $f\,=\,0$
for perfect disorder. Some basic models to calculate the available
exchange energy were discussed in the previous sections. For even
more complicated situations when the AF consists of AF grains
and/or AF domains the exchange energy can be further estimated as
described in Ref. ~\cite{takano:1997}.

With these notations we write the system of equations resulting
from the minimization of the Eq.~\ref{SG1} with respect to the
angles $\alpha$ and $\beta$:

\begin{eqnarray}
& &h \,  \sin(\theta-\beta) +\frac{(1-f)}{f} \, \sin(2\,
(\beta-\gamma))+  \sin(\beta-\alpha)=0 \nonumber\\
& &R\, \sin(2\, \alpha)- \sin(\beta-\alpha)=0
\label{SG2b}
\end{eqnarray}
where, $$h=\frac{H}{{-\frac{J_{eb}^{eff}}{\mu_0 \, M_F \, t_F
\,}}} \,=\frac{H}{{-\frac{f\, J_{eb}}{\mu_0 \, M_F \, t_F \,}}}$$
is the reduced applied field and $$R\equiv\frac{K_{AF} \,
t_{AF}}{J_{eb}^{eff}}=\frac{ K_{AF} t_{AF}} { f\, J_{eb}}$$ is the
$R$-ratio defining the strength of the AF layer.

The system of equations above can easily be solved numerically,
but it does provide simple analytical expressions for the exchange
bias. Numerical evaluation provides the $\alpha$ and $\beta$
angles as a function of the applied magnetic field $H$. The
reduced longitudinal component  of magnetization along the field
axis follows from $m_{||}=cos(\beta-\theta)$ and the transverse
component from $m_{\perp}=sin(\beta-\theta)$.

With the assumptions made above the absolute value of the exchange
bias field is directly proportional to $f$. The parameter $f$ can
be called {\it{conversion factor}}, as it describes the conversion
of interfacial energy into coercivity. For example, in the M\&B
phase diagram in the region II  and III corresponding to reduced
$R$-ratios, the exchange bias field is zero and the coercive field
is enhanced as a result of such a conversion of the interfacial
energy into coercive field. 

The idea of reduced interfacial anisotropy at the interface can be
traced back to the  the N\'eel weak ferromagnetism at the surface
of AF particles. N\'eel~\cite{neel:1967} discussed the training
effect as a tilting of the superficial magnetization of the AF
domains. Later, Schlenker ~\textit{et al.}~\cite{schlenker:1986}
suggested that  successive reversals of the F magnetization could
lead to changes of the interface uncompensated AF magnetization
and therefore provide means of going from one ground state to
another. Such multiple interface configurations are similar to a
spin glass system. The spin-glass interface is further discussed
by other
authors~\cite{stoecklein:1988,michel:1988,rubinstein:1999,krivorotov:2001,wang:2004,koch:2005,boubeta:2006,li:2006,ali:2007}.
Exchange bias has been observed recently for spin-glass/F
system~\cite{westerholt:2003}. The interfacial magnetic disorder
was observed through hysteresis loop widening below a critical
temperature point~\cite{berger:2000}. Non-collinearity have been
observed at the AF/F even in saturation~\cite{radu:2002:1,
radu:2002:2, brems:2005}. The frozen spins at the interface were
also observed by MFM~\cite{Kappenberger:2005}. Using element
specific techniques such as soft x-ray resonant magnetic dichroism
(XMCD) and of x-ray resonant magnetic scattering (XRMS),
 both frozen and rotatable AF spins can be
studied~\cite{ohldag:2001,hanke:2001,zaharko:2002,camarero:2003,ohldag:2003,engel:2004,roy:2005,radu:jmmm:2006,ohldag:2006}.
The frozen-in spins appear as a shift of the hysteresis loop along
magnetization axis, whereas the AF rotatable spins exhibit a
hysteresis loop. Moreover, an evidence for SG behavior is recently
reported in thin films~\cite{gruyters:2005} and AF
nanoparticles~\cite{salabas:2006}. Therefore, we believe that
there is enough experimental evidence to consider the interface
between the AF/F layer as a disordered state behaving similar to a
spin-glass system.

\subsection{Hysteresis loops as a function of the conversion factor $f$}

If Fig.~\ref{figSG1} we show simulations of several hysteresis
loops as a function of the conversion factor $f$. We  assume a
strong antiferromagnet in contact with a ferromagnet, where the
interface has different degrees of disorder depicted in the right
column of Fig.~\ref{figSG1}. For the R ratio we  assume the
following value: $R=\frac{K_{AF} \, t_{AF}}{f J_{eb}}=62.5/f$
which corresponds to a 100~\AA~thick CoO antiferromagnetic layer.
The field cooling direction and the measuring field direction are
parallel to the anisotropy axis of the AF. The anisotropy of the
ferromagnet is neglected in the simulations below. For the
interface we have chosen a SG anisotropy oriented 10 degrees away
from the unidirectional anisotropy orientation ($\gamma=10°$). On
the abscissa  the reduced exchange bias field
$h=H/|H_{eb}^{\infty}|$ ($H_{eb}^{\infty} \, \equiv \,
-\frac{J_{eb}^{eff}}{\mu_0\, M_F \, t_F} $)is plotted, which then
can  easily be compared to the M\&B model. With this assumption
the system of Eqs.~\ref{SG2b} was solved numerically.

\begin{figure}[!ht]
\begin{center}
\includegraphics[clip=true,keepaspectratio=true,width=1\linewidth]{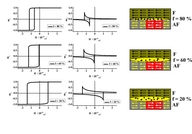}
\caption[Longitudinal ($m_{||}=\cos(\beta)$) and transverse
($m_{\perp}=\cos(\beta)$) components of the magnetization (SG
model).]{Longitudinal ($m_{||}=\cos(\beta)$) and transverse
($m_{\perp}=\sin(\beta)$) components of the magnetization for a
F($K_F=0$)/AF($R=62.5/f, \gamma=10° $,$\theta=0$) bilayer for
different values of the conversion factor $f$ ($f=80~\%, 60~\%,
20~\%$). We observe, that when the AF layer is strong ($R\gg1$)
the hysteresis loops are symmetric when measured along the field
cooling direction. The hysteresis loops are simulated by solving
numerically the Eqs.~\ref{SG2b}. In the right column  is
schematically depicted the layer structure, here the emphasis is
given to the disorder state at the interface. The AF layer is
depicted as consisting  of  magnetic domains which  also
contribute to interface disorder. \label{figSG1} }
\end{center}
\end{figure}

The left column shows the longitudinal component of the
magnetization (parallel to the measuring field direction)
($m_{||}= \cos(\beta)$) whereas the middle column shows the
transverse component of the magnetization ($m_{\perp}=
\sin(\beta)$).

We observe that with decreasing conversion factor $f$ the exchange
bias vanishes linearly. The reduction of the EB field is
accompanied by an increased coercivity. The shape of the
hysteresis loop is close to the results found in literature. For
instance the hysteresis loop with $f=60 \%$ is similar to the data
shown in Refs.~\cite{gokemeijer:2001,gruyters:2000}. The
hysteresis loop with $f=20 \%$ is similar to the data shown in
Refs.~\cite{radu:2003:1,berger:2000}. The longitudinal and
transverse components of the magnetization show that the reversal
mechanism is symmetric. The symmetry is directly related to the
strength of the AF layer, when no training effect is involved. For
the examples depicted in Fig.~\ref{figSG1}, the R-ratio is much
larger than 1 ($R\gg 1$), and therefore the hysteresis loops are
symmetric when measured along the field cooling direction and
along anisotropy axis of the AF layer. The asymmetry of the
hysteresis loops is discussed further below.

\subsection{Phase diagram of exchange bias and coercive field within the spin glass model}

In this section we discuss the phase diagram for the exchange
field and coercive field as a function of the $R$-value within the
SG-model. The additional parameter is the conversion factor $f$.
In Fig.~\ref{figSG3} phase diagrams are shown for reduced exchange
bias and reduced coercive fields as a function of the $R$-ratio
for four different values of the conversion factor $f$. This
allows us to compare directly the behavior of exchange bias fields
as predicted in the SG model and the M\&B model. The reduced
exchange bias field plotted in Fig.~\ref{figSG3} (left panel) is
defined:
$$h_{eb}\, = \frac{H_{eb}}{\frac{J_{eb}^{eff}}{\mu_0 \,  M_F \, t_F}}=\frac{H_{eb}}{f\, H_{eb}^{\infty}} ,$$
where the $H_{eb}$ is the absolute value of the exchange bias
within the SG model and the denominator term $\frac{J_{eb}}{\mu_0
\, M_F \, t_f}$ is the exchange bias field within the ideal M\&B
model.

\begin{figure}[!h]
\begin{center}
\includegraphics[clip=true,keepaspectratio=true,width=1\linewidth]{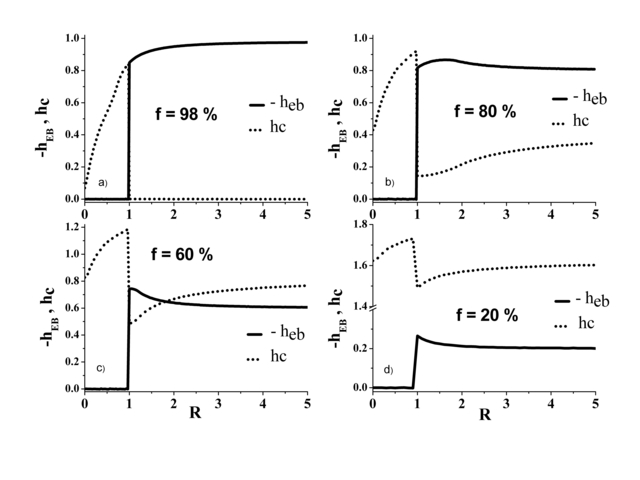}
\caption[Phase diagrams of exchange bias and coercive fields (SG
model). ]{\label{figSG3} The dependence of the reduced exchange
bias field $h_{eb}$ and the reduced coercive field $h_{c}$ as a
function of the R-ratio for four  different values of the
conversion factor $f$. }
\end{center}
\end{figure}
The reduced coercive field shown in Fig.~\ref{figSG3} is defined:
$$h_{c}\, = \frac{H_{c}^{SG}} { \frac{J_{eb}^{eff}}{\mu_0 \, M_F \, t_F}},$$
where $H_{c}^{SG}$ is the absolute value of the coercive field
within the SG model. It has no relation to the coercive field of
the M\&B model because the coercive field within the M\&B model is
considered to be a constant when the exchange bias is finite.

\subsection{Azimuthal dependence of exchange bias and coercive field
within the spin glass model.}

In this section, the azimuthal dependence of exchange bias and
coercive fields within the SG model are discussed and compared
with experimental results of polycrystalline $ Ir_{17}Mn_{83} (15
\, nm)/Co_{70}Fe_{30}(30 \, nm)$ exchange bias
system~\cite{radu:jpcm:2006}.

In Fig.~\ref{figSG4hys} calculated magnetization components are
plotted together with the experimental data points, and in
Fig.~\ref{figSG4azi}b the azimuthal dependence of the coercive
field and exchange bias field are plotted and compared to the
experimental data in Fig.~\ref{figSG4azi}a. The hysteresis loops
were calculated by numerical minimizing the expressions in
Eq.~\ref{SG2b}. The parameters used in the simulation  $f=80\% ,
\, R=5.9/f, \, \gamma= 20^\circ$  do best reproduce  the
experimental data. Furthermore, it is assumed that the AF layer
has a uniaxial anisotropy.  The MCA anisotropy of the F layer is
neglected ($K_F=0$). Therefore, the coercivity which appears in
the simulations is not related to the F properties, but to the
interfacial properties of the F/AF bilayer. 

\begin{figure}[!ht]
\begin{center}
\includegraphics[clip=true,keepaspectratio=true,width=1\linewidth]{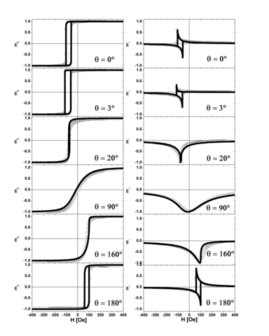}
\caption[Simulated hysteresis loops for different azimuthal angles
(SG model) .]{\label{figSG4hys} Experimental (open circles)and
simulated hysteresis loops (black lines) for different azimuthal
angles. The simulated curves are calculated by the Eq.~\ref{SG2b}
with the following parameters: $f=80\% , \, R=5.9/f, \, \gamma=
20^\circ$~\cite{radu:jpcm:2006}. }
\end{center}
\end{figure}

First we discuss the hysteresis loops shown in
Fig.~\ref{figSG4hys}. The system is cooled down in a field
oriented parallel to the AF anisotropy direction. The hysteresis
loops (solid lines) are simulated for different azimuthal angles
$\theta$ of the applied field in respect to the field cooling
orientation. In Fig.~\ref{figSG4hys} representative hysteresis
curves are shown for the longitudinal ($m_{||}$) and transverse
($m_{\perp}$) magnetization. At $\theta=0°$, the magnetization
curves are symmetric and shifted to negative fields. At
$\theta=180°$, the magnetization curves are also symmetric but
shifted to positive fields. At $\theta=3°$, however, the
longitudinal hysteresis loop becomes asymmetric. The first
reversal at $H_{c1}$ is sharp and the reversal at $H_{c2}$ is more
rounded. This asymmetry is also seen in the transverse component
of the magnetization. The F spins rotate asymmetrically: the
values of the $\beta$ angle depend on the external field scan
direction, being different for swaps from negative to positive
saturation as compared with swaps from positive to negative
saturation. As the azimuthal angle increases, the coercive field
becomes zero. For instance, at $\theta=20°, 90°$~and~$160°$ there
is almost no coercivity. Also, the transverse component of the
magnetization shows that the F spins do not follow a 360° path,
but they rotate within the 180° angular space.

\begin{figure}[!h]
\begin{center}
\includegraphics[clip=true,keepaspectratio=true,width=1\linewidth]{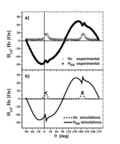}
\caption[Azimuthal dependence of the exchange bias as a function
of the $\theta$ angle (SG model).]{\label{figSG4azi}a) The
experimental coercive field (open symbols ) and exchange bias
(filled symbols) field as a function of the azimuthal angle
$\theta$.d) Simulated coercive field (dotted line) and exchange
bias field (continuous line) as a function of the azimuthal
angle~\cite{radu:jpcm:2006}.}
\end{center}
\end{figure}

In Fig.~\ref{figSG4azi}  the coercive field and the exchange bias
field are extracted from the experimental and simulated hysteresis
loops using Eq.~\ref{EXPRHcEB}. We distinguish the following
characteristics of the $H_c$ and $H_{eb}$: the unidirectional
behavior ($\approx \cos(\theta)$) of the $H_{eb}$ as a function of
the azimuthal angle is (see Fig.~\ref{figMBAngDep}) clearly
visible; additionally, the behavior of the $H_{eb}$ as a function
of the azimuthal angle shows sharp modulations close to the
orientation of the AF uniaxial anisotropy;  the coercive field
$H_c$ has a peak-like behavior close to the orientation of the AF
uniaxial anisotropy, at $\theta=0^\circ$ and $\theta=180^\circ$.
In all cases we find an astounding agreement between calculated
curves and experimental data. It is remarkable, that  the EB field
and the coercive field are completely reproduced by the SG model

Experimentally the azimuthal dependence of the exchange bias field
was first explored for NiFe/CoO bilayers~\cite{ambrose:1997}. It
was suggested that  the experimental results can be better
described with a cosine series expansions, with odd and even terms
for $H_{eb}$ and $H_c$ , respectively, rather than being a simple
sinusoidal function as initially suggested by Meiklejohn and
Bean~\cite{bean:1956,bean:1957}.

The simulations shown in this section are different with respect
to the previous reports on the angular dependence of exchange bias
field~\cite{ambrose:1997,geshev:2001,
mewes:2002,steenbeck:2004,yoo:2005}. One difference is that the
MCA anisotropy of the F is supposed to be negligible when compared
with the coercive fields obtained experimentally, and the sharp
features of the $H_{eb}$ are reproduced numerically rather then
being described by cosine series expansions. Recently, Camarero
et. al~\cite{camarero:2005} reports  on very similar azimuthal
dependent hysteresis loops as shown here. There, an elegant way
based on asteroid curve is used to describe the intrinsic
asymmetry of the hysteresis loops close to the 0° and 180°
azimuthal angle. A unidirectional anisotropy displaces the
asteroid critical curve from the origin. Therefore, if the applied
field is not parallel to the unidirectional anisotropy, the field
sweep line does not pass through the symmetry center of the
asteroid critical  curve leading to inequivalent switching fields
and consequently asymmetric reversals~\cite{camarero:2005}.

\subsection{Dependence of exchange bias field  on the thickness of the antiferromagnetic layer \label{depexbi}}

After a short inspection of the phase diagram of EB and coercive
field (Fig.~\ref{figSG3}) we notice that there is a critical value
for the R-ratio at which the exchange bias vanishes and the
coercive field is enhanced.  This critical value R=1 depends on
four parameters: the anisotropy of the antiferromagnet, the
interfacial exchange coupling parameter, the thickness of the
ferromagnet, and the conversion factor. The conversion factor
further depends on the AF domain  and/or AF grain size, if any.

\begin{figure}[!h]
\begin{center}
\includegraphics[clip=true,keepaspectratio=true,width=1\linewidth]{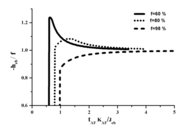}
\caption[AF thickness dependence of exchange bias field (SG
model)]{\label{figSG4}  The normalized exchange bias
$h_{eb}/f=H_{eb}/H^{\infty}_{eb}$ as a function of the $t_{AF}
K_{AF}/J_{eb}$ for different values of the conversion factor $f$.
The anisotropy constant $K_{AF}$ and $J_{eb}$ are assumed to be
constant as the AF thickness is being varied for each $f$. An
asymmetric peak like behavior of the exchange bias field develops
for high values of the conversion factor.}
\end{center}
\end{figure}

 In Fig.~\ref{figSG4} the normalized exchange bias field is
plotted  as a function of the AF thickness and for several
conversion factors. The anisotropy of the AF layer (K$_{AF}$) and
the J$_{eb}$ parameter are assumed to be constant.We notice two
main characteristics of the EB field dependence on the AF
thickness: when $f$ has high values close to unity, the EB field
decreases with decreasing AF thickness.  However, when $f$ is
reduced, a completely different behavior of the EB is observed.
The EB field increases as the thickness of the AF decreases,
developing a peak-like feature. This peak-like behavior for the EB
field at critical AF thickness is a result of enhanced coercivity
which is accounted for by the f-factor. Also, an essential
parameter is the $\alpha$  angle, which describes the rotation of
the AF spins during the magnetization reversal.  The critical
thickness is  preserved by the SG model, but it differs in
magnitude as compared to the M\&B model. Since some interfacial
coupling energy is dissipated as coercivity, the critical
thickness within SG model is lower as compared to the
corresponding one given by the M\&B model. The critical AF
thickness within the SG model follows from the condition R~=~1:
\begin{equation}
t^{SG}_{AF, cr}=\frac{f \, J_{eb}}{K_{AF}}=f \ t^{MB}_{AF,cr},
\end{equation}
where $t^{SG}_{AF, cr}$ and $t_{AF,cr}^{MB}$ are the AF critical
thickness predicted by the SG and M\&B models, respectively.

Experimentally the R-ratio can be tuned by changing the thickness
of the AF and keeping the other three parameters constant. As a
result one observes a critical thickness of the AF layer for which
the EB
disappears~\cite{jungblut:1994,mauri:1987:2,binek:2001,kohlhepp:2006,gredig:2002:2,coehoorn:2002,ali:2003}.
This AF critical thickness can be qualitatively understood within
the MB model. When the hardness of the AF layer is reduced, the AF
spins will rotate under the torque created by the F layer trough
the interfacial coupling constant. The shape of the EB as function
of AF thickness , however, can be different from one system to
another depending on the other three parameters. The most
prominent experimental feature of the EB dependence on the AF
thickness is the development of a peak  close to the critical
thickness. Several proposals were made to describe the peculiar
shapes of EB field dependence on AF thickness. According to the
Malozemoff model, a change in the AF domain size as function of AF
thickness results in a change of exchange bias magnitude.
 Another influences on the AF dependence of
the EB and coercive field are of structural
origin~\cite{jungblut:1994, kohlhepp:2006, kuch:2006}. It has been
shown by Kuch \textit{et. al}~\cite{kuch:2006} that at the
microscopic  level, the coupling between the AF an F layers
depends on the  atomic layer filling and on the morphology of the
interface. The AF-F coupling was observed to vary by a factor of
two between filled and half-filled interface. Moreover,  islands
and vacancy islands at the interface lead to a quite distinct
coupling behavior. Therefore, structural configurations are indeed
contributing to the EB-dependence as function of AF thickness.

The peak-like behavior of the EB field as function of the AF
thickness is strongly dependent on temperature. An almost complete
set of curves, showing a monotonous development of the AF peak
from high to low temperatures was measured by Ali ~\text{et
al.}~\cite{ali:2003}. The data is reproduced in
Fig.~\ref{figSG4ali} together with the simulations based on DS
model. Although the DS model does describe well some experimental
observed features, some discrepancies still exists. For instance
the development of the AF peak as well as the critical AF
thickness as function of temperature are more pronounced in the
experimental data as compared to the DS simulations.

\begin{figure}[!h]
\begin{center}
\includegraphics[clip=true,keepaspectratio=true,width=1\linewidth]{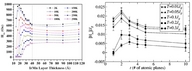}
\caption[AF thickness dependence of exchange bias field (SG
model)]{\label{figSG4ali} Left: IrMn thickness dependence of the
exchange bias field H$_{eb}$  for a number of temperatures. Lines
between the points are a guide to the eye. Right: Prediction of
the DS model for the AF thickness dependence of the exchange bias
field from the stability analysis of the interface AF domains at
different temperatures. (from Ref.~\cite{ali:2003})}
\end{center}
\end{figure}

The SG model, through the conversion factor f, appears to be able
to describe  the evolution of the critical AF thickness(see
Fig.~\ref{figSG4} and Fig.~\ref{figSG4ali}~(left)). Also the shape
of the EB dependence on the AF thickness, from an almost ideal
M\&B type at T=290~K to a pronounced peaked curve at T=2~K is
qualitatively reproduced. Although not considered so far, the
conversion factor seems to be temperature dependent. This can be
understood if we consider the basic assumption of the SG model,
namely the frustration at the interface. Temperature fluctuations
acting on
 metastable spin states cause a variation of the SG anisotropy
as function of temperature. We will not further speculate on the
exact temperature dependence of the f-factor, but we mention that
a numerical analysis for the EB dependencies shown in
Fig.~\ref{figSG4ali} would allow to untangle all the parameters in
the R-ratio. The conversion factor is given by the shape of the EB
curves,  $f J_{eb}$ can be extracted from the temperature
dependence of the EB field at high AF thicknesses, and finally,
the anisotropy constant will be deduced from the critical AF
thickness. Note also, that the SG model has the potential to even
describe the different temperature dependent shapes of the EB
field, namely linear dependence versus more rounded shape: for an
AF thickness close to the critical region, the temperature
dependence of the EB bias will be clearly steeper (linear-like) as
compared to the corresponding one at higher AF thicknesses(more
rounded).

\subsection{The blocking temperature for exchange bias }

Experimentally it is found that the temperature where the exchange
bias effect first occurs is usually lower than the N\'eel
temperature of the AF layer ($T_N$)~\cite{nogues:1999}. This lower
temperature is called  blocking temperature ($T_B$). For thick AF
layers $T_B\leq T_N$, whereas for thin AF layers $T_B\ll
T_N$~\cite{nogues:1999}. Furthermore, the coercive field increases
starting just below $T_N$ (with some exceptions) in contrast to
the EB field, which appears only below $T_B$.

These three experimentally observed  characteristic features can
qualitatively be explained within the M\&B and SG models. In order
to have a non-vanishing EB field in the region with $R\ge 1$, the
following condition has to be be fulfilled:
\begin{equation}
 K_{AF}>\frac{f J_{eb}}{t_{AF}}=K_{AF, crit},\label{SGtb}
\end{equation}
where $K_{AF, crit}$ is the critical AF anisotropy for the onset
of the EB field. For a fixed AF layer thickness, the condition
above sets a critical value for the AF anisotropy for which the EB
can exist. Considering that the AF anisotropy increases steadily
below $T_N$, for large AF layer thicknesses the condition of
Eq.~\ref{SGtb} is fulfilled just below $T_N$, whereas for thinner
AF layers this condition is fulfilled at a correspondingly lower
temperature $T_B$. It is clear from the phase diagrams in
Fig.~\ref{figSG3} and in Fig.~\ref{figSG4} that there is a region
of anisotropy $0<K_{AF}<K_{AF}^{crit}$ where the EB field is zero
and the coercive field is enhanced. It follows that the
enhancement of the coercive field  should be observed  above the
blocking temperature and below the  N\'eel temperature of the AF.
This situation is indeed observed experimentally. For the case of
CoO(25~\AA)/Co layers the coercive field increases starting from
the $T_N^{CoO}=291~K$, whereas the exchange bias field first
appears below $T_B=180~K $~\cite{radu:2003:1}.

Further possible causes for a reduced blocking temperature and for
the behavior of the EB and coercive fields as a function of
temperature are discussed elsewhere: finite size
effects\cite{lederman:1993}, stoichiometry~\cite{tsuchiya:1997} or
multiple phases~\cite{tsunoda:1997}, AF grains~\cite{hu:2004} and
diluted AF~\cite{scholten:2005}.

\section{Training effect}

The training effect refers to the dramatic change of the
hysteresis loop when sweeping consecutively the applied magnetic
field of a system which is in a biased state. The coercive fields
and the resulting exchange bias field versus $n$, where $n$ is the
$n^{th}$ measured hysteresis loop, displays a monotonic
dependence~\cite{paccard:1966,schlenker:1967,schlenker:1967,schlenker:1986}.
The absolute value of $H_{c1}$ and of the EB field decreases from
an initial value at $n=1$ to an equilibrium value at $n=\infty$.
The absolute value of the coercive field $H_{c2}$, however,
displays an opposite behavior, i.e. it increases with $n$. These
features of the training effect is referred to as Type I by Zhang
et. al.\cite{zhang:2002}. The other case when both $|H_{c1}|$ and
$|H_{c2}|$ decrease is called Type. II. In this section we deal
only with the so-called Type I training effect. Several mechanisms
were suggested as a possible cause of the effect. While it is
widely accepted that the training effect is related to the
unstable state of the AF layer and/or F/AF interface prepared by
field cooling procedure, it is not yet well established what
mechanisms are dominantly contributing to the training effect.

N\'eel~\cite{neel:1967} discussed the training effect as a tilting
of the superficial magnetization of the AF domains. This would
lead to a  Type I training effect. N\'eel also discussed that a
creeping effect could lead to a Type II training effect.

Micromagnetic simulations within the DS
model~\cite{miltenyi:2000,nowak:2002} show that the hysteresis
curve is not closed after a complete loop. The lost magnetization
is directly related to a partial loss of the superficial
magnetization of the AF domains, which further leads to a
decreased exchange bias.

Zhang et. al.~\cite{zhang:2001} suggested that the training effect
can be explained by incorporating into the Fulcomer and Charap's
model~\cite{fulcomer:1972:2} positive and negative exchange
coupling between the grains constituting the AF layers.

In Ref.~\cite{hochstrat:2002}, the authors found  direct evidence
for the proportionality between the exchange bias and the total
saturation moment of the heterostructure. The findings were
related  to the prediction of  the phenomenological M\&B approach,
where a linear dependence of the exchange bias on the AF interface
magnetization is expected.

Binek~\cite{binek:2004} suggested that the phenomenological origin
of the training effect is a deviation of the AF interface
magnetization from its equilibrium value. Analytical calculations
in the framework of non-equilibrium thermodynamics leads to  a
recursive relation accounting for the dependence of the $H_{eb}$
field on $n$.

Hoffmann~\cite{hoffmann:2004} argues that only  biaxial AF
symmetry can lead to training effects, reproducing important
features of the experimental data, while simulation with uniaxial
AF symmetry show no difference between the first and second
hysteresis loops.

\begin{figure}[!ht]
\begin{center}
\includegraphics[clip=true,keepaspectratio=true,width=1\linewidth]{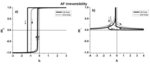}
\includegraphics[clip=true,keepaspectratio=true,width=1\linewidth]{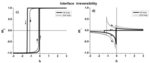}
\caption[Simulations of training effect within SG
model]{\label{figSG5} Simulations of training effect within the SG
model.  Longitudinal a)  and transverse b) components of
magnetization for a scenario (see text) involving irreversible
changes of the AF domain angle $\theta_{AF}$ during the
magnetization reversal. Longitudinal c)  and transverse d)
components of magnetization for the case when only the interface
disorder parameter $\gamma$ increases and the AF state remains
unchanged. }
\end{center}
\end{figure}

Experiments performed by PNR, AMR and Kerr Microscopy
~\cite{radu:2002:1,gredig:2002, radu:2003:1,mccord:2003,
welp:2003,brems:2005,temst:2006,gredig:2006} also support the
irreversible changes taking place at the F/AF interface and in the
AF layer. It has been observed that during the very first reversal
at $H_{c1}$, interfacial magnetic domains are formed and they do
not disappear even in positive or negative "saturation". The
interfacial domains serve as seeds for the subsequent
magnetization reversals. These ferromagnetic domains at the
interface have to be intimately related to the AF domain
state~\cite{nolting:2000}. Therefore, the irreversible changes of
the AF domain state  are responsible for the training effect.
Furthermore measurements have detected  out-of-plane magnetic
moments~\cite{radu:2003:2,welp:2003} hinting at the existence of
perpendicular domain walls in the AF layer, as originally
suggested by Malozemoff. Therefore, irreversible changes of the AF
magnetic domains and  of the interfacial domains during the
hysteresis loops play an important role for the training effect.

\subsection{Interface disorder and the training effect}

In the following we analyze  the AF domains and interface
contributions to the training effect. We assume that a gradual
increase of the interfacial disorder of the F/AF system leads  to
a training effect. Within the SG model, the magnetic state of the
F/AF interface  can be mimiced   through a unidirectional induced
anisotropy $K_{eff}$, which is allowed to have an average
direction $\gamma$, where $\gamma$ is related to the spin disorder
of the interface. Also, we will consider the influence of a
progressive rotation of an AF domain anisotropy during the
reversal. Both situations will be treated below.

In Fig.~\ref{figSG5}a) and b) we show first and second hysteresis
loops (longitudinal and transverse components of magnetization)
calculated with the help of Eq.~\ref{SG2b}. In these calculations
we set the  conversion factor to $f=60\%$ and $R=62.5/f$. We
consider  a drastic change of an AF domain which progressively
rotates its anisotropy axis during the magnetization reversal.
This situation can be accounted for in the SG model  by replacing
the $\alpha$ angle in Eq.~\ref{SG2b} with $\alpha-\theta_{AF}$,
where the $\theta_{AF}$ is the orientation of the AF domain
anisotropy. We also set the $\gamma$ angle to be almost zero.

Following closely the experimental observations~\cite{radu:2002:1,
radu:2003:1}, before the first reversal $\theta_{AF}$ is zero, and
just after the first reversal $\theta_{AF}$ increases towards an
equilibrium value. The first branch of the hysteresis loop appears
rather sharp, therefore we assume that the AF spins and F spins
are  collinear immediately after cooling in a
field~($\theta_{AF}=0°$). For the second branch  of the 1st
hysteresis loop  we consider that $\theta_{AF}=20°$, therefore the
second leg appears more rounded. The transition from
$\theta_{AF}=0°$ to $\theta_{AF}=20°$ is assumed to happen right
after or during the first reversal at $H_{c1}$.  This is in
accordance with the observation that for thin CoO
layers~\cite{radu:2002:1, radu:2003:1} where the disordered
interface appears after the first reversal at $H_{c1}$. Now, the
first branch of the second hysteresis loop is simulated with
$\theta_{AF}=20°$. At the third reversal, we again assume that the
AF domain angle further increases. Therefore, the second branch of
the hysteresis loop is simulated assuming a new value of
$\theta_{AF}=30°$.

The hysteresis loops bear all the features observed
experimentally~\cite{schlenker:1986,zhang:2001,zhang:2002,velthuis:2000,
gruyters:2000,radu:2002:1,radu:2002:2, hochstrat:2002}. More
strikingly,  the transverse component of magnetization shows a
small step increase at $H_{c1}$ and larger increase at $H_{c2}$,
on the reverse path. These are typical features observed
experimentally by PNR~\cite{radu:2002:1, radu:2003:1},
AMR~\cite{gredig:2002,gredig:2006,brems:2005} and
MOKE~\cite{mccord:2003}. Moreover, the transverse component at
saturation behaves very close to recent observations of Brems
\textit{et al.}~\cite{brems:2005} on Co/CoO bilayers and
lithographically nanostructured wires. After field cooling and
before passing through the first magnetization reversal in the
descending field branch, the resistance in saturation (which is
proportional to the square of the orientation of transverse
component of magnetization) reaches its maximum because all spins
are oriented along the cooling field. After going through a
complete hysteresis loop, the resistance at saturation is reduced,
indicating that spins in the F are rotated away from the cooling
field. After reversing the field back to positive saturation the
resistance does not recover its initial value. Moreover, the
untrained state can be partially reinduced by changing the
orientation of the applied magnetic field~\cite{brems:2005} which
can be interpreted as a further indication of AF domain rotation
during the reversal.

Next, we consider only the interface disorder through a
progressive change of $\gamma$ angles. In Fig.~\ref{figSG5}b) and
c) we show first and second hysteresis loops calculated with the
help of Eq.~\ref{SG2b}. In these calculations we consider that the
AF is strong, $R=62.5/f$. For the conversion factor we take a
value of $f=60\%$. Also, we assume the average AF orientation to
be parallel to the field cooling orientation ($\theta_{AF}=0$).
Following closely the experimental observations, before the first
reversal $\gamma$ is zero, while just after the first reversal,
$\gamma$ increases towards an equilibrium value. The first branch
of the hysteresis loop appears rather sharp, therefore we assume
that the AF spins and F spins are  collinear immediately after
cooling in a field~($\gamma=0$). For the second branch  of the 1st
hysteresis loop  we consider that $\gamma=10°$, therefore the
second leg appears rather rounded. The transition from $\gamma=0$
to $\gamma=10°$ is assumed to happen right after or during the
first reversal at $H_{c1}$.   The first branch of the second
hysteresis loop is simulated with $\gamma=10°$. At the third
reversal , we again assume that the disorder of the interface
increases. Therefore, the second branch of the hysteresis loop
(and right after during during the reversal at $H_{c1}$ of the
second loop) is simulated assuming a new value of $\gamma=20°$.

The simulations above implies a viscosity-like behavior  to the
disordered interface~\cite{fulcomer:1972:3}. For example, when the
F magnetization acquires an angle  with respect to the
unidirectional anisotropy, the torque exerted on the interfacial
 spins  will drag them away from the initial direction
set by the field cooling. Reversing the magnetization back to
positive directions the $K_{eff}$ spins will not follow
(completely), they remain close to this position (viscosity). This
is because the maximum torque exerted by the F spins was already
acting at negative fields, while for positive fields it is much
reduced. When measuring again the hysteresis loop, at negative
coercive field, the $K_{eff}$ spins will rotate even further and
so on. Therefore, the angle of the $K_{eff}$ anisotropy and/or of
the AF domains  increases after each hysteresis loop, similar to a
rachet, causing a decreased exchange bias field.

Comparing the hysteresis curves shown in Fig.~\ref{figSG5}a) with
the ones in Fig.~\ref{figSG5}c) one notices the same qualitative
characteristics. This is not the case for the transverse
magnetization curve. The F spins rotate only on the positive side
for the first case, whereas for the second case  the F
magnetization rotation covers the entire 360° angular range. The
anisotropic magnetoresistance (AMR) and PNR hides the chirality of
the ferromagnetic spin rotation as they provide $\sin^2(\beta)$
information, whereas MOKE is sensitive to the chirality as it
provides $\sin(\beta)$ information. Therefore, measuring both
hysteresis components by MOKE, can help to distinguish between the
dominant influence on the training effect: AF domain (or/and
grain) rotation versus SG interface instability. The training
effect is discussed furthermore in section~\ref{treff}.

\subsection{Empirical expression for the training
effect\label{treff}}

The very first empirical expression for
training~\cite{paccard:1966,schlenker:1967} effects suggested a
power law dependence  of the coercive fields and the exchange bias
field as a function of cycle index $n$:
\begin{equation}
H^n_{eb}=H^{\infty}_{eb} + \frac{k}{\sqrt{n}}, \label{SG5tr1}
\end{equation}
where k is an experimental constant. This expression follows well
the experimental dependence of the EB field for $n\ge2$, but when
the very first point is included to the fit, then the agreement is
poor.

\begin{figure}[!ht]
\begin{center}
\includegraphics[clip=true,keepaspectratio=true,width=1\linewidth]{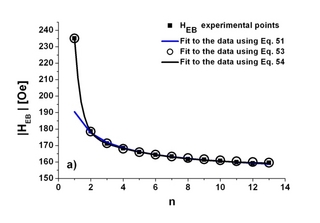}
\includegraphics[clip=true,keepaspectratio=true,width=1\linewidth]{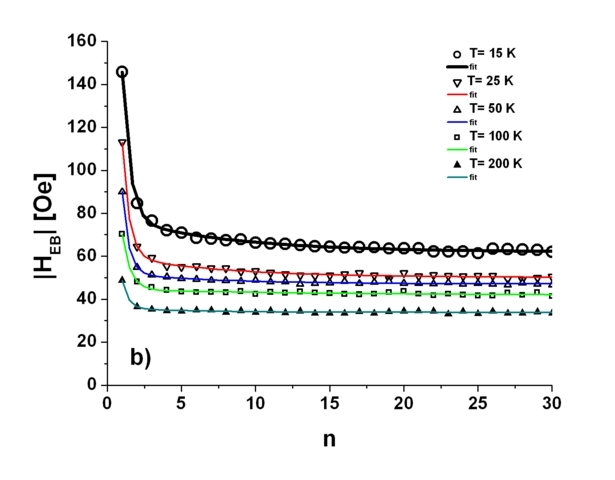}
\caption[Exchange bias as function of the loop index
n.]{\label{figSG6}  a) Exchange bias as function of the loop index
$n$. The gray line is the best fit to the data using
Eq.~\ref{SG5tr1}. The open circles are the best fit~\cite{binekPC}
to the data using Eq.~\ref{SG5tr2007}.  The black line is the best
fit to the data using Eq.~\ref{SG6}. b)Temperature dependence of
the training effect. The lines are the best fit to the data using
Eq.~\ref{SG6}}
\end{center}
\end{figure}

 Binek~\cite{binek:2004} has shown by using
non-equilibrium thermodynamics that using a recursive relation,
the evolution of the EB field as a function of $n$, can be well
reproduced for all cycle indexes ($n\ge1$). The recursive
expression reads:

\begin{equation}
H^{n+1}_{eb}-H^n_{eb}=-\Gamma \, (H^n_{eb} -H^{\infty}_{eb})^3,
\label{SG5tr2}
\end{equation}
where $\Gamma$ is a physical parameter which, for $n\gg1$, was
directly related to the $k$ parameter of Eq.\ref{SG5tr1}. It was
shown that a satisfactory agreement between the Eq.~\ref{SG5tr1}
and Eq.~\ref{SG5tr2} is achieved for $n\ge 3$. Therefore the
approach of Binek appears to provide the phenomenological origin
of the hitherto unexplained power-law decay of the EB field with
increasing loop index $n>1$. The analytic expression
(Eq.~\ref{SG5tr2}) was further tested for temperature dependent
training effect~\cite{binek:2005}.

More recently the Eq.~\ref{SG5tr2} has been further refined by
extending the free energy expansion with a correction of the
leading term. The new equation for training effect
reads~\cite{sahoo:2007}:
\begin{equation}
H^{n+1}_{eb}-H^n_{eb}=-\gamma_b \, (H^n_{eb}
-H^{\infty}_{eb})^3-\gamma_c \, (H^n_{eb} -H^{\infty}_{eb})^5,
\label{SG5tr2007}
\end{equation}
where the new $\gamma_c$ parameter results from the higher order
expansion of the free energy and hence $\gamma_c \ll \gamma_b$.
The $\gamma_b$ parameter is similar to the $\Gamma$  in
Eq.~\ref{SG5tr2}. Both  parameters $\gamma_b$ and $\gamma_c$
exhibits a exponential dependence on the sweep rate for measuring
the hysteresis loop. A number of three fit parameters is required
for both Eq.~\ref{SG5tr2} and Eq.~\ref{SG5tr2007}, but the last
equation provides better fitting results for moderate sweep rates.

In the following we analyze another type of expression which
reproduces the dependence of the coercive field and exchange bias
field as a function of the loop index $n$ and for different
temperatures. It is based on the simulations shown in the previous
section. There it was argued that the training affect is related
to the interfacial spin disorder. With each cycle the spin
disorder increases slightly, thereby decreasing the exchange bias
field. Additional effects are related to the  AF domain size that
also affects the magnitude of the EB and $H_c$ fields. Both
contributions cause a gradual decrease of exchange bias as a
function of cycle $n$. They can be treated probabilistically. We
suggest the following expression to simulate the decrease of the
EB as a function of $n$:

\begin{equation}
H^n_{eb}=H^{\infty}_{eb} + A_f\exp (-n/P_f)+A_i\exp(-n/P_i),
\label{SG6}
\end{equation}
where, $H^n_{eb}$ is the exchange bias of the $n^{th}$ hysteresis
loop, $A_f$ and  $P_f$ are parameters related to the change of the
frozen spins, $A_i$ and $P_i$ are  parameters related to the
evolution of the interfacial disorder. The $A$ parameters have
dimension of Oersted while the $P$ parameters have no dimension
but they are similar to a relaxation time, where the continuous
variable "time" is replaced by a discrete variable $n$. We expect
that the interfacial contribution  sharply  decreases with $n$ as
the anisotropy of the interfacial spins is reduced (low AF
anisotropy spins), while the contribution from the "frozen" AF
spins belonging to the AF domains ("frozen-in" uncompensated spins
) appear as a long decreasing tail as they are intimately embedded
into a much stiffer environment.

In the following we show fits to the "trained" exchange bias
field. In Fig.~\ref{figSG6}a) The EB field of thirteen consecutive
hysteresis loops  were measured at T~=~10~K and are plotted as a
function of loop index for a CoO(40~\AA)/Fe(150~\AA)/Al$_2$O$_3$
bilayer. Three fits are shown: one using the empirical relation
Eq.~\ref{SG5tr1}, the second one is a fit performed by
Binek~\cite{binekPC} using the Eq.~\ref{SG5tr2007}, and the third
one using Eq.~\ref{SG6}. We observe that the fit using the
Eq.~\ref{SG5tr1} ($H^{\infty}_{eb}=146.6 \, Oe, k=44 \,Oe$)
follows well the experimental curve for $n\ge 2$.  However, the
best fits are obtained using Eq.~\ref{SG5tr2007} and
Eq.~\ref{SG6}. The best fit parameters using Eq.~\ref{SG5tr2007}
are~\cite{binekPC}: $H^{\infty}_{eb}=148.632 \, Oe$,
$\gamma^{b}_{eb}=0.00029472$, and $\gamma^{c}_{eb}= \, -2.772 \,
10^{-8}$. The parameters obtained from  fits to the data using the
Eq.~\ref{SG6} are: $H^{\infty}_{eb}=158 \, Oe, A_f=25.87\, Oe,
P_i=4.33, A_i=739.14\, Oe, P_i=0.39$. Within the SG approach, we
distinguish, indeed, a sharp contribution due to low anisotropy AF
spins at the interface and a much weaker decrease from the
"frozen-in" uncompensated spins.

The temperature dependence of the training effect for a
monocrystalline Fe(150~\AA)(110)/CoO(300~\AA)(111)/Al$_2$O$_3$
bilayer~\cite{gnowak:2004,gnowak:2007,RaduDiss} is shown in
Fig.~\ref{figSG6}b). The sample was field cooled in saturation to
the measuring temperature where 31 consecutive hysteresis loops
were measured. The fits to the data using Eq.~\ref{SG6} are shown
as continues lines  in Fig.~\ref{figSG6}b).

We distinguish three main characteristics related to the
temperature dependence of the training effect:
\begin{itemize}
\item{each curve shows two regimes, a fast changing one and a
slowly decreasing tail;} \item{the "relaxation times" ($P_i$ and
$P_f$) do not visibly depend on the temperature; } \item{the
interface transition towards the equilibrium state is
approximatively ten times faster then the transition of the
"frozen" spins towards their stable configuration.}
\end{itemize}


\section{Further characteristics of EB-systems}

\subsection{Asymmetries of the hysteresis loop\label{expassym}}

A curios   characteristic of EB systems is the often observed
asymmetry between the two branches of the hysteresis loop for
descending and ascending magnetic fields. The hysteresis loop
shape of an isolated ferromagnet is with no exceptions symmetric
with respect to the field and magnetization axis.  This is not the
case for  an exchange bias system, where the unidirectional
anisotropy and the stability of the AF can result in asymmetries
of the hysteresis loops and of the magnetization reversal modes.
One can distinguish two different classes of the hysteresis loop
asymmetries. One of them can be assigned to intrinsic properties
of the EB systems which lacks training effects, and the another
one is intimately related to irreversible changes of the AF domain
structure during the magnetization reversal.\\
i) In the first category we encounter four different situations of
asymmetryic magnetization reversal
all related to a stable interface without training effect: \\
a) the first branch of the hysteresis loop is much extended
compared to the ascending branch (see Fig.~\ref{asymmNik}). This
asymmetric hysteresis loop was observed in FeNi/FeMn
bilayers~\cite{nikitenko:2000}. The underlying mechanism is
related to a Mauri type mechanism for exchange bias  where a
parallel domain wall (exchange spring ) is formed in the AF layer.
The reversal is understood in terms of domain wall pinning in the
antiferromagnet~\cite{nikitenko:2000,kim:2005,gornakov:2006}.

\begin{figure}[!h]
\begin{center}
\includegraphics[clip=true,keepaspectratio=true,width=1\linewidth]{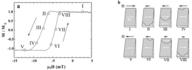}
\caption[Asymmetry PyFeMn Nikitenko]{\label{asymmNik} a)
Asymmetric hysteresis loop (a)  of a NiFe/FeMn bilayer and
schematics of domain structure at different stages of
magnetization reversal~\cite{nikitenko:2000}. }
\end{center}
\end{figure}

b) coherent rotation during the first reversal at H$_{c1}$, domain
wall nucleation and propagation at H$_{c2}$ (see
Fig.~\ref{asymFitz}). Such asymmetric magnetization reversal  has
been observed by PNR in Fe/FeF$_2$ and Fe/MnF$_2$
systems~\cite{fitzsimmons:2000}. This  asymmetry
  depends on  the relative orientation of
the field cooling direction  with respect to the twin structure of
the AF layer. The reversal asymmetry mentioned above takes place
when the FC is parallel oriented with a direction bisecting the
anisotropy axes of the two AF structural domains. When the field
cooling orientation is parallel to the anisotropy axes of one AF
domain, the reversal mechanism is symmetric, i. e.  for both
branches of the hysteresis loop magnetization rotation prevails.

\begin{figure}[!h]
\begin{center}
\includegraphics[clip=true,keepaspectratio=true,width=1\linewidth]{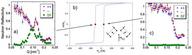}
\caption[Asymmetry PyFeMn Nikitenko]{\label{asymFitz} a)
Asymmetric hysteresis loop (b)  of a Fe/MnF$_2$ bilayer and and
the corresponding neutron reflectivity curves: a) the PNR curves
recorded at the first coercive field, H$_{c1}$ and c) the PNR
curve recorded at H$_{c2}$. The lack of SF reflectivity at
H$_{c2}$ suggests that the reversal proceeds by domain wall
nucleation and propagation, whereas at H$_{c1}$ the magnetization
reverses by rotation~\cite{fitzsimmons:2000,fitzsimmons:2001}. }
\end{center}
\end{figure}

c)  sharp reversal on the descending branch and rounded reversal
on the ascending one (see Fig.~\ref{asymjpcm}).  This asymmetry
has recently been  clarified by studies of  the azimuthal
dependence of exchange bias in IrMn/F
bilayers~\cite{camarero:2005, radu:jpcm:2006,mccord:unpbl}. It is
an intrinsic property of the EB bilayer systems  and it takes
place whenever the measuring external field is offset with respect
to the field cooling orientation.  By simply analyzing  the
geometrical asteroid solutions, it becomes obvious that  the sweep
line does not symmetrically cross the shifted asteroid when the
field is not parallel to the unidirectional  anisotropy. Actually,
this is a peculiar case of asymmetry which can be understood even
within the phenomenological model for exchange
bias~\cite{camarero:2005} which assumes a rigid AF spin structure.
Within the SG and M\&B model~\cite{radu:jpcm:2006} such asymmetric
reversals can be simulated over a wide range of AF thicknesses and
anisotropies. Also, within  the DS
model~\cite{beckmann:2003,beckmann:2006}, the effect of an offset
measuring field axis with respect to the anisotropy axes of the AF
and F layer can result in  asymmetric reversal modes.
\begin{figure}[!h]
\begin{center}
\includegraphics[clip=true,keepaspectratio=true,width=1\linewidth]{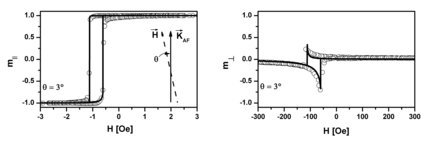}
\caption{\label{asymjpcm}  Asymmetric hysteresis loop   of a
IrMn/CoFe bilayer along an offset $\theta=3$°
angle~\cite{radu:jpcm:2006}. }
\end{center}
\end{figure}

d)  the descending part is  steeper, while the ascending branch is
more rounded (see Fig.~\ref{figSG2}). This asymmetry of the
hysteresis  loop needs to be distinguished from the previous ones,
since it occurs when the external field is oriented parallel  with
respect to the anisotropy axis. It is observed in EB bilayers with
thin antiferromagnetic layers or for systems containing low
anisotropy AF layers. We call these antiferromagnets  weak
antiferromagnets and characterize them by the R-ratio. When the
R-ratio is slightly higher than one (weak AF layers), then the
asymmetry of the hysteresis loop can be reproduced within the SG
model. When the R-ratio is much higher than one (strong AF
layers), then the hysteresis loops are symmetric  as shown in
Fig.~\ref{figSG1}. To account for the asymmetry
\begin{figure}[!h]
\begin{center}
\includegraphics[clip=true,keepaspectratio=true,width=1\linewidth]{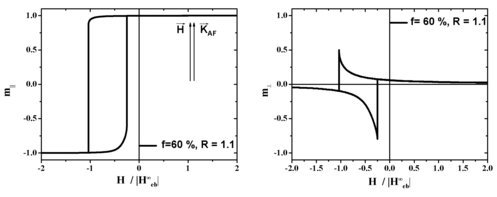}
\caption[Longitudinal  and transverse components component of the
magnetization for a weak antiferromagnet (SG
model).]{\label{figSG2}Longitudinal (left) and transverse
components (right)  of the magnetization vector for an week
antiferromagnet: $R=1.1$. The hysteresis loop is asymmetric: the
descending part is steeper than the ascending part. The asymmetry
is clearly seen also in the transverse component of the
magnetization. }
\end{center}
\end{figure}
we consider the following example where it is essential  that the
AF layer is weak but the R ratio is higher than 1: $R\, =\, 1.1$,
$f\, =\, 60 \%$ , $\gamma\, =\, 5°$, and $\theta=0°$. For these
values the minimization of the free energy is evaluated
numerically. The results are plotted in Fig.~\ref{figSG2}. The
 longitudinal and transverse components of the magnetization vector
is shown as a function of the reduced field
$h=H/|H_{eb}^{\infty}|$. We clearly recognize  that the hysteresis
loop is asymmetric: steeper on the descending leg and more rounded
on the ascending leg. The asymmetry is due to the large rotation
angle of the AF spins during the F magnetization reversal. This
asymmetry has not received experimental recognition so far,
therefore it remains a prediction of the SG model. \\
ii) The second class of  asymmetric hysteresis loop is directly
related to the training effect, and therefore to the stability of
the AF layer and AF/F magnetic interface during the magnetization
reversal.

a)sharp reversal  at H$_{c1}$ and more rounded reversal on the
ascending branch, at H$_{c2}$ (see Fig.~\ref{asymPNR}). Measuring
subsequent hysteresis loops, the rounded reversal character  does
not change but appears also at H$_{c1}$. This type of asymmetry,
being related to the training effect, is frequently reported in
different exchange bias systems~\cite{paccard:1966,schlenker:1986,
gruyters:2000,velthuis:2000,
zhang:2001,hellwig:2002,zhang:2002,radu:2002:1,radu:2002:2,
hochstrat:2002,mccord:2003,welp:2003,temst:2003,girgis:2005,temst:2006}.
Its underlying microscopic origin, has  been recently demonstrated
to stem in irreversible changes that occurs in the AF layer.
Polarised Neutron Reflectivity measurements have revealed that at
H$_{c1}$ the reversal proceeds by domain wall nucleation and
propagation~\cite{radu:2002:1, gierlings:2002,
radu:2002:2,lee:2002, welp:2003,
temst:2003,girgis:2005,temst:2006}. At the second coercive field ,
magnetization rotation  is the reversal mechanism. Moreover, by
analyzing the AF/F interface~\cite{radu:2002:1, radu:2002:2}, it
has been observed experimentally that a transition from a
collinear state to an non-collinear disorder state occurs. It
suggests that the AF layer in (CoO thin layer)/F evolves from a
single AF to a multiple AF domain state. In a  AF layer that
exhibits a domain state the anisotropy orientation in different
domains is laterally distributed causing a reduced coercive and
exchange bias field. Note that a thick CoO film is suggested to be
already in a domain state~\cite{nowak:2002} after field cooling,
therefore one would expect a variation of the loop asymmetry (and
training effect) with respect to the thickness of the AF layer.

\begin{figure}[!h]
\begin{center}
\includegraphics[clip=true,keepaspectratio=true,width=1\linewidth]{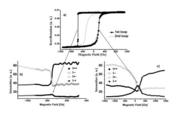}
\caption[Asymmetry PNR]{\label{asymPNR}(a) MOKE hysteresis loop of
a CoO/Co bilayer after field cooling to 50~K in an external field
of 2000~Oe. The black dots denote the first hysteresis loop, the
dotted line the second loop. Any further loops are not
significantly different from the second. (b) and (c) Hysteresis
loops recorded by polarized neutrons from the same sample but at
10 K. I$+\,+$, I$-\,-$, I$+\, -$ and I$-\,+$ refer to non-spin
flip and spin-flip intensities as a function of external magnetic
field. (Ref.~\cite{radu:2002:1})}
\end{center}
\end{figure}



\subsection{Temperature dependence of the rotatable  AF spins (coercivity)}

We discuss here the temperature dependence of the interfacial
properties. Experimentally this can best be studied by an element
selective
method~\cite{hanke:2001,zaharko:2002,engel:2004,roy:2005,radu:jmmm:2006}
to distinguish between the hysteresis of the F and AF layer.
Element specific hysteresis loops have been studied for
Fe/CoO~\cite{radu:jmmm:2006}, which highlights the behavior of the
rotatable interfacial AF spins.



\begin{figure}[!h]
\begin{center}
\includegraphics[clip=true,keepaspectratio=true,width=1\linewidth]{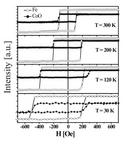}
\end{center}
\caption[The temperature dependence of the exchange biased
hysteresis loops]{\label{xrmsfig4} The temperature dependence of
the exchange biased hysteresis loops measured at $L_{3}$
absorption edges of Co (E=780 eV, closed symbols) and Fe (E=708.2
eV, open symbols). Scattering angle is
2$\theta=~32^{\circ}$~\cite{radu:jmmm:2006}. }
\end{figure}

The exchange bias hysteresis loops measured at the $L_{3}$
absorption edges of Co (E=780 eV, closed symbols) and Fe (E=708.2
eV, open symbols) and for different temperatures are shown in
Fig.~\ref{xrmsfig4}. After FC to 30~K, several hysteresis loops
were measured in order to eliminate training effects.
Subsequentially the temperature was raised
stepwise, from low to high T. For each temperature an
element-specific hysteresis loop at the energies corresponding to
Fe and CoO, respectively, was measured. The hysteresis loops of Fe
as a function of temperature show a typical behavior. At low
temperatures an increased coercive field and a shift of the
hysteresis loop is observed. As the temperature is increased, the
coercive field and the exchange bias decrease until the blocking
temperature is reached. Here, the exchange bias vanishes and the
coercive field shows little changes as the temperature is further
increased.

A ferromagnetic hysteresis loop corresponding to the CoO layer is
observed for all temperatures, following closely the hysteresis
loop of Fe , with some notable differences. It appears that the
ferromagnetic components of CoO develop higher coercive fields
than Fe below the blocking temperature. This is an essential
indication that the AF rotatable spins mediate coercivity between
the AF layer and the F one, justifying the conversion factor
introduced in the SG model.  After careful analysis of the element
specific reflectivity data~\cite{freeland:1997,radu:jmmm:2006},
one can conclude that  a positive exchange coupling across  the
Fe/CoO interface. The ferromagnetic moment of CoO is present also
above the N\'eel temperature. Here, the AF layer is in a
paramagnetic state, therefore the coercive fields for the Fe and
CoO rotatable spins are equal.

\subsection{Vertical  shift of magnetization curves (frozen AF spins)}

\begin{figure}[!h]
\begin{center}
\includegraphics[clip=true,keepaspectratio=true,width=1\linewidth]{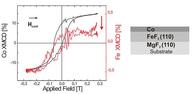}
\caption[Vertical Shift of the magnetization]{\label{figOhldag}
Element-specific Co (black) and Fe (gray) hysteresis loops
acquired at T=~15~K after field cooling in +200 Oe along the
FeF$_2$~[001] axis, parallel to the AF spin axis. The direction of
the cooling field and the vertical shift of the Fe loop at
T~=~15~K is indicated by arrows. Reproduced
from~\cite{ohldag:2006}.}
\end{center}
\end{figure}

A vertical shift of magnetization has been observed frequently
~\cite{kagerer:2000,nogues:2000,hochstrat:2002,keller:2002,
nowak:2002,
liu:2004,ohldag:2003,papusoi:2006,ohldag:2006,iglesias:2006,nogues:2005,iglesias:2006,salabas:2006,mumtaz:2006}
and is considered to have several origins related to the different
mechanisms for exchange bias. Within the M\&B model a AF monolayer
in contact to the F layer is assumed to be uncompensated, but
still being part of the AF lattice. At most one could expect a
contribution to the macroscopic or microscopic  magnetization
equal to that of the net magnetization of an AF monolayer and this
only by probing an AF layer consisting of an odd number of
monolayers. The Mauri mechanism for exchange bias is not likely to
result in a vertical shift of the hysteresis loop, since the AF
interface is compensated. Within the SG, Malozemoff, and DS models
for EB, a small vertical  shift is intrinsic. At the interface
between the AF and F layer, a number of  frozen AF  spins will be
uncompensated due to the proximity of the F layer. Their
orientation is either parallel or antiparallel oriented with
respect to the F spins, depending on the type of coupling(direct
or indirect exchange). They contribute to  the magnetization of
the system. In case of an indirect exchange coupling mechanism,
the hysteresis loop should be shifted
downwards~\cite{nogues:1996,nogues:2000}, whereas in case of
direct exchange coupling the magnetization curve should be shifted
upwards. The other part of the interface magnetization, namely the
rotatable AF spins, do not cause any shift since they rotate in
phase with the F layer. Within the DS model  AF domains
 cause an
additional shift to the macroscopic magnetization curve along the
magnetization axis~\cite{nowak:2002,keller:2002,papusoi:2006}.

To demonstrate the shift of the magnetization we discuss a recent
experiment by Ohldag~\textit{et al.}~\cite{ohldag:2006} using
XMCD~\cite{stohr:book}. The evolution of the dichroic signal as
function of the magnetic field, providing the element specific
hysteresis loops, is shown Fig.~\ref{figOhldag}.

The sample structure is
Pd(2~nm)/Co(2.8~nm)/FeF$_2$(68~nm)(110)/MgF$_2$(110)/substrate and
has been grown by via molecular beam epitaxy. The F layer is a
polycrystalline Co whereas the AF layer is a FeF$_2$(110) untwined
single crystalline layer. The system exhibits a positive exchange
bias at large cooling fields~\cite{nogues:1996}. For weak cooling
fields the exchange bias curve is shifted to the negative fields,
as usual for all  EB systems.  The microscopic origin of the
positive exchange bias is an antiferromagnetic coupling at the
F/AF interface~\cite{nogues:1996}. Although, the mechanism was
clearly demonstrated by magnetometery measurements, the
microscopic investigation of the AF interface provides more
detailed information.

The hysteresis loop of Co is typical and appears symmetric with
respect to the magnetization axis (see Fig.~\ref{figOhldag}). This
is not the case for the AF interface magnetization which shows a
twofold behavior: a) some interfacial AF spins are parallel
oriented with respect to the F spins and both are rotating almost
in phase; the AF hysteresis loop is shifted downwards with respect
to the magnetization axis. This seems to be a direct proof of the
preferred antiparallel coupling between the F Co and the
 AF Fe magnetization at the AF-F interface. The
interfacial AF Fe moments are aligned during FC by the exchange
interaction which acts as an effective field on the uncompensated
spins, most likely confined to the interface~\cite{ohldag:2006}.
An AF uncompensated magnetization in FeF$_2$ can also occur due to
piezomagnetism, which is allowed by symmetry in rutile-type AF
compounds and may be induced by stresses occurring below the
N\'eel temperature~\cite{binekbook,binekPC}.

These experimental observations can also be  described by the SG
model. During the field cooling procedure the frozen-in spins
depicted  in Fig.~\ref{figModel} will be pointing opposite to the
F spins as to fulfill the indirect exchange condition at the F/AF
interface. The rotatable ones remain unchanged in the figure,
being parallel aligned with the F spins. The orientation of the
frozen spins is governed by the exchange interaction at the F/AF
interface, whereas the rotatable spins are align by the F layer
and the external field. Upon reversal the rotatable spins will
follow the ferromagnet whereas the frozen spins remain pinned,
leading to a shift downwards of the interfacial AF hysteresis
loop. As the system described above does not exhibits training
effects,  no irreversible changes occurs during the magnetization
reversal. In high enough cooling fields, the frozen-in spins will
align parallel to the cooling field becoming also parallel with
the rotatable AF and F spins. This will cause a positive shift of
the hysteresis loop (positive exchange bias), since the
orientation of the AF frozen spins is negative with respect to the
coupling sign~\cite{nogues:1996}.

The experimental results described above  could most likely be
described also by DS model and the Malozemoff model if  magnetic
domains in the AF layer will be confirmed.

A relative vertical shift of the magnetization  related to
training effect is described  by Hochstrat~\text{et
al.}~\cite{hochstrat:2002} for a NiO(0001)/Fe(110) exchange bias
system. The antiferromagnet is a single crystal NiO whereas the
ferromagnetic material is an Fe layer deposited under ultrahigh
vacuum condition. Upon successive reversals of the F layer a
decreased of the total magnetization   was observed by SQUID
magnetometery. The variation of the vertical shift as a function
of the hysteresis loop index n was correlated to a decrease of the
AF magnetization. A linear correlation between the AF
magnetization, deduced from the vertical shift, and the exchange
bias field during training was found, suggesting that the training
effect may be related to a reduced AF magnetization along the
measuring field axis and as a function of the loop index.

In Fig.~\ref{vshit1} we show another situation where a vertical
shift is visible in  macroscopic magnetization curves measured by
SQUID magnetometery~\cite{brems:2006:unpubl}. The system is a
polycrystalline CoO(2.5 nm)/Co(15 nm) bilayer grown by magnetron
sputtering~\cite{brems:2005}. The hysteresis loop at room
temperature  and along the easy magnetization axis of the F layer
is symmetric with respect to the magnetization axis and shows a
low coercive field. Upon field cooling the system through the
N\'eel of the CoO layer ($T_N=291~K$) to the measuring temperature
T=4.2~K the system is set in an exchange bias state. Then, a
hysteresis loop is measured by sweeping the field from 2000~Oe to
-2000~Oe and back. By comparing this hysteresis loop with the one
measured at room temperature one clearly observes a vertical shift
up along the magnetization axis. Moreover, this up-shift is due to
the ferromagnetic Co spins which do not fully saturate at -2000
Oe. Previous studies have shown that after the reversal at
H$_{c1}$, the AF CoO layer breaks into AF domains exhibiting an
anisotropy distribution. Due to the strong coupling between the F
and the AF domains, the F layer cannot easily be  saturated. Next,
after repeating the field cooling procedure, another hysteresis
loop was measured  by sweeping the external field to much larger
negative value, namely to -9000~Oe. Now one observes that the
hysteresis loop becomes more symmetric with respect to the
magnetization axis. The saturation field, where the hysteresis
loops closes, is about -8000~Oe.

The example above shows that  a vertical shift of magnetization
can be related to non a homogeneous state of the F layer due to
non-collinearities  at the AF/F interface. The relation to the
training effect is clearly seen as different coercive fields
H$_{c2}$ depending on the strength of the applied fields. When a
stronger field is applied in the negative direction, the exchange
bias decreases due to a larger degree of irreversible changes into
the AF layer.

\begin{figure}[!h]
\begin{center}
\includegraphics[clip=true,keepaspectratio=true,width=1\linewidth]{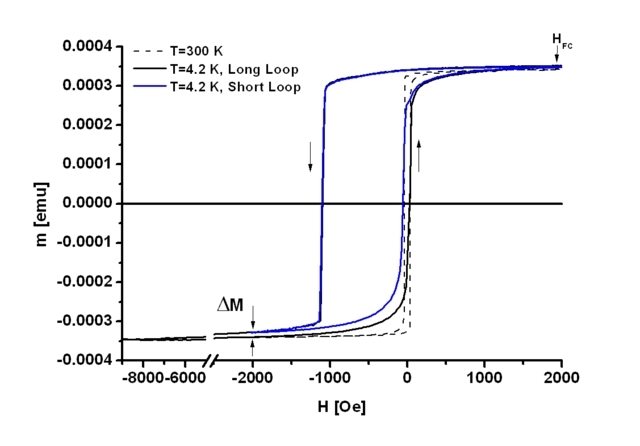}
\caption[Vertical Shift of the magnetization]{\label{vshit1}
Hysteresis loops of a CoO(25\AA)/Co(180\AA). The dashed line is
the hysteresis loop recorded at 300~K which is above the N\'eel
temperature of the CoO layer. After field cooling the system to
4.2~K, a hysteresis loop is recorded sweeping the applied field
from 9000 to -9000~Oe and back to 9000~Oe (black line, long loop).
A second loop is measured under the same conditions as the
previous one, but the field range is shorter, namely between
2000~Oe to -2000~Oe and back to 2000~Oe. The cooling field was
positive. A vertical shift of the hysteresis loop is clearly
visible for the short loop, whereas the long loop appears to be
centered  with respect to magnetization
axis.(Ref.~\cite{brems:2006:unpubl})}
\end{center}
\end{figure}

A distinct class of vertical shifts is found in exchange bias
systems where a diluted
antiferromagnet~\cite{nordblad:1995,keller:2002,nowak:2002,papusoi:2006}
acts as a pinning layer. As function of dilution  of a CoO layer
by Mg impurities, as well as, partial oxygen pressure during the
deposition, the AF layer acquires an excess magnetization seen as
a vertical shift of the hysteresis loop. This strongly supports
the fact that the domain state in the AF layer as well as the EB
effect is caused and controlled  the
defects~\cite{keller:2002,papusoi:2006}.

Vertical shifts in nanostructures and nanoparticles are often
observed due to uncompensated AF spins. For a detailed discussion
we refer to  recent reviews  by Nogues~\textit{et
al.}~\cite{nogues:2005} and Iglesias~\textit{et
al.}~\cite{iglesias:2006} and also recent papers on AF
nanoparticles~\cite{salabas:2006,mumtaz:2006}.


\section{Further evidence for Spin-Glass like behavior observed in finite size systems}

Although nanoparticles are not covered in this review, we,
nevertheless, discuss two recent instances which in AF
nanoparticles confirm the SG  behavior. One is Co$_3$O$_4$
nanowires~\cite{salabas:2006} and the other is CoO granular
structure~\cite{gruyters:2005}.

\subsection{AF Nanoparticles}
\begin{figure}[!ht]
\begin{center}
\includegraphics[clip=true,keepaspectratio=true,width=1\linewidth]{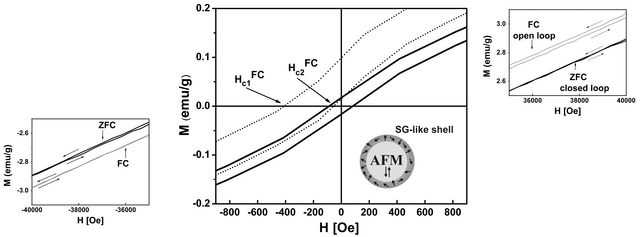}
\caption[Simulations of training effect within SG
model]{\label{nanoLo} Field cooled and zero field cooled
hysteresis loops of Co$_3$O$_4$ nanowires. From Ref.
~\cite{salabas:2006}}
\end{center}
\end{figure}

Nanoparticles of antiferromagnetic materials have been predicted
by N\'eel~\cite{neel:1962:afm} to have a small net magnetic moment
due to an unequal number of spins on the two sublattices as a
result of the finite size~\cite{kodama:1999}. Hysteresis loops of
AF nanoparticles have been observed and several suggestions were
made to account for their weak ferromagnetism~\cite{kodama:1999}.
One important finite size effect of AF magnetic nanoparticle is
the breaking of a large number of exchange bonds for surface
atoms. This can have a particularly strong effect on ionic
materials, since the exchange interactions are  superexchange
interactions. The deficit of exchange bonds could lead  to a spin
disordered shell exhibiting spin-glass like behavior. The last
effect is demonstrated most recently by Salabas~\textit{et
al.}~\cite{salabas:2006} and discussed further below.

In Fig.~\ref{nanoLo} two hysteresis loops of Co$_3$O$_4$
nanoparticles (8~nm diameter) prepared by a nanocasting route  are
shown~\cite{salabas:2006}. Both curves were measured at T~=~2~K
after cooling the system in zero field (ZFC hysteresis loop) and
in an external applied field of +4~T(FC hysteresis loop). Whereas
the ZFC loop shows  typical weak ferromagnetic properties specific
to AF particles, after FC  a completely different behavior is
observed. The hysteresis loop is exchange biased, it is vertically
shifted and shows training effect. Moreover, the temperature
dependence (not shown) of the coercive field and exchange bias
field increases with decreasing temperature, which is also a usual
behavior of  EB systems. More strikingly the FC hysteresis loop
does not close on the right side at positive fields. This open
hysteresis loop is a direct indication of a spin-glass like
behavior, similar to the loops observed in  pure spin glass
systems~\cite{binder:1986}. This behavior appears to support the
SG model(Fig.~\ref{figModel}), where a reduced anisotropy is
assumed to occur at the F/AF interface. Both frozen-in spins and
rotatable spins are directly seen  in the FC hysteresis loop.
Moreover, the irreversible changes of the surface AF spins are
causing the open loop.  A similar open loop is also predicted by
the DS model. Within the DS model the hysteresis loop of the AF
interface layer is not closed on the right hand side because the
DS magnetization is lost partly during F reversal due to a
rearrangement of the AF domain structure. The AF particles,
however, are supposed to be single AF domains, therefore
 irreversible changes are due to surface effects rather than  caused by  AF domain kinetics.
 Certainly, at very large cooling fields, the bulk
structure of the AF particle should be also affected. The
spin-glass like behavior was also recently observed for cobalt
ferrite nanoparticles~\cite{mumtaz:2006}, for (Mn,Fe)$_2$O$_{3-t}$
nanograins~\cite{passamani:2007}, and for a $\varepsilon$\-Fe$_3$N\-CrN nanocomposite system~\cite{gajbhiye:2007}. 

\subsection{Extended granular AF film}

\begin{figure}[!ht]
\begin{center}
\includegraphics[clip=true,keepaspectratio=true,width=1\linewidth]{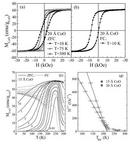}
\caption{\label{gruyters} a) ZFC  hysteresis loops of [CoO [x
\AA]/Au (60 \AA)] multilayers. b) FC hysteresis loop of a [CoO [20
\AA]/Au (60 \AA)] multilayer. c) ZFC and FC magnetization curves
as a function of increasing temperature T in different fields for
a [CoO [15 \AA]/Au (60 \AA)] multilayer. d) Field dependence of
T$_{irr}$ raised to the 2/3 power for two different
multilayers.(Ref.~\cite{gruyters:2005})}
\end{center}
\end{figure}

Another direct experimental  evidence of a spin-glass like
behavior was observed by Gruyters~\cite{gruyters:2005}  on CoO/Au
multilayers, CoO/Cu/Fe trilayers, and CoO/Fe bilayers with
granular structure. In Fig.~\ref{gruyters}a) the ZFC hysteresis
loop of a CoO/Au multilayer and for different temperature is
shown. One notices that the coercive field is enhanced at low
temperatures, but no hysteresis shift develops.  The FC hysteresis
loop (see Fig.~\ref{gruyters}b)), however, is almost completely
shifted to one side of the field axis. These observations are
explained by Gruyters as an effect of the uncompensated AF spins
of the granular film structure. The saturation magnetization of
this granular CoO film  is about 60-62 emu/g$_{CoO}$ which would
result in an enormous amount of uncompensated spins equivalent to
 22\% for the observed particles volume. No stoichiometry
influences are considered to contribute to this value. Although
the uncompensation level is unclear, the evidence of a spin-glass
behavior is well demonstrated. In Fig.~\ref{gruyters}c) ZFC and FC
magnetization curves are measured as function of temperature and
for different external fields. Two main characteristic features
are observed for these curves: an irreversibility temperature
T$_{irr}$, where the ZFC and FC branches of M$_{CoO}$(T) coalesce,
and a pronounced peak due to superparamagnetic blocking in the ZFC
magnetization. The direct evidence of a spin-glass behavior is the
field dependence of T$_{irr}$ shown in Fig.~\ref{gruyters}d).The
existence of critical lines spanned by the variables temperature
and magnetic field can be explained by mean-field theory. One of
these lines has been predicted by de Almeida and Thouless  for
Ising spin systems~\cite{gruyters:2005}. The  T$_{irr}$ raised to
2/3 power as function of field exhibits a linear dependence in
agreement with the predictions Almeida and Thouless.

\subsection{Dependence of the exchange bias field on lateral size of the AF domains}

The relation between the exchange bias and the reduced size
effects due nano-structuring of the AF-F systems is important from
both fundamental and technological points of view. From a
fundamental point of view, the reduced lateral size of both F and
AF objects induces significant changes of the exchange bias field,
coercive fields and also the asymmetry of the hysteresis
loops~\cite{fraune:2000,yu:2000,liu:2001,shen:2002,guo:2002,hoffman:2003,
girgis:2003,baltz:2004,scholl:2004,paul:2004,eisenmenger:2005,girgis:2005,temst:2005}.
In some systems an increased exchange bias field occurs for
reduced lateral sizes, whereas in other cases an opposite behavior
is reported, the exchange bias field decreases with decreasing the
lateral length scales~\cite{nogues:2005}. We refer in the
following to the last situation.

In Fig.~\ref{latstrct} are depicted schematically several lateral
systems which are commonly used to study the influences of the
nano-structuring onto the exchange bias properties. A reduced
lateral size of the ferromagnet (Fig.~\ref{latstrct}b) and
Fig.~\ref{latstrct}c) ) gives rise to additional shape
anisotropies for the ferromagnet leading to a change of both
coercive  and exchange bias fields as well as a change of the
hysteresis shape. In Fig.~\ref{latstrct}a) these additional
anisotropies are minimized and therefore the dependence of
exchange bias as function of the AF lateral size is more
transparent. For all three situations we can assume that at the
borders defined by the geometrical nanostructures, there is an
additional disorder extending to the interface. Even for the case
b) where the F is nano-structured we may expect that the
lithographic process will not always be stopped exactly at the
interface but also affecting the AF layer around the dot.

 Due to nano-structuring it is natural to expect that at the
edges of the dot there are AF spins with reduced anisotropy. These
spins will contribute to the coercivity at the expense of the
interfacial exchange energy. The effective interfacial exchange
energy can be written as: $J_{eb}^{eff}=F f J_{eb}$, where F is a
conversion factor related to  size effects, similar to the f
defined for the interface. It is easy to estimate the
$F$-parameter, as the fraction of the outer shell area divided by
the total dot area: $ F \approx A1/A2\approx (\pi (D-d)^2)/(\pi
D^2)=1-2d/D +d^2/D^2$, where  $d$ is the lateral thickness of the
outer shell and $D$ is the diameter of the dot itself. Assuming
that $d<<D$ and that $d$ is constant, we obtain the general
expression for the variation of the exchange bias field as D:
\begin{equation}
\frac{-H_{eb}+H_{eb}^{D\rightarrow \infty}}{H_{eb}^{D\rightarrow
\infty}}\approx 1-F \approx  \frac{1}{D}, \label{SGAFdo}
\end{equation}
where $H_{eb}^{D\rightarrow \infty}$ is the exchange bias of an
extended film. Within this simple model the exchange bias field
decreases as the lateral size of the AF structure decreases. This
1/D dependence is often observed
experimentally~\cite{nogues:2005}.

\begin{figure}[!ht]
\begin{center}
\includegraphics[clip=true,keepaspectratio=true,width=1\linewidth]{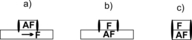}
\caption{A schematic view of different nanostructered systems used
to study the finite size effects on the exchange bias properties.}
\label{latstrct}
\end{center}
\end{figure}



The general behavior of the hysteresis loops as a function of the
$f$-parameter is depicted in Fig.~\ref{figSG1}. We see that for
high $f$ corresponding to a large diameter AF dot, the coercive
field is small and the exchange bias is high. When, however, the
lateral size of a AF object becomes smaller, the $f$ ratio
decreases leading to an increased coercive field and reduced
exchange bias field. Transition from the top hysteresis  to the
bottom hysteresis  of Fig.~\ref{figSG1} are usually observed due
to nanostructuring of the AF
layers~\cite{fraune:2000,baltz:2004,eisenmenger:2005,nogues:2005}.





\section{Conclusions}

The results presented in this chapter provides an overview of  the
physics of a F/(Interface)/AF exchange bias systems. Fundamental
properties of the unidirectional anisotropy are considered and
discussed. The Meiklejohn and Bean model as well as the Mauri
model are considered in detail and compared, both analytically and
numerically. The N\'eel approach, Malozemoff model, Domain State
model and a model of Kim and Stamps are shown as they provide
novel and fundamental ideas on the EB phenomenon.  The  Spin Glass
model is discussed in even more details. Further experimental
 results touching the fundamentals of exchange bias
are described.

We distinguish several outcomes of our overview:

\begin{itemize}
\item{ The exchange bias is an interface effect, as clearly proved
by the 1/t$_F$ dependence. Deviations from this law were observed
in the literature, but fundamentally this expression is clearly
well established.}

 \item{
The AF anisotropies in the bulk of the AF layer and at  the
interface to a  soft ferromagnetic layer  give rise to an
impressively rich behavior of the magnetic properties: the
hysteresis loops can be shifted along both field and magnetization
axis and in both positive and negative directions, the azimuthal
dependence of exchange bias exhibits non-intuitive behavior such
as a shift of its maximum with respect to the field cooling
orientation, the hysteresis loops are asymmetrically shaped, etc.
Some of the exchange bias characteristics mentioned above are  not
fully and consistently revealed experimentally which leaves an
open window for further quests.}

\item{The analytical formulae for the exchange bias within M\&B
model   depends mainly on  the properties of the AF layer, namely
on its anisotropy and thickness.}

\item{The peak-like increase of the EB close to the
critical thickness of the AF layer can be reproduced by the SG
model. This effect is also accounted for by the DS model.}

\item{Azimuthal dependence of exchange bias could help to distinguish
between two ideal mechanisms for exchange bias: Mauri versus M\&B
models.}

\item{ The SG model describes the conversion of  coupling into coercivity. }
\end{itemize}



{\Large{List of acronyms}}\\
\\
\\
\\
\makebox[30mm][l]{AMR} Anisotropic Magneto-Resistance\\
\makebox[30mm][l]{FC} Field Cooling\\
\makebox[30mm][l]{ZFC} Zero Field Cooling\\
\makebox[30mm][l]{EB} Exchange Bias\\
\makebox[30mm][l]{DS} Domain State Model\\
\makebox[30mm][l]{SG} Spin Glass Model\\
\makebox[30mm][l]{SW} Stoner-Wohlfarth Model\\
\makebox[30mm][l]{M\&B} Meiklejohn and Bean Model\\
\makebox[30mm][l]{MCA} Magneto-Crystalline Anisotropy\\
\makebox[30mm][l]{MOKE} Magneto-Optical Kerr Effect\\
\makebox[30mm][l]{PNR} Polarized Neutron Reflectivity\\
\makebox[30mm][l]{SQUID} Superconducting Quantum Interference Device\\
\makebox[30mm][l]{AF} Antiferromagnet\\
\makebox[30mm][l]{F} Ferromagnet\\
\makebox[30mm][l]{MFM} Magnetic Force Microscopy\\
\makebox[30mm][l]{MRAM}  Magnetic Random Access Memory\\
\makebox[30mm][l]{DAFF}  Diluted Antiferromagnets in an External Magnetic Field\\
\makebox[30mm][l]{XMCD} Soft X-ray Resonant Magnetic Dichroism\\
\makebox[30mm][l]{XRMS} Soft X-ray Resonant Magnetic Scattering\\


\section*{Acknowledgments}
We gratefully acknowledge support by the DFG
Sonderforschungsbereich 491: Magnetic Heterostructures: spin
structure and spin transport, and by BESSY.

We have benefitted from discussions with: Werner Keune, Wolfgang
Kleemann, Kurt Westerholt, Ulrich Nowak, Klaus D. Usadel,
Christian Binek, Kristiaan Temst, Ivan K. Schuller, Robert L.
Stamps.


 \end{document}